\documentstyle[12pt, aaspp4]{article}
 
\newcommand{\lsun}{log $L/L_{\odot}\,$}
\newcommand{\msun}{$M/M_{\odot}\,$}
\newcommand{\dydz}{$\Delta Y/\Delta Z\,$}

\begin{document}
 
\title{EVOLUTIONARY AND PULSATIONAL CONSTRAINTS FOR SUPER-METAL-RICH
STARS WITH Z=0.04}
 
\author{Giuseppe Bono}
\affil{Osservatorio Astronomico di Trieste, Via G.B. Tiepolo 11,
34131 Trieste, Italy; bono@oat.ts.astro.it}
 
\author{Filippina Caputo}
\affil{Osservatorio Astronomico di Capodimonte, Via Moiariello 16,
80131 Napoli, Italy; caputo@astrna.na.astro.it}
 
\author{Santi Cassisi\altaffilmark{1}}
\affil{Osservatorio Astronomico di Teramo, Via M. Maggini,
64100 Teramo, Italy; cassisi@astrte.te.astro.it}
 
\author{Vittorio Castellani\altaffilmark{2}}
\affil{Dipartimento di Fisica, Univ. di Pisa, Piazza Torricelli 2,
56100 Pisa, Italy; vittorio@astr1pi.difi.unipi.it}
 
\and
\author{Marcella Marconi}
\affil{Dipartimento di Fisica, Univ. di Pisa, Piazza Torricelli 2,
56100 Pisa, Italy; marcella@astr1pi.difi.unipi.it}
 
\altaffiltext{1}{Dipartimento di Fisica, Universit\`a de L'Aquila,
Via Vetoio, 67100 L'Aquila, Italy}
 
\altaffiltext{2}{Osservatorio Astronomico di Teramo, Via M. Maggini,
64100 Teramo, Italy}

\normalsize
\vspace*{5.0mm}
\begin{abstract}
 
\noindent
We investigate the evolutionary behavior of stellar structures 
with metallicity Z=0.04 in order to disclose theoretical
expectations for both evolutionary and pulsational behaviors 
of Super-Metal-Rich (SMR) objects, which are found in the solar
neighborhood, in the Galactic bulge and in elliptical galaxies.
A suitable set of stellar models is presented for the given
metallicity value but for two alternative assumptions about the amount
of original He, namely Y=0.34 and Y=0.37.
Theoretical isochrones for H burning evolutionary phases are
presented for ages ranging from 18 to less than 1 Gyr. The evolutionary
behavior of He burning structures is  discussed for
suitable assumptions about the mass of the progenitors and the amount of
mass loss. For both quoted assumptions of original He abundance we confirm
that at metal contents larger than the solar value the luminosity of the
Horizontal Branch (HB) at the RR Lyrae gap increases as the metal content
increases, a direct consequence of the expected simultaneous increase of
original He. We find that at the exhaustion of central helium SMR stars
definitely undergo the gravonuclear instabilities previously
found in some He burning structures with solar metallicity (Bono et
al. 1997b). 
 
\noindent
On the basis of such an evolutionary scenario, we investigate the expected
pulsational behavior of He burning SMR stars for suitable assumptions on
the pulsators evolutionary parameters.
Linear blue boundaries for pulsational instability in the
fundamental and in the first overtone modes are derived and their
dependence on stellar mass and chemical composition is investigated.
Nonlinear, nonlocal, and time-dependent convective models are
discussed, the modal stability is investigated for the first two modes, 
and the theoretical predictions about the period distribution inside the 
instability strip and the shape of both light and velocity curves are 
presented.
Full amplitude, nonlinear envelope models  show that the range of
effective temperatures where SMR RR Lyrae variables present a stable limit
cycle is smaller in comparison with pulsators characterized by lower metal
abundances. In fact, the width of the instability strip at the Zero Age 
Horizontal Branch (ZAHB) luminosity level decreases from 1400 K to 1100 K. 
Taking also into account the peculiar narrow mass range characterizing
SMR pulsators we estimate that these two factors alone cause a decrease in
the occurrence of RR Lyrae pulsators by a factor of
seven in comparison with metal-poor, globular cluster-like stellar
populations.
We find that canonical analytical relations connecting the nonlinear
periods of metal-poor variables to their luminosity, mass and effective
temperature cannot be safely extrapolated to the range of SMR pulsators.

\noindent
We show that gravonuclear instabilities largely increase the lifetimes of
stars crossing the instability strip at luminosity levels higher than the HB
luminosity, thus increasing the expected occurrence of luminous
low-mass variables.
We show that both periods and light curves of different groups of type II 
Cepheids with periods shorter than six days, presented by 
Diethelm (1983,1990), 
can be all reproduced by suitable variations in the effective temperature 
or in the luminosity level of our SMR post-HB models, supporting 
evidence of a substantial homogeneity of these variables. On the basis of 
both evolutionary and pulsation findings we finally
predict the rate of period change for a typical field, metal-rich type II
Cepheid across the instability strip and discuss an observational test for 
validating the present theoretical scenarios. In the Appendix we discuss 
in detail the physics of gravonuclear instabilities, which appear as
a surprisingly exact confirmation of the theoretical predictions given by 
Schwarzschild and Harm as early as 1965.
\end{abstract}
 
\noindent
{\em Subject headings:} galaxies: stellar content -- field stars -- stars:
evolution -- stars: horizontal branch -- stars: variables: RR Lyrae, 
Type II Cepheids

\pagebreak
\section{INTRODUCTION}
 
\noindent
Almost thirty years ago Spinrad \& Taylor (1969), discussing a large sample
of narrow-band spectral indices for field and cluster G and K stars,
identified a class of giants characterized by very strong line absorption
and by an excess of UBV blanketing and suggested that similar features are
the signature of an overabundance of metals up to 4 times larger than the
solar metal content.  Accordingly, they revealed the existence of a 
population of "Super-Metal-Rich" (SMR) stars, a name that we adopt 
here for all stars more metal-rich than the Sun.
Much more recently, Rich (1988) first disclosed that the Galactic
bulge population does contain a substantial fraction of SMR giants.
Sadler, Rich and Terndrup (1996),
discussed an impressive amount of spectroscopic observations of
K and M giants in the Baade's Window, finding that more than 50\% of their
sample was characterized by metal abundances in the range $[Fe/H] \sim
0.3-0.4$. By relying on Str\"omgren photometry,
Rocha-Pinto \& Maciel (1996) have recently obtained a quite
similar metallicity distribution for G dwarfs in the solar neighborhood.
Thus SMR stars appear to be a rather common stellar component of our Galaxy.
 
\noindent
Of course the occurrence of SMR stars is not restricted to our Galaxy.
As a matter of fact, the presence of SMR globular clusters in external
galaxies is becoming a well-established observational evidence.
Recent ground-based and HST integrated photometry 
of a large sample of globular clusters belonging either to a peculiar 
(Minniti et al. 1996) or to an elliptical
galaxy  (Geisler, Lee \& Kim 1996, hereinafter referred to as GLK)
has firmly brought out the existence of a radial metallicity gradient
between the inner and the outer regions of these galaxies,
with the tail of the metallicity distribution which may reach very large 
metal abundances ($[Fe/H]\approx 0.5-1.0$).
 
\noindent
According to this evidence, theoretical constraints on the evolutionary
behavior of SMR stars have received increasing attention. After some
pioneering papers (Torres Peimbert 1971; 
Caloi, Castellani \& Di Paolo 1974) the attention of theoreticians
has been mainly devoted to investigate the behavior of SMR
He burning stars in connection with the problem of UV excesses in elliptical
galaxies. The first exploratory papers on this matter (Brocato
et al. 1990; Castellani \& Tornamb\'e 1991) were followed by
more extensive investigations of HB structures by Horch, Demarque \&
Pinsonneault (1992, hereinafter referred to as HDP) and, more recently and
more detailed, by Dorman, Rood \& O'Connell (1993,
hereinafter referred to as DRO).
 
\noindent
However, SMR stellar tracks which cover not only helium burning
but also the previous H burning evolutionary phases are a fundamental
ingredient for approaching the problem of the age distribution in
both bulge and solar neighborhood stars and in turn for constraining the
formation history and the chemical evolution of the Galaxy. Moreover,
and maybe more relevantly, theoretical constrains concerning the
whole evolutionary history of SMR stars will disclose the evolutionary
scenario of similar stellar populations in external galaxies,
allowing for the evaluation of integrated colors of SMR globular
clusters and supplying useful clues for the spectral evolution
codes which provide the
integrated spectral energy distributions of multipopulation models
(Magris \& Bruzual 1993; Bressan et al. 1994; Weiss, Peletier \&
Matteucci 1995; Pozzetti, Bruzual \& Zamorani 1996).
 
\noindent
In this context, it is worth noting that detailed estimates of the
metallicities of extra galactic SMR globular clusters based on the
Washington index ($C-T_1$) may still present problems.
In fact, even though GLK evaluated an internal accuracy of
metallicity estimates roughly of the order of 0.2 dex, the
empirical metallicity calibration of the Washington index
relies on a sample of Galactic clusters with ($-2.25 < [Fe/H] <-0.25$).
Therefore extrapolations at much larger metallicities could be risky.
As an alternative approach, the problem of empirical calibrations could be
overcome by evaluating the predicted integrated colors of a template globular
cluster sequence by using evolutionary population synthesis models
(Brocato et al. 1990; Buzzoni 1995; Cellone \& Forte 1997) for
which full coverage of the main evolutionary phases is obviously
needed.
 
\noindent
In order to provide such a theoretical framework for the evolution of
SMR stars, in this paper we present an evolutionary investigation
performed at fixed metal abundance, namely Z=0.04.
In order to properly settle
the chemically homogeneous models on their ZAMS, a suitable
assumption concerning the ratio between the helium and the
heavy elements enrichment (\dydz) has to be provided as well.
The evaluations of the parameter \dydz based on observations of HII
regions present large uncertainties. On the basis of observations of
emission lines in HII galaxies, Pagel et al. (1992) obtained
\dydz $\approx 6$, whereas Maeder (1992)
by adopting a homogeneous and complete library of evolutionary models
suggested for this parameter a value between 1.1 and 1.8 for
metal-poor stars and a value between 1.5 and 2.2 for metal-rich stars.
However, as Peimbert \& Torres-Peimbert (1974) first claimed, the value
of the enrichment parameter should be cautiously treated, since at present
both theoretical and observational data seem to suggest the possible
lack of an universal linear regression between Y and Z (see also
Bertelli et al. 1996). Therefore the evaluation of \dydz seems to be
a really thorny problem. The reader interested in recent discussions
on this matter is referred
to Peimbert (1993), Carigi et al. (1995), Traat (1995), and Dorman,
O'Connell \& Rood (1995 hereinafter referred to as DOR).
 
\noindent
On the basis of the photometric database collected by the OGLE microlensing
survey in the Baade's Window, Renzini (1994) has recently discussed the
parameter $R_c$, i.e. the ratio between the number of HB clump stars and
Red Giant Branch (RGB) stars, and suggested on this basis that in bulge stars 
the helium abundance should range from Y=0.30 to Y=0.35. Similar
results were given by Terndrup (1988) and more recently by
Minniti (1994, 1995). However, it should be mentioned that similar
evaluations of the  enrichment parameter take into account the "mean
metallicity" of a  stellar sample, so that no firm evaluations can be
provided close to the tail of the metallicity distribution
and in particular at very high metallicities. According to this
contradictory situation, we choose Y=0.34 as a reasonable guess
for the original amount of He, which allows for a direct comparison
with previous SMR computations by DOR. A second alternative choice,
Y=0.37, has to be regarded as a temptative but again reasonable
upper limit for the amount of He, as produced by a value of \dydz
approximately of the order of 5.
 
\noindent
In \S 2 we present the physical and chemical assumptions adopted for
constructing the evolutionary tracks. Section 2.1 deals with
 H burning stellar models and with their comparison
with similar data already appeared in the literature. Selected results
of these H burning models, namely the He core mass at the He ignition
($M_{cHe}$) and the extra-helium ($\Delta{Y_{du}}$) dredged up at the
base of the RGB, allow for a proper evaluation of He burning models
presented in section 2.2, where we discuss the occurrence of
gravonuclear  instabilities affecting the phase of central helium exhaustion.
He burning models of SMR stars will be used for providing some fundamental
input parameters such as the mass/luminosity ratio for both linear and
nonlinear pulsation investigations which are presented in \S 3.
The modal behavior of SMR RR Lyrae pulsators at the ZAHB luminosity
level and the dependence of the width of the instability strip on the
metal content are presented.
 
\noindent
The pulsation properties and the modal behavior of envelope models
for post-HB stars are compared with the observed
scenario of field, metal-rich type II Cepheid stars (Diethelm 1990
and references therein) for which up to now there is no detailed
theoretical investigation available in the literature. The dependence of
the light curves of fundamental pulsators on effective temperature and
on the appearance of the Hertzsprung progression in this group of
variable stars are discussed in section 3.2.
A brief summary of the results obtained in this investigation and the
overall conclusions concerning both evolutionary and pulsational 
properties of SMR stellar populations are reviewed in \S 4.
 
\noindent
Finally, the physical parameters which govern the appearance of
"gravonuclear loops" -GNL- and a thorough analysis of the physical 
mechanisms which drive and/or inhibit the appearance of the gravonuclear 
instability are presented and discussed in the Appendix.
 
\section{THEORETICAL STELLAR MODELS}
 
\noindent
Theoretical stellar models
have been computed by adopting the FRANEC evolutionary code,
whose structure has been discussed in Chieffi \& Straniero (1989).
As for the input physics,  OPAL radiative opacity tables (Iglesias,
Rogers \& Wilson 1992; Rogers \& Iglesias 1992) were adopted
for temperatures higher than 10,000 K, while for lower temperatures were 
adopted the molecular opacities provided by Alexander \& Ferguson (1994).
Both high and low temperature opacity tables assume
a solar scaled heavy element distribution (Grevesse 1991).
The equation of state has been taken from Straniero (1988),
supplemented by a Saha EOS at lower temperatures. The outer boundary
conditions have been evaluated by assuming the $T(\tau)$ relation
provided by Krishna-Swamy (1966). In the superadiabatic region of the
stellar envelope a mixing length value of {\sl ml}=2.25 $H_p$ has been 
assumed, according to the calibration on the solar standard model
(Salaris \& Cassisi 1996).

\noindent
Moreover, near the core helium exhaustion we inhibited the so-called 
"breathing pulses" (Castellani et al. 1985) by adopting the procedure 
suggested by Caputo et al. (1989).
In this approach the onset of "breathing pulses" is avoided on the basis
of a straightforward assumption concerning the size of the convective
core. In other words, the extent of the convective core is checked so that
the central helium abundance cannot increase between consecutive models. 
 
\subsection{H BURNING EVOLUTIONARY PHASES}
 
\noindent
Evolutionary tracks for the H burning phase have been computed
for a fixed metal content (Z=0.04) and for the two quoted initial helium
abundances, namely Y=0.34 and 0.37. For each given chemical composition
calculations have been provided for a set of stellar masses ranging
from \msun =0.7 to \msun =3.0. Table 1 reports the masses of all computed
models and from left to right, for each mass, the luminosity of the model
igniting He, the lifetime in H burning phase (i.e. the time at the onset
of He burning) and the mass of the He core and the extra-helium brought
to the surface respectively. The last two parameters are the ingredients 
needed to evaluate the behavior of our models during the subsequent phase 
of He burning evolution.
 
\noindent
As it is well known, for each given stellar population (i.e. for each assumed
value of chemical composition and age) we can define a critical mass value,
i.e. the transition mass $M_{tr}$, as the upper mass limit of stars which
experience strong electron degeneracy into the He core during the phase
of H shell burning and which, in turn, ignite He through one or more
violent off-center He flashes at the tip of the RGB.
Around this value, in a mass range of only few tenths of solar mass, 
stars undergo a sudden variation in the maximum luminosity reached by 
H shell burning structures known as the {\sl Red Giant
Transition Phase} (RGTP, for a detailed discussion see Sweigart, Greggio \&
Renzini 1989, hereinafter referred to as SGR, Sweigart, Greggio \&
Renzini 1990; Castellani, Chieffi \& Straniero 1992).
Note that the RGTP also plays a key role in theoretical
studies of binary evolution (Tornamb\`e 1982) and in the age evaluation of
populous stellar clusters.
 
\noindent
The fine grid of evolutionary H burning sequences computed up to 
the helium ignition allows for a detailed investigation of the RGTP
for a SMR stellar population. Figure 1 shows the behavior of the He core
mass, of the luminosity and of the age at the He ignition as a function of the
total mass of the evolving star. We find that the RGTP occurs, for Y=0.34,
at an age of t$\approx7.95\times10^8$ yrs. At this age the stellar mass 
of the stars at the RGB tip is  $M_{tr}\approx2.1M_{\odot}$ and the He core 
mass is $M_{cHe}\approx0.40M_{\odot}$.
For the larger helium content, Y=0.37, we find $M_{tr}\approx2.0M_{\odot}$
with once again a He core mass at the RGB tip $M_{cHe}\approx0.40M_{\odot}$
and an age $\approx8.12\times10^8$ yrs. Figure 2 compares the present
evaluation
of the transition mass (Y=0.34) with previous evaluations obtained for
lower metallicities, as given by Cassisi \& Castellani (1993),
Cassisi, Castellani \& Castellani (1997), and Bono et al. (1997b,
hereinafter referred to as Paper I). Data plotted in this figure
confirm that $M_{tr}$ is not a monotonic function of the metal content.
For metallicity values larger than Z=0.01 the transition
mass decreases as the metal content increases, a feature to be connected
with the corresponding increase in He abundance which plays a major role
in decreasing the degree of electronic degeneracy inside the core of RG
stars, decreasing in turn the value of the transition mass
at a given metallicity. This occurrence provides, at the same time, a plain
explanation for the increase shown by the ZAHB luminosity level at high
metal contents (for a discussion see Paper I and {\em infra}).
 
\noindent
Present results about the transition mass can be usefully compared
with similar values provided
by SGR for the same metallicity but for different assumptions
on the initial helium abundance, namely Y=0.20 and Y=0.30.
SGR found for Y=0.20 a $M_{tr}\approx 2.75 M_{\odot}$ 
and a RGTP age  approximately equal to 540 Myr, whereas
for Y=0.30 and the same evolutionary parameters they give 
$M_{tr}\approx 2.33 M_{\odot}$ and an age $\approx$ 510 Myr, respectively. 
The large difference being
easily attributed to the different assumptions about the helium abundance.
As a result, we find that with our assumptions about the correlation
between He and metals, a SMR stellar population is expected to spend a
longer time interval before its integrated light
becomes red.

\noindent
Data in Table 1 show that the extra-helium
brought by the first dredge up into the stellar envelope
attains a minimum for  $M\approx2M_{\odot}$ and increases again
when the stellar mass is further increased. The behavior of the
extra-helium in the range of low mass stars and, in particular,
the sudden decrease of such an evolutionary parameter at the
RGTP have been recently discussed by Castellani \& Degl'Innocenti
(1995) (but see also Cassisi, Castellani \& Castellani 1997) and
they will be not discussed further. However, here is  
interesting to underline that with a further increase of the stellar mass
the amount of extra-helium starts increasing again as a consequence
of the deepening convective zone in structures whose H
burning evolution was already constantly dominated by CNO burning,
but in which external convection succeeds in dredging up the He formed
in the original ZAMS convective core.
 
\noindent
Evolutionary tracks covering H burning phases have been
used for computing H burning isochrones, which are presented
in Figure 3 for selected assumptions on the stellar age.
Data concerning these isochrones are reported in Table 2, in which,
from left to right, are listed the age (Gyr), the mass of the
stellar structure at the bluest point of the isochrone, and both luminosity
and effective temperature of this point. Columns (5) and (6) give
the $M_V$ magnitude and the $(B-V)$ color of the same point
respectively, evaluated by adopting bolometric corrections and
color-temperature relations provided by Kurucz (1992),
with $M_{Bol,\odot}=4.75$ mag.
 
\noindent
Figure 3 quite clearly shows the RG giant
clump marking the encounter of the H burning shell with the
chemical discontinuity produced in the stellar interior by the first
dredge up. This figure reveals a rather curious feature: a decrease
of the stellar mass implies, as expected (see for example Castellani,
Chieffi \& Norci 1989), a decrease of the bump luminosity which appears
to move in such a way that the effective temperature of the bump keeps
roughly constant over the whole range of the investigated isochrones.
A similar feature can be also detected in the isochrones for
Z=0.01 and Z=0.02 presented in Paper I. As a further point, we find that
the clump is barely notable in the age range from 2 to 4 Gyr. By recalling
that the clump is originated from the chemical discontinuity at the
bottom of the mixed envelope, we can easily relate such a behavior
to the already discussed occurrence of a minimum in the surface extra-helium,
since a lower extra-helium implies a reduced chemical discontinuity and,
in turn, a smaller evidence for the RG clump.

\subsection{HE BURNING EVOLUTIONARY PHASES}
 
\noindent
On the basis of the evolutionary values of both the He core mass, $M_{cHe}$,
and the surface He abundance at the RGB tip obtained by means of the H burning
stellar models discussed in the previous section, we now investigate the
evolution of He burning stars for selected assumptions about the RG
progenitor mass, i.e. about the age of the stellar population.
In the case Y=0.34 we choose RGB progenitors with masses
\msun=1.0,1.8,2.0 which correspond to ages (see Table 1) ranging from
11.8 Gyr to 985 Myr. However, data listed in Table 1 point out that
RG stars with masses smaller than 1.5$M_{\odot}$
have all electron degenerate cores, so that
the mass of the He core at the He ignition presents rather negligible
variations. As a consequence,  He burning models originated
from a 1.5 $M_{\odot}$ progenitor can be considered representative of 
He burning stars with ages larger than about 2.5 Gyr, provided that the 
variation of the progenitor mass with age is taken into account and the
small variations in the surface extra-helium are neglected (see Table 1).
 
\noindent
For each given progenitor mass, i.e. for each assumed value of the He core
mass at the He ignition, a suitable set of ZAHB
models  has been computed  for different assumptions about the mass of
the H rich stellar envelope (i.e. about the efficiency of mass loss during 
the RGB phase). Figure 4a shows the HR diagram location of these
ZAHB models, whereas Table 3 gives related quantities for
the computed models.
In the same Table we report the expected difference in V magnitude between
the magnitude of the ZAHB at the effective temperature typical of the RR Lyrae
instability strip (log $T_e=3.85$) and the bluest point of the isochrone.
 
\noindent
The computations have been extended up
to very hot (i.e. with thin stellar envelopes) ZAHB models
in order to investigate the evolutionary behavior of the expected progenitors
of hot AGB-manqu\'e and Early post-AGB stars (Greggio \& Renzini
1990; DRO) which, according to current suggestions (see e.g. DOR) are
regarded as the main sources of ultraviolet emission in
elliptical galaxies. Similar computations have been performed under the
alternative assumption Y=0.37.
Figure 4b shows the ZAHB models referred to the quoted helium abundance, 
while the leading physical quantities are listed in Table 3.
Figure 5 shows the comparison between
the two ZAHB sequences characterized by the same RG progenitor mass but
with alternative assumptions about the original He content. For a
discussion of the differential effect of the He abundance on the ZAHB
location the interested reader is referred to Caloi, Castellani
\& Tornamb\'e (1978) and to DRO.

\noindent
As already known, the ZAHB effective temperatures of metal-rich stars 
are strongly
dependent on the ratio between the mass of the envelope and the total
mass of the star: $q=M_{env}/M_{tot}$ (see also HDP and DRO). For SMR
stars we find that this
dependence is further enhanced, so that a variation  of only 0.02-0.03
solar masses is sufficient for shifting a ZAHB model from a low
effective temperature region ($T_e\approx5500$ K) to a very high
effective temperature region ($T_e\approx22,000-25,000$ K).
As a consequence, in a metal-rich stellar population
the number of stars in the intermediate range of effective
temperatures and thus the expected number of RR Lyrae variables
should be quite small, as already discussed in Paper I
(see also the discussion in \S 3.2). At the same time, if SMR stars are 
old enough (and therefore with low evolving masses), we expect the easy 
occurrence of Extremely Hot HB (EHHB) stars with their progeny 
AGB-manqu\'e stars.
 
\noindent
Since EHHB objects spend a relevant fraction
of their lifetime as very luminous blue stars, they are presently
considered (see e.g. Greggio \& Renzini 1990 and references therein;
DOR) the dominant source for the phenomenon of UV
upturn observed in elliptical galaxies (Code 1969) and in galactic bulges.
As a matter of fact, a high frequency of EHHB stars and of their progeny
in SMR stellar populations is expected  on twofold grounds, namely: i) the
increased helium abundance speeds up the evolution and therefore less
massive stars, for a given age, are evolving off the main
sequence;
ii) metal-rich stars should lose their mass more efficiently than
metal-poor ones (D'Cruz et al. 1996). In fact, at fixed stellar mass
and luminosity level the metal-rich stars present larger radii and
therefore lower gravities and lower escape velocities in comparison
with metal-poor structures. As a consequence, the classical Reimers's
(1975) relation and/or other empirical estimates (Goldberg 1979;
Nieuwenhuijzen \& de Jager 1990 and references therein) predict larger
mass loss rates for metal-rich stars.

\noindent
Present off-ZAHB computations have been extended either to the onset of 
thermal pulses for more massive models, or until the luminosity along 
the cooling sequence of white dwarfs decreases to \lsun $\approx-2.5$ 
for less massive models.
The evolutionary path in the HR diagram of these models during
their He burning evolution is reported in Figures 6 and 7 for the labeled
assumptions on the RG progenitor mass.
The main evolutionary properties of He burning stars with metallicities
equal and/or larger than the solar value have been already discussed
in several investigations (Brocato et al. 1990; Castellani \&
Tornamb\'e 1991; HDP; DRO; D'Cruz et al. 1996; Paper I) and
they not deserve detailed discussion here.
As expected, the computed set of models allow for a rather
detailed investigation of the various evolutionary phases which follow the
exhaustion of the central He burning. Three different cases are presented: 
i) Extremely Hot Horizontal Branch stars which do not reach the
Asymptotic Giant Branch (AGB) and evolve as AGB-manqu\'e stars;
ii) Post Early AGB stars, which leave the AGB before the onset of thermal
pulses, and iii)  {\em bona fide} AGB stars, i.e. stars
which are massive enough to experience the evolutionary phase of thermal
pulses along the AGB.
 
\noindent
For each given assumption about chemical composition and age, we report again
in Table 3 the value of the stellar mass ( $M^{AGB}$) separating
AGB-manqu\`e models from models approaching the AGB, together with
the value ($M^{TP}$) for the less massive stars
experiencing thermal pulses.
Thanks to the fine grid of stellar models, we estimate that the formal
uncertainty for both these
parameters is of the order of $0.003M_{\odot}$.
As expected, we confirm that in SMR stellar populations the maximum stellar mass
of AGB-manqu\'e stars is remarkably increased, an occurrence to be
related to the increased efficiency of the H burning shell during the
central He burning phase. As a consequence, a larger range of stellar
structures exhausts hydrogen during central He burning, being forced to
spend all their He burning phase near the He main sequence (i.e. at very
high effective temperatures) and therefore as hot UV sources.
 
\noindent
A sound comparison between our predictions on $M^{AGB}$ and the
results recently obtained by HDP for  the same
metallicity is made difficult by the different values adopted
by the quoted authors for the He abundance and the progenitor
mass. However, we can compare the value of $M^{AGB}=0.62M_{\odot}$
as given by HDP for Y=0.37, $M_{cHe}=0.455M_{\odot}$ with our value
$M^{AGB}= 0.477M_{\odot}$ obtained for Y=0.37 and $M_{cHe}= 0.457M_{\odot}$ 
as given by the $1.8M_{\odot}$ progenitor.
We attribute  the discrepancy  mainly  to the coarse
step in the mass value adopted by HDP, with only a minor influence of the
restricted difference in both the adopted chemical and physical inputs.
A more meaningful comparison can be made with a model by DRO, who give
for $M_{cHe}=0.454M_{\odot}$ and for $Y_{HB}=0.356$ the value
$M^{AGB}\approx0.480M_{\odot}$. Our model with $M_{cHe}=0.454M_{\odot}$,
$Y_{HB}=0.347$  predicts $M^{AGB}\approx0.473M_{\odot}$.
This remarkable agreement, with a difference of the order of only 0.007 solar
masses, can be regarded as an evidence of the accuracy achieved in the
investigation of the evolutionary properties of He burning stellar models.
 
\noindent
As discussed by DRO, a further relevant evolutionary parameter
is given by the maximum stellar mass $M^{TP}$ which skips the phase of
thermal pulses before leaving the AGB. By definition this parameter marks
the transition between stars which evolve as Post-Early
AGB and stars which climb up the AGB experiencing recursive
flashes in both H and He burning shells. The values of $M^{TP}$ listed
in Table 3 show that this parameter attains values of the order of
$0.54-0.55M_{\odot}$ for all models. Quite similar values have been
found by DRO even for quite different assumptions about RG progenitor masses
and initial helium abundances. We conclude that $M^{TP}$ appears fairly 
independent of evolutionary parameters other than the metal content.
 
\noindent
As a relevant point, Figures 6 and 7 show that after the exhaustion of 
central He many of our models undergo the gravonuclear instabilities  
we shortly discussed
in Paper I for a few "young" solar metallicity HB stars.
The evolutionary tracks plotted in the quoted figures show that 
gravonuclear instabilities are
a rather common feature of the off-ZAHB evolution of less massive SMR HB
models. As a matter of fact, for Y=0.34 we find that
all the models with mass lower than \msun $\approx 0.52\div 0.53$,
either AGB-manqu\`e or evolving toward the AGB, undergo this kind of
instability. Accordingly, gravonuclear instabilities appear for Y=0.37 
in all models less massive than \msun $\approx 0.51\div 0.52$.
This small difference in the limiting stellar mass which experience 
gravonuclear instabilities can be easily
connected with the parallel evidence that moving from Y=0.34 to Y=0.37
the He core masses decrease of approximately 0.01 $M_{\odot}$.
This evidence indicates that the mass of the stellar envelope plays a
key role in the onset of gravonuclear instabilities.
 
\noindent
Owing to the intrinsic relevance of this finding, the physics of 
gravonuclear instabilities 
will be discussed with some details in the Appendix to this paper.
Here we briefly note that, as a consequence of gravonuclear instabilities, 
many HB models experience GNL (see also Paper I), which slow down
the off-ZAHB evolutionary phases. As already discussed in Paper I, such
an occurrence obviously appears relevant for the evaluation of UV radiation
from hot He burning stars. However, both Figs. 6 and 7 show that
during the blueward and/or the redward excursions GNL also succeed in pushing
some HB models into the region of pulsation instability at luminosities
much larger than the original HB luminosity. The expected
observational relevance of such an occurrence is disclosed for a helium
abundance Y=0.34 in Figure 8, in which we report the time behavior of both
effective temperature and luminosity for a "cool" model with mass equal to
0.510\msun and $M_{cHe}= 0.454$\msun during the GNL phase.
 
For the sake of this discussion, let us
assume here the  interval $3.806<$log$T_e<3.690$ (see below) as the region
where stars experience pulsational instabilities. As shown in this figure,
we find that the model along its "natural" trajectory toward the AGB
spends about 1.6 $10^5$ years in that region. Taking into account that
for such model the He central burning lifetime is about 1.6 $10^8$ yrs, this
occurrence implies that only 1 star out of about 1000 stars on the HB should
be expected to be a pulsator.
However, due to GNL, the star will spend further 2 $10^6$ yrs crossing
the instability region, and the probability to detect a pulsator
increases by more than one order of magnitude. According to such
evidence, in the next section the expected pulsational behavior of these
luminous post-HB pulsators will be investigated.

\section{SMR VARIABLE STARS}
 
\noindent
According to theoretical predictions we foresee that SMR He burning stars 
could suffer
pulsational instabilities either on their major phase of central helium
burning or while crossing the instability strip above the ZAHB luminosity.
In terms of a well established
pulsational scenario, this suggests the occurrence of RR Lyrae and/or
type II Cepheids produced by HB and/or post-HB stars, respectively.
Curiously enough, current observational surveys of variable stars give scarce
evidence, if any, of SMR RR Lyrae stars either in the Galactic field
(Suntzeff et al. 1991; Layden 1995 and references therein) or in the solar
neighborhood (Preston 1959; Kemper 1982). Most interesting, RR Lyrae stars
in the Baade's Window present a narrow metallicity distribution with a
mean value close to [Fe/H]=-1 (Walker \& Terndrup 1991), whereas the K
giants have a metallicity distribution which ranges from [Fe/H]$\approx$-1
to 0.5 (Terndrup, Sadler \& Rich 1995 and references therein).
 
\noindent
On the other hand, Harris \& Wallerstein (1984) and Diethelm (1986)
have recently
reported  the evidence for the occurrence of SMR type II Cepheids.
Taken at its face value, this observational evidence appears as a rather
strange occurrence. In fact, HB evolutionary times are in all cases much
longer than post-HB times and therefore {\em prima facie} we should
expect a larger evidence for RR Lyrae pulsators. We expect that the
small range of masses producing RR Lyrae plus the occurrence of GNL should
play a role in this matter. However, we will show in the following that there
are further intrinsic pulsational reasons which reduce the probability
for the occurrence of SMR RR Lyrae variables.
 
\noindent
Several thorough analyses have been already devoted to type II Cepheids
belonging to Galactic globular clusters (Wallerstein \& Cox 1984; Harris 1985;
Sandage, Diethelm \& Tammann 1994 hereinafter referred to as SDT;
Bono, Caputo \& Santolamazza 1997 hereinafter referred to as BCS).
Under this term are commonly classified old,
low-mass variable stars with periods ranging from 1 to roughly 30 days.
On the basis of their evolutionary properties, type II Cepheids
are often divided into two different groups. The short-period group,
called BL Herculis stars, is characterized by periods between 1 and
approximately 10 days. Their progenitors are low-mass HB stars which,
during their evolution, cross the instability strip moving from the blue
tail of the HB toward the AGB. The long-period group, called W Virginis
stars, is characterized by periods longer than 10 days with low-mass AGB
progenitors which experience one or more loops inside the instability strip
when starting their AGB evolution. As a whole, the evolutionary and 
pulsational properties for type II Cepheids in globular clusters emerging 
from the observational scenario appear in fairly
satisfactory agreement with current theoretical predictions
(see SDT, BCS and references therein).
 
\noindent
The scenario for type II Cepheids in the field appears
much less clear. The distinction between classical Cepheids and
type II Cepheids is -first of all- a thorny
problem since these two groups of variables present similar
periods and effective temperatures. A detailed discussion concerning
the limits of both photometric and spectroscopic classifications has
been given by  Harris (1985) and Diethelm (1986, 1990). According
to the accepted evidence, field type II Cepheids are found over a very
wide range of metallicities, from metal-poor up to SMR stars, with
the period distribution for metal-rich type II Cepheids which
differs from the distribution shown by similar variables in globular
clusters. This feature, together with several observed properties of
field, old-disk type II Cepheids still lacks a comprehensive and
homogeneous explanation, indicating that moving from low to high metal
contents the theoretical scenario will face some intrinsic differences.
 
\noindent
In order to investigate on a quantitative basis the pulsational behavior 
of the evolutionary models presented in the previous sections,
we computed a large set of both linear and nonlinear pulsation models
for suitable assumptions about the stellar mass and the effective temperature
of the stars. The results of these
calculations together with some simple clues concerning the evolutionary
status of these variable stars will be discussed in the following subsections.

\subsection{LINEAR NONADIABATIC MODELS}
 
\noindent
During the last fifteen years several theoretical investigations on the
pulsation properties of BL Her stars have been presented
(Carson \& Stothers 1982; Hodson, Cox \& King 1982;
Buchler \& Buchler 1994; Bono, Castellani \& Stellingwerf 1995 and
references therein). They appear in good agreement with observational
constraints concerning
the modal stability as well as
secondary features of both luminosity and velocity curves.
However, the quoted surveys were mainly devoted to type II Cepheids in
Galactic globular clusters with metal-poor chemical
compositions. Up to now no investigation on the dependence
of the pulsational behavior of BL Her stars on metal content and
stellar masses has been provided. In order to approach the problem
we investigate the pulsational behavior of He burning SMR variables at 
fixed chemical
composition (Z=0.04, Y=0.34) and by assuming three suitable stellar mass
values, namely \msun =0.520, 0.505 and 0.485, chosen as representative
of He burning stars populating the instability strip either during
their HB or their post-HB phases.
 
\noindent
The linear modal stability was investigated by exploring a wide range
of effective temperatures at three different luminosity
levels (\lsun =1.56, 2.0, 2.2), again chosen in such a way to cover theoretical
expectations concerning HB and post-HB SMR variable stars.
In order to provide a detailed analysis of the pulsation properties of SMR
RR Lyrae variables we also computed, for \msun =0.505, a sequence of 
envelope models located at \lsun =1.48.  
Models located close to the blue boundary of the instability strip were
computed with a temperature step of 100 K in order to reach
a sufficiently detailed information on the effective temperature of the 
linear blue edge. As it is well known, no information on the red edge can be
obtained from linear radiative computations. Both linear and nonlinear 
calculations
have been performed by assuming the same set of radiative opacity tables
adopted in the evolutionary calculations, even though the method adopted
for handling the tables is different (for more details
see Bono, Incerpi \& Marconi 1996 hereinafter referred to as BIM).
According to the prescriptions suggested by Bono \& Stellingwerf (1994
hereinafter referred to as BS),
for each envelope model the number of zones is of the order of 200.
The boundary conditions -i.e. the optical depth at the outer
boundary and the depth of the envelope- were chosen as in previous
investigations (Bono et al. 1997c hereinafter referred to as BCCM).
 
\noindent
Tables 4, 5, and 6 summarize the results of the linear nonadiabatic survey
for the three different assumptions on the stellar mass value and for the
labeled values of the luminosities. Columns (1) and (2) report the surface
effective temperature (K) and the static gravity, whereas columns (3),
(4) and (5)  list, in the order,  the period, the pulsation constant, and the growth
rate of the fundamental mode. The growth rate - i.e. the fractional amount
of energy added or subtracted to the pulsation in each period- is defined
in such a way that negative values imply the pulsational stability of
the stellar envelope for the given mode. In the same Tables,
columns (6), (7) and (8) report the same quantities listed in the previous
three columns, but referred to the first overtone mode.
No attempt has been made for evaluating the modal stability of the second
overtone since at large luminosities it is expected to be
stable on the whole instability strip (Bono et al. 1997a).
 
\noindent
As a first result, we find that models at luminosity
levels close to \lsun=2.0  reproduce with fairly good accuracy
the effective temperatures, gravities and periods observed
in field BL Her stars (Diethelm 1990). By a quick inspection of Figures 
6 and 7 we note that the quoted luminosity level falls inside the 
luminosity interval where GNL tend to accumulate stars. 
On the contrary, in the case of a quiescent
evolution such a luminosity would be attained only by hot HB stars in their
final and rapid crossing of the instability strip (see Figure 3 in DRO).
As a result, we find that the occurrence of GNL which push a larger amount of
post-HB models inside the instability strip at large luminosity levels
allows for a simple and homogeneous explanation of this group of variable
stars. If this is the case, at the mean time we would also understand why field,
metal-rich type II Cepheids present a larger period spread in comparison
with the cluster type II Cepheids (Harris 1985), the instability
strip of the former variables being populated by HB models characterized by
a larger range of stellar masses.
 
\noindent
Figure 9 shows the location of the blue boundaries of the instability
strip in the HR diagram for the three different assumptions on the
stellar masses, disclosing
that in this mass range the modal stability of both fundamental and first
overtone presents only a negligible dependence on stellar mass.
As a consequence, the luminosity  of the "intersection point", i.e. of
the point where the First Overtone Blue Edge (FOBE) intersects the Fundamental
Blue Edge (FBE) moves only from \lsun =2.0 for the models with \msun =0.485
to \lsun =2.05 for the models with \msun =0.505.
As it is well known, only fundamental pulsators are allowed above the
quoted luminosities. On this basis, it has been found (see BCS) that in
galactic globular clusters blue ZAHB models  evolving across the
instability strip should scarcely produce  first overtone, metal-poor
type II Cepheids (see BCS). On the contrary, we find that the location
inside the HR diagram of SMR post-HB models during the phase of GNL should
allow for the occurrence of both fundamental and first overtone pulsators.
 
\noindent
The dependence of the blue boundaries on the chemical composition
has been studied by computing selected linear models with
a given  mass  (\msun =0.505) and for two different "metal-rich"
compositions, namely Z=0.02, Y=0.28, and Z=0.04, Y=0.34.
However the investigation has been extended to models representative
of metal-poor, globular cluster type II Cepheids, by investigating the case
\msun =0.58, Y=0.24, Z=0.001. The blue instability boundaries
for these three assumptions are given in Figure 10. We find that
moving from solar to super solar chemical compositions the increase in
helium
and metal content affects only marginally the location of the FBE
at high luminosity levels, whereas  FOBE moves toward large effective
temperatures. Moreover, we find that the blue boundaries of metal-poor 
pulsators, and in particular the FBE, move toward hotter
effective temperatures and hence toward larger gravities.
Even tough the modal stability can be safely evaluated only through
a nonlinear approach, the above linear analysis provides useful
insights into the expected behavior of the dependence of blue edges on
physical parameters. In particular, we find that the behavior depicted
in Figure 10 would provide a sound explanation for the observational
evidence, originally suggested by Diethelm (1990), according to which 
both temperatures and gravities of type II Cepheids decrease as soon as 
the metallicity increases.
 
\noindent
The quoted linear computations allow for the analysis of the dependence
of the modal stability of metal-rich pulsators on the chemical composition.
Figure 11 shows the linear total work curves ($W$ is the work integral
per logarithmic temperature in units of the total kinetic energy) as a
function of the
logarithmic temperature for two models characterized by the same
luminosity (\lsun =2.0), temperature (6400 K) and stellar mass
(\msun =0.505) but with different chemical compositions, namely Y=0.28,
Z=0.02 and Y=0.34, Z=0.04. The arrows mark the location of the driving
regions, whereas the symbols indicate, according to the linear
eigenfunctions, the location of the first overtone nodes.
Note that Figure 11 displays $d W / d (log T)$ rather than the canonical
$d W / d (log Me)$ ($Me$ = exterior mass). Similar plots
provide useful information concerning the sources which drive (positive
values) or quench (negative values) the pulsation instability since
they can be closely
connected with the opacity features. At the same time this approach
maintains the relative areas of both driving and quenching regions under
the total work curve (Stellingwerf 1979).
 
\noindent
As expected on the basis
of the topology of the instability strip, the driving regions of the
fundamental mode, caused by the Hydrogen Ionization Region
(HIR, $T\approx1.3\times 10^4$ K) and the Helium Ionization Region (HEIR, 
$T\approx5.0\times 10^4$ K) present only a negligible change moving from
Y=0.28 to Y=0.34. The additional driving region located at
$T\approx2.0\times 10^5$ K is caused by the iron opacity bump.
The SMR model presents in this
region a slight enhancement in comparison with the model at solar
composition, but its contribution to the total driving is negligible.
In envelope models characterized by larger stellar masses this driving
region becomes more effective since the iron bump shifts toward the
nonadiabatic region of the stellar envelope and therefore can supply
a larger amount of net positive work.
 
\noindent
The total work curves referred to the first overtone disclose
a stronger dependence on the chemical composition and indeed the driving
region caused by the HEIR is significantly larger in the SMR model than
in the model with solar composition. In the latter model the driving region
caused by the iron bump disappears almost completely since it is located
quite close to the node of the luminosity eigenfunction and therefore
an obvious reduction of the amount of driving takes place.
A detailed discussion of the effects caused by this feature on modal
stability of RR Lyrae and $\delta$ Scuti stars is given in BIM and
Bono et al. (1997a), respectively.
The strong dependence of the first overtone instability on the helium
content explains why an increase of metals and helium moves the FOBE
toward higher effective temperatures.

\subsection{NONLINEAR LIMITING AMPLITUDE MODELS}
 
\noindent
To further investigate the behavior of SMR variables and, in particular, 
to attain information on both the red edge of the instability and the 
pulsational amplitudes, nonlinear,
nonlocal and time-dependent convective models have been computed by
assuming the same input physics adopted in
the linear analysis. The envelope models were integrated in time until
the radial displacements approach their asymptotic behavior, i.e. the
nonlinear limit cycle stability, and the pulsation amplitudes attain
a periodic similarity of the order of or lower than $10^{-4}$ over two
consecutive full pulsation cycles.
The initial velocity profiles were imposed by perturbing the linear
radial eigenfunctions with a constant velocity amplitude ranging
from 10 to 20 km$s^{-1}$.
 
\noindent
We first explored the luminosity level of SMR HB stars by computing
envelope models for \msun=0.505, \lsun=1.48 and for selected choices
of stellar effective temperatures. As a relevant result, we found that the
fundamental red edge slightly moves toward higher effective temperatures,
whereas the first overtone blue edge moves in the opposite direction in
comparison with RR Lyrae variables belonging to globular clusters.
As a matter of fact, the fundamental red edge referred to SMR RR Lyrae
variables is located at $T_e\sim$ 5950 K, whereas the first overtone blue
edge is located at $T_e\sim$ 7050 K. As a whole we find that the instability
strip in the SMR case covers a smaller temperature range in comparison with
pulsators characterized by lower metal contents, since
from the canonical value of $\Delta T_e\sim$1400 K
(see BCCM and Bono et al. 1997d) the width of the instability strip narrows 
down to $\Delta T_e\sim$1100 K. Such a result adds of course a further
theoretical motivation to the observed paucity of SMR RR Lyrae stars, since
the width of the instability strip remains almost constant from Z=0.0001
(Y=0.24) up to the solar chemical composition (Z=0.02, Y=0.28), whereas an
increase of a factor of two of the metallicity (Z=0.04) together with a
decrease of the order of 10\% in the H abundance causes a narrowing of the
intability strip of the order of 20\%.
 
\noindent
An estimate of the combined effects of metallicity on the
occurrence of RR Lyrae pulsators has been obtained by distributing
single models with a mass step $\Delta{M}=0.005M_{\odot}$ all over the 
ZAHB and evaluating the ratio between the sum of evolutionary times of
models falling within the instability strip and the sum of evolutionary
times for all ZAHB models (as far as the evolutionary time is concerned, we 
have taken into account the total central He burning lifetime). Figure 12 
shows the ratio $\tau(RR Lyrae)/\tau(HB)$ as a function of the metal content.
Data plotted in this figure show
that moving from metal-poor to metal-rich stellar populations
the number of RR Lyrae variables, due to the narrowing of the mass range
which populates the instability strip and the narrowing of the instability
edges, decreases of a factor of seven.
An occurrence which could play a not negligible role in the observed
lack of metal-rich variables in the Galactic bulge (Walker \& Terndrup 1991).
 
\noindent
On theoretical grounds, as a final remark, one can notice that
the origin of the narrowing of the instability
strip can be found in the prediction already given in BS and BIM concerning
the dependence of the boundaries of the RR Lyrae instability strip on the
chemical composition and in particular the dependence of the fundamental
red edge on the H abundance. If this is the
case, we can predict that a further increase in the helium content over
the adopted value Y=0.34 would cause a further narrowing of the strip and
therefore a further decrease of the probability to detect HB pulsators.
As for "theoretical" SMR RR Lyrae variables we report in Table 7 selected 
quantities depicting the pulsational behavior. We find that the pulsation
amplitudes and the shape of both light and velocity curves do not present
any peculiar feature in comparison with canonical RR Lyrae pulsators and
therefore they are not further discussed. Detailed data are available
upon request.
 
\noindent
To explore the pulsational behavior of our theoretical BL Her candidates, 
a sequence of nonlinear fundamental models was computed by exploring at 
fixed stellar mass (\msun=0.505), luminosity level (\lsun= 2.0), and chemical 
composition (Y=0.34, Z=0.04) a wide range of effective temperatures. A fine 
temperature step ($\Delta T_e \leq 200$ K) has been adopted to investigate 
the appearance of secondary features like bumps on light and velocity curves.
Few selected models have been also computed at a larger luminosity level 
(\lsun=2.2) to explore the dependence of both light and velocity curves
on this physical parameter. Selected results
of the nonlinear survey of fundamental models located at \lsun=2.0
and 2.2 are also reported in Table 7.
The pulsation amplitudes reported in this table are only for
pulsators which show a pure single nonlinear limit cycle stability.
The only exception is the model located at $T_e$=5700 K, \lsun= 2.0 which
at full amplitude presents a  permanent mixture of fundamental and
higher overtone(s) (mixed-mode pulsator).
Columns (1) and (2) report the effective temperature (K) and the
nonlinear period (days). The comparison between nonlinear periods and the
periods listed in Table 5 shows that nonlinear effects only marginally
affect this parameter. Columns (3), (4), and (5) list
the fractional radial oscillation, the surface velocity amplitude,
and the bolometric amplitude, respectively. Columns (6) and (7) present
the static and the effective surface gravities, whereas columns (8) and
(9) report the variation of surface and effective temperature throughout
a full pulsation cycle.
The overall trend of the pulsation amplitudes of BL Her fundamental
pulsators present a peculiar feature.  Unlike fundamental RR Lyrae
variables the amplitudes are not strictly decreasing when moving from
the blue to the red edge of the instability region.
A relevant consequence of this new behavior will be discussed further on.
 
\noindent
Figures 13 and 14 show the light curves at limiting amplitude for the sequence
of fundamental pulsators computed at fixed luminosity level. A quick
inspection of these figures shows that moving from the hot to the cool 
edge of the instability strip the shape of the bolometric light curves 
presents remarkable differences worth being discussed in detail.
We find that the light curve of the model located close
to the FBE ($T_e$=6400 K) shows almost a smooth sinusoidal luminosity
variation throughout the pulsation cycle, whereas the light curves
of  models located at lower effective temperatures show the occurrence of
bumps both before and after the main luminosity maximum.
Moving from $T_e$=6300 K to $T_e$=6100 K
the bump located at the base of the rising branch becomes more and
more evident. At the same time, a second bump appears along the
decreasing branch which moves toward earlier pulsation phases as
the temperature decreases. 

\noindent
The light curve of the models located
around $T_e$=5900 K presents even a different shape, with the
luminosity that after the main maximum is strictly decreasing, 
and the disappearance of the second bump. Interesting enough, the model
located at $T_e$ =5700 K is a mixed-mode pulsator since the pulsation
amplitudes of this model -followed over 3000 periods- fluctuate around
a mean value over consecutive periods.
The light curves of  models located at even lower effective
temperatures change once again: the second bump now takes place
before the luminosity maximum and becomes the absolute luminosity maximum
at effective temperatures lower than 5500 K. In this temperature
interval the first bump becomes less and less evident, and almost disappears
for temperatures lower than 5200 K. Finally, the light curves of the models
located between this temperature and the red edge of the instability strip
are characterized by a smooth decreasing branch and by a small bump located
between the flat "true" luminosity maximum and the phase of minimum
radius.
 
\noindent
This analysis of the dependence of the light curves on effective 
temperatures brings out some relevant results we will discuss in detail:
 
\noindent
i) the varying shape of the light curve  across the instability
strip provides the theoretical evidence that, as originally
suggested by Diethelm (1990), variable stars classified in the
{\em General Catalog of Variable Stars} (GCVS, Kholopov et al. 1981
and references therein) as CW and BL Her may belong to the same
family of radial pulsators. The two quoted classes have been indeed
separated on the basis of the bump location, and now we find that,
decreasing  the star temperature, the bump appears  first along the
rising branch and then on the decreasing branch of the light curve.
Due to the incomplete coverage of the possible pulsators parameters
presented in this exploratory paper, we do not wish to push
such a comparison too far. However, we note that bolometric
curves in Figures 13 and 14 do show a remarkable similarity with
the photoelectric V light curves reported in figures 2 and 3 of
Diethelm (1983).
 
\noindent
ii) The regular light curves of models located close to the red edge of the
instability strip ($T_e \leq 5200$ K) supplies the intriguing
suggestion that some of the variable stars classified as "Classical
Cepheids", (C$\delta$) in GCVS, when characterized by
relatively short periods (2-6 days) could have been misclassified.
As a matter of fact, both the range of periods and the shapes of the light
curves reported in figure 4 of Diethelm (1983) appear in fairly good
agreement with theoretical results for the reddest pulsators.
If this is the case, similar radial pulsators would belong to the
group of low-mass  metal-rich type II Cepheids instead of belonging
to the group of intermediate-mass classical Cepheids.
Note that a similar suggestion has been already given by Petersen
(1981) in discussing  the light curves of classical
Cepheids originally collected by Payne-Gaposchkin (1961).
 
\noindent
iii) Finally we also note that the sample of "Sinusoidal Cepheids"
(SA in the nomenclature of the GCVS) provided by Connolly (1980) and
Diethelm (1990) presents  periods and light curves which appear in
close agreement with the models located near the blue or the red edge
of our BL Her pulsating models. Thus once again we could speculate
about the possible misclassification of at least a fraction of variable
stars belonging to this group.
Unfortunately, here as well as in the previous cases, observational
light curves are characterized by low photometric accuracy and poor time
resolution to allow for a detailed comparison with theoretical predictions.
Even though several methods have been proposed for discriminating
among the pulsational properties of short period Cepheids
(Simon 1986; Morgan 1995), a general
and meaningful classification scheme of these objects needs further
spectroscopic and photometric data which could cast light on their
pulsational behavior.
 
\noindent
The regular progression of the bump disclosed by the light curves of
BL Her stars across the instability strip presents a close
similarity to the well known Hertzsprung (1926) progression of classical
Cepheids. This is the first time such a phenomenon appears
in  full amplitude nonlinear pulsation models. We suggest that this
finding could be connected with the discussed theoretical
evidence according to which the bolometric amplitudes of fundamental BL Her
pulsators attain an absolute maximum right at the period
($\approx 1.7$ d) in which the second bump is in phase with the "true"
luminosity maximum. The physical mechanisms governing the appearance of
such a phenomenon and a detailed analysis of the region inside
the instability strip where it appears will be addressed in a forthcoming
paper (Bono et al. 1997e).
 
\noindent
The appearance of the Hertzsprung progression among type II Cepheids
supports the theoretical prediction by Christy (1970) for a
period-phase-of-bump relation in this group of pulsators, as well as
the results presented  by Stobie (1973) for a small sample of
halo/old disk Cepheids. More recently Petersen (1981) investigated the
linear period limits, predicting a bump progression closely connected
with the results of our nonlinear theoretical scenario. Finally we note
that our mixed-mode pulsator located at $T_e$=5700 K presents a linear
period ratio P(FO)/P(F) = 0.687, smaller than the ratios observed for
TU Cas and U TrA
(P(FO)/P(F)=0.710), the only presently known mixed-mode
pulsators belonging to the group of classical Cepheids with
fundamental periods shorter than three days (Hoffmeister, Richter \& Wenzel
1985). The agreement could be possibly achieved by tuning the
luminosity and/or the mass of the pulsators.
 
\noindent
Figures 15 and 16 show the radial velocity curves of the sequence of 
fundamental pulsators presented in Figs. 13 and 14. An interesting 
feature is that the BL Her pulsators
which present the second bump along the decreasing branch of
the light curve {\em attain the luminosity maximum almost at the
same phase of the velocity maximum}, i.e. luminosity and velocity
variations are correlated throughout the pulsation cycle.
On the contrary, models located at effective temperatures
lower than 5600 K which have the second bump along the increasing branch
of the light curve {\em attain the "true" luminosity maximum {\em before}
the phase of velocity maximum}. In particular,  models located between
5400 K and 5200 K reach the absolute luminosity maximum at the phase of
minimum radius, whereas for cooler models this maximum takes place
{\em after} the phase of minimum radius and {\em before} the phase of
maximum outward excursion. All these features suggest that the
Hertzsprung progression in BL Her stars should be tightly connected
with a substantial variation in the physical properties of the outermost
layers which cause in turn a variation in both the thermal and the
dynamical timescales of these pulsators.
 
\noindent
The dependence of the light curve on the assumed luminosity level
has been investigated by constructing two models at \lsun =2.2
and for two given effective temperatures, namely 5700 K and
5500 K. Figure 17 shows both the light (top panels) and the velocity
(bottom panels) curves for these models. Comparison with Figures 
13 and 14 shows a large similarity with less luminous models. 
As a main difference, we find that the first bump precedes
the rising branch, whereas the same feature appears along the rising branch
of models located at lower luminosities. However, this result appears
in agreement with the discussion given in BCCM, where we found that
a small increase in the luminosity level of RR Lyrae pulsators
involves almost a rigid shift of the pulsation behavior toward higher
effective temperatures, i.e. the same shape of the light curve can be
found at higher luminosity but in hotter models. If this is the case,
now we expect that while the luminosity of  BL Her stars increases, the
Hertzsprung progression is only shifted toward larger temperatures.

\noindent
As already discussed, on the basis of linear evaluations (and thanks to the
occurrence of GNL) we expect the occurrence of BL Her first overtone pulsators.
In order to verify the occurrence of a stable FO limit cycle at the same
luminosity level of F pulsators, we computed three
first overtone models located at $T_e$=6300, 6000, and 5700 K
respectively. However, the radial motions of these models, soon after an initial
transient phase, during which slow low-amplitude modes appear,
undergo a sudden transition toward the fundamental mode. As soon as
the mode switching takes place the oscillations rapidly approach pure
fundamental periodic motions and eventually the nonlinear limit cycle
stability. Figure 18 shows for a first overtone model the changes in
period (top panel), bolometric amplitude (middle panel), and total
kinetic energy (bottom panel) as a function of the integration time.
The time behavior of these quantities shows quite clearly both the mode
switching shown by the steplike adjustment of the period at $t=0.3$ yrs
($P(FO)\,=\,1.2354 d\; \Rightarrow \;P(F)\,=\,1.7952 d$), and
the approach of dynamical motions to the fundamental asymptotic amplitude
(see the arrow plotted in the bottom panel).
At least for this luminosity level we conclude that predictions
of linear computations are not supported by the nonlinear approach,
and that only the fundamental mode is a stable attractor of the system.
However, the occurrence of FO pulsators cannot be ruled out at luminosity
levels lower than the explored one.

\subsection{PERIODS AND PERIOD CHANGES}
 
\noindent
The period of radial pulsators is one of the most common parameters adopted
in the comparison between theoretical predictions and
observations. Accordingly, here we will investigate
{\em in primis} the difference between linear and
nonlinear periods and {\em in secundis} the dependence of nonlinear
periods on the chemical composition.
 
\noindent
Figure 19 shows both linear (solid line) and nonlinear
(dotted line) periods as a function of the effective temperature
for the labeled values of stellar masses and luminosities.
The nonlinear periods refer to the sequence of fundamental models
reported in Figs. 13 and 14, whereas the linear ones come from data listed
in Table 5.  We find that the difference between linear and nonlinear
periods can be  barely noted. Only close to the red edge of the
instability strip  nonlinear periods become systematically shorter
due to the nonlinear effects introduced by the increased efficiency
of convective motions on the "average" density profile (see BS).
 
\noindent
However, the same figure shows the  periods evaluated according to
the analytical relation given by BCCM for metal-poor cluster
variables (dashed line), which reproduce the results of
nonlinear models with an accuracy better than 1 percent.
A similar formula but based on linear, nonadiabatic models was originally
derived by van Albada \& Baker (1971). The discrepancy between periods
from the present investigation and from the analytical
relation appears almost constant and of the order of 15 percent in the
logarithm of period. This difference has to be obviously attributed to
the remarkable difference in the chemical composition. Thus it appears that
a general fitting formula which can safely represent the actual
periods of both metal-poor and metal-rich variable stars would require
an extra term also taking into account the dependence on the helium content.
 
\noindent
Among the plentiful astrophysical legacy left by Eddington (1918), one of
the most simple and weighty insights connecting  stellar evolution
and stellar pulsation was the evaluation of period changes
due to evolutionary effects. Period changes can provide
not only useful clues about the variation of the density distribution
caused by the evolution, as suggested by Eddington, but can be also connected
with the evolutionary rates inside the HR diagram. Even though the
measurement of this quantity is quite simple in principle, reliable
estimates are often a tantalizing
problem due to the systematic errors and in turn to the spurious effects
introduced by different sets of old photographic data.
Several investigations have been devoted to both theoretical and
observational aspects connected with the evaluation of the evolutionary
period changes (Sweigart \& Renzini 1979;  Wehlau \& Bohlender 1982;
Lee, Demarque \& Zinn 1990; SDT).
 
\noindent
To cast light on the effects of different evolutionary scenarios for 
low-mass He burning stars, we computed the evolutionary rate of fundamental
periods for selected  physical structures, taking into account three
different prototypes:
a metal poor RR Lyrae variable (\msun=0.65, Y=0.23, Z=0.001), a globular
cluster type II Cepheid (\msun=0.58, Y=0.23, Z=0.001), and a field
metal-rich type II Cepheid (\msun=0.51, Y=0.34, Z=0.04).
Note that the RR Lyrae prototype is given by a moderately blue ZAHB model
which crosses the strip moving to the red during its off-ZAHB
evolutionary phases.
 
\noindent
Figure 20 shows the predicted rates of period change for
the three prototypes. The top panel discloses the large and repeated
variation of the rate of period change caused by the GNL. Positive values
denote a redward evolution, whereas the negative ones mark the evolution
in the opposite direction.
The asymmetry between
positive and negative values prompts a substantial difference in the
time spent by this evolutionary track during the redward and the blueward
excursions, showing that the latter is a factor of two faster in
comparison with the former one.
The curves plotted in the middle and in the bottom panels show
that the rate of period change for both RR Lyrae and globular
cluster type II Cepheid prototypes keeps increasing, as expected,
across the instability strip. According to data in Figure 20,
one would predict that type II Cepheids should show rates of
period change as a function of time two orders of magnitude larger
in average when compared to canonical RR Lyrae stars, as
already suggested by SDT (and references therein)
on the basis of the HB evolutionary tracks provided by Dorman (1992).
However, the rate of period change of a BL Her star would be
at least a further order of magnitude larger  in comparison
with type II Cepheids. As a result, if GNL are at work the rate of period
change in BL Her stars should be much easier to detect, thus providing
a test for the suggested evolutionary scenario.

\section{SUMMARY AND CONCLUSIONS}
 
\noindent
In this paper we present a homogeneous theoretical scenario covering
both evolutionary and pulsation
properties of a SMR stellar population. In order to provide a proper
investigation of the main evolutionary phases two extensive sets of H burning
evolutionary tracks have been computed by adopting different assumptions
on the initial helium content. The results of these calculations
allow for the  evaluation of detailed grids of hydrogen burning isochrones
covering a wide range of stellar ages, as well as for a sound estimate
of several astrophysical parameters such as the transition mass,
the He core mass at the He ignition and the amount of extra-helium dredged
up during  RGB evolution.
 
\noindent
On the basis of these results we investigated the
He burning evolutionary phases. The HB promenade has been investigated
assuming different initial helium abundances (Y=0.34, Y=0.37)
and adopting different values of the RG progenitor mass (i.e. different 
cluster ages).
Evolutionary ZAHB  have been populated assuming an increasing amount of
mass loss until the bluest ZAHB models have been
almost completely peeled of the stellar envelope. This approach supplied
a fine coverage of the various evolutionary phases which follow the
exhaustion of the central He burning, allowing  for a sound evaluation
of several astrophysical parameters connected with the ZAHB and the off-ZAHB
evolutionary phases.

\noindent
As a relevant result, we found widespread appearance of
gravonuclear instabilities at the ignition of the He shell burning,
according to a mechanism already found in some "young" HB models
at solar metallicity. Since the occurrence of gravonuclear loops
may play a crucial role in understanding the evolutionary status
of field, metal-rich type II Cepheids, in the Appendix to this paper we
discuss the physical mechanisms which drive their appearance.  
 
\noindent
Evolutionary results for SMR He burning models have been used
to investigate the pulsational behavior of models crossing the instability
strip, either as HB or post-HB structures. Linear nonadiabatic
computations disclose a reasonable agreement between the present estimates
of periods and  gravities  of stars at the luminosity level of GNL
with the observational value for BL Her variables.
 
\noindent
The dependence of the blue boundaries  on stellar mass
and chemical composition was approached again in the frame of
linear nonadiabatic computations, investigating the modal stability
of both fundamental and first overtone modes for solar and super solar
metallicities. We find that the HR diagram location of the instability
boundaries presents a negligible dependence on the stellar mass
at least in the range of stellar masses typical of SMR He burning stars.
A sequence of linear nonadiabatic models representative
of metal-poor type II Cepheids was also computed. The results provide
a sound explanation for the observed decrease of both effective
temperature and gravities for increasing metal abundances. 
 
\noindent
In order to better investigate the expected properties of SMR stars
we have also carried out the computation of a  sequence of
nonlinear, nonlocal and time-dependent convective models at given
stellar mass and chemical composition: Z=0.04, Y=0.34.
The computations were mainly aimed at predicting both the
modal stability and the shape of both light and velocity curves across
the instability strip. We found that the range of effective temperatures
covered by the instability strip for SMR RR Lyrae pulsators is sensibly
reduced in comparison with less metallic pulsators, an occurrence which can
play a not negligible role in the observed lack of SMR RR Lyrae variables.
 
\noindent
The theoretical investigation of a selected sample of pulsational models
suitable for BL Her pulsators revealed some important findings
concerning the nature of these objects. The agreement between the
shape of bolometric light curves and observational data is quite
reasonable. The appearance of the bump after the luminosity maximum in  
models located close to the blue edge and its shift at pulsation
phases which precede the luminosity maximum in models characterized
by lower effective temperatures suggests that the group of variables
classified as CW and BL Her stars in the GCVS
may refer to very similar stellar structures. The close similarity
between the light curves of  models located close to the red edge
and the light curves observed for classical Cepheids with periods lower
than six days suggests that out of this sample at least some variables 
could be  low-mass, metal-rich type II Cepheids.
 
\noindent
The nonlinear analysis has also qualitatively suggested a theoretical
framework concerning the appearance of the Hertzsprung progression among
present BL Her models, due to a substantial difference in the dependence
of both thermal and dynamical timescales on the effective temperature.
Further theoretical investigations aimed at producing an extended
survey of limiting amplitude, nonlinear models covering a wider range of
luminosities and stellar masses are needed to establish an exhaustive
pulsational scenario. However, in view of the rather encouraging and
plausible results given by the adopted theoretical approach, we suggest
that observational data for the rate of period change could be a
relevant test for the theory. In fact, we predict that for SMR BL Her this
rate should be at least an order of magnitude larger
in comparison with the corresponding values for metal-poor type II Cepheids
in globular clusters. Moreover, we predict that BL Her would show
both positive and negative rates, due to the occurrence of GNL,
whereas only positive values  are expected in metal-poor Cepheids,
due to the evolution of these metal-poor, hot HB stars toward the Hayashi
track.
 
\noindent
Theoretical luminosity and radial velocity curves reveal a plethora
of interesting features that cannot be soundly compared with 
observational data
due to the paucity of extensive and accurate photometric data presently
available. We expect that the impressive database of photometric
data collected by recent international collaborations (MACHO, EROS, OGLE)
devoted to the search for microlensing events in the bulge and in the
halo of the Galaxy will not only improve the accuracy of the light curves
but also increase the present sample of BL Her stars.
However, the problem in which the data of microlensing experiments
will certainly play a role of paramount importance is the comprehensive
evaluation of the evolutionary rate of period change among type II
Cepheids. In fact, the long set of observational runs which characterize
these databases will provide useful insights on the detection of this
effect which can be barely accomplished, in spite of its inherent
simplicity, on the basis of data available at present.
 
\noindent
It is a real pleasure to thank L. Rusconi for a detailed reading of an 
early draft of this paper and A. Tornamb\'e and O. Straniero for several 
interesting discussions on this topic. We wish also to acknowledge the 
referee, R. T. Rood, for his 
careful reading of the manuscript and for the pertinence of his suggestions
which substantially improved the content and the readability of the appendix.
The evolutionary tracks, the isochrones, the ZAHBs, as well as the linear
nonadiabatic blue boundaries and both light and velocities curves can be 
found in http://terri1.te.astro.it/oact-home/cassisi.html.  
This research has made use of NASA's Astrophysics Data System Abstract
Service and of SIMBAD database operated at CDS, Strasbourg, France. 
This work was partially supported by MURST, CNR-GNA and ASI.

\appendix
 
\section{GRAVONUCLEAR LOOPS AND SCHWARZSCHILD-HARM INSTABILITIES}
 
\noindent
Even though a large amount of theoretical investigations has
been recently devoted to the  evolution of HB structures
(Castellani, Chieffi \& Pulone 1991; HDP; DRO; Bertelli et al. 1996),
only in Paper I we found that in some HB models with solar metallicity 
the He shell ignition is affected by what we named gravonuclear 
instability, an occurrence which appears to be a rule for super-metal-rich
He burning structures. According to this evidence, the related 
appearance of {\em gravonuclear loops} -GNL- in the HR diagram
is dependent on the metal content and on the mass of the H
rich envelope. In this Appendix we discuss in more detail the physical 
mechanisms which govern the appearance of this phenomenon
in metal-rich HB models and its dependence on astrophysical
parameters.

\noindent
As reported in Figure 8 the time behavior of surface luminosity and 
effective temperature for the "cool" 0.51 $M_{\odot}$ model shows
that at the ignition of the He shell this model spends roughly 3 
million years showing periodic drops in luminosity accompanied by a 
contraction of the structure which drives the increase in the effective 
temperature. Eventually, the model  succeeds in quietly burning He 
in the shell as a red Asymptotic Branch structure. Figure 21, where we 
report the same data but for the "hot" 0.48 $M_{\odot}$ model, shows that 
the phenomenon is remarkably similar even in much hotter models: in both 
cases, before quietly burning He in a shell, the models approximatively 
experience 20 loops with quite similar timescales. In order to better 
define the occurrence of this phenomenon, Figure 22 discloses the 
energetics  of the "hot" model during the GNL phase. We find that the 
loops are driven by a 
sequence of periodic switching off/on of both H and He burning 
shells, which in turn drives the expansion of the structure: the 
majority of the energy produced by nuclear reactions is absorbed by
the expansion which succeeds in cooling the stellar interior,
switching off the shells. This phase is obviously followed by a 
contraction which raises internal temperatures, reigniting the shells
and reiterating the process.

\noindent
In order to disentangle the effects of gravitational energy, $\epsilon_g$, 
from the nuclear energy sources, on the development of GNL we performed,
as suggested by the referee, a numerical experiment in which the 
$\epsilon_g$ term in the equation for energy conservation has been 
artificially switched off throughout the stellar structure. 
A similar approach was originally suggested by Sweigart (1971)
for suppressing the thermal instability in helium shell burning stars.
The consequence of assuming a vanishing $\epsilon_g$ term on the
appearance of GNL are shown in Fig. 22 for the "hot" 0.48 $M_{\odot}$
model previously discussed. The energetics of this model (dashed
lines) show quite clearly that after the ignition of helium shell burning
the gravitational energy rules the onset of gravonuclear instability, and
indeed this model does not present any cyclic change in the energy sources.
At the same time, it turns out that this model does not experience the
luminosity and temperature excursions typical of GNL and hence the
evolutionary track moves along its canonical phases. 

\noindent
Moreover, we also note that the occurrence of GNL does not depend on the 
chemical composition profile of the region located between the hydrogen 
and helium burning shells, since as pointed out by Sweigart (1971) the 
SH instabilities are not affected by the abundance profile of this region. 

\noindent
The smooth variation of all physical quantities along the evolutionary 
phases during which the model undergoes the loop can be taken as 
a plane evidence that we are dealing with a real phenomenon rather than with
a numerical instability. Initially we thought about a possible interaction 
between the two shells, producing an oscillatory overstability. However,
we early found the evidence that the H shell can play quite a 
negligible role on that matter, as shown in Figure 23 where we report
the energetics of GNL in a less massive 0.457 $M_{\odot}$ HB model, where the
H shell burning plays a negligible contribution to the onset and development 
of GNL. This difference in the efficiency of the CNO cycle is mainly caused by 
the location of the H burning shell, and indeed in hot models is 
closer to the surface than it is in the cool models. As a consequence 
we reach the conclusion that GNL deal with the occurrence of a thermal 
instability in the He burning shell. However,
inspecting the literature on this subject we disclose that
this is not an unexpected finding, since it was predicted
by Schwarzschild \& H\"arm in a beautiful paper dating back to 1965 
(hereinafter referred to as SH), in which they discuss the instability of 
a \msun=1.0 model {\it in the phase of He burning shell}. In that paper 
the authors demonstrated that non-degenerate shells 
containing a highly temperature-sensitive nuclear energy source (as triple
alpha reactions are) can undergo thermal runaways provided that  the
temperature drop across the shell  is  sufficiently large (eq. [12] in SH). 

\noindent
It is not necessary to repeat here the elegant discussion given 
by SH in their pioneering paper. The philosophy
is that, in that  case, the ignition of the shell is not efficiently
counteracted by the local expansion, inducing an increase in the local
temperature which leads to a thermal runaway. We find that this 
prediction is nicely confirmed by the behavior of the model shown in 
Figure 24. This figure shows the time behavior of temperature and density 
at both the stellar center and the He burning shell. At the same time,
these quantities are compared with gravitational and neutrino luminosities 
of the structure. We find that the exhaustion of the central He is 
followed by an overall contraction phase, where 
the increase of central temperature is partially dumped by the increased 
efficiency of neutrino cooling. The ignition of the shell is marked by
the sequence of peaks in the shell temperature, which points out the
positive feed back affecting this parameter until the expansion of the 
structure first cools and then switches off the shell. As a consequence, 
we find that our models follow closely SH predictions. However, we find 
that stellar matter at the shell location appears affected by moderate 
electron degeneracy, which contributes with a further 40\% to the
perfect gas pressure, possibly playing a role in favoring
the onset of gravonuclear instabilities. Without further details, 
it is worth underlining that the scenario outlined by SH 
casts light only on the appearance of GNL in metal-rich stars, since the 
increased opacity steepens the temperature gradients in the stellar 
interior, which, as already discussed, is the parameter governing the  
onset of this instability.

\noindent
SH only speculated about the consequence of these
thermal instabilities, suggesting that the thermal perturbation would likely
increase the He burning rates by more than a factor of 10 over its normal 
value, possibly inducing deep mixing and, although not likely, dynamical
effects. However, their best guess was that since the runaway:  
{\em "would probably
not alter the star basically and hence may be repetitive, one might guess
that a thermal instability of the type here considered may lead to 
a kind of relaxation oscillation, a thermal flicker. Whether the convective
mixing could reach an extent sufficient to have substantial
evolutionary consequences seems at present an entirely open question"}.
We find that such a scenario appears soundly confirmed by our detailed 
computations. In fact, He burning rates reach a factor of 10 over 
their normal values and the phenomenon is indeed repetitive, producing
a sequence of relaxation oscillations. However, convection is damped since 
energy is converted into radial expansion, and no substantial mixing occurs. 
As shown in the same Figure 24, we can only add that the transient
flickering succeeds in affecting the  whole stellar structure, and indeed  
central conditions also flicker as a function of time until the steady 
burning in the He shell is reached.  

\noindent
As a whole, we note that helium burning shells, due to both the strong
dependence of triple alpha reactions on local temperature and the sudden
decrease of the temperature across the shell, cause the onset of SH
instability during the AGB thermal pulses (AGB-TP) and the post-HB phases
(GNL). However, the GNL are tightly connected with the envelope opacity
and indeed the appearance of this phenomenon is a common feature of
metal-rich HB models. On the other hand, both metal-poor and metal-rich
stars which evolve along the AGB experience the phase of thermal pulses. 
Moreover, we find that during the GNL the helium burning shells are
characterized by a moderate amount of electron degeneracy, whereas the
AGB-TP are only marginally affected by this phenomenon. Even though the
hydrogen and helium burning shells govern the morphology of GNL, the
helium shell, in contrast with the AGB-TP, never approaches the hydrogen
shell. Finally, we note that the timescale of AGB-TP is ruled by the
efficiency of H burning shell, whereas the GNL are ruled by the thermal 
timescale of the envelope. 

\noindent
We have already outlined the role played by envelope opacities on the 
appearance of GNL (see section 2.1 in Paper I). However, in order to assess
in more detail the effects of new radiative opacities on this phenomenon,
Figure 25 shows the opacity distribution in the stellar layers located
above the H burning shell for several envelope models computed at different
effective temperatures (see labeled values) but at fixed luminosity level
(\lsun =1.90) and stellar mass (\msun =0.48). According to the data
plotted in this figure we find that the iron bump is the main opacity
source of the stellar envelopes located in the high temperature region of
the SMR horizontal branch stars.  
At effective temperatures lower than log $T_e =4.6$ the opacity bump due
to the second HEIR appears and it becomes a non-negligible opacity source
for effective temperatures of the order of 25,000 K. At even lower effective
temperatures the opacity bumps due to both hydrogen and first helium
ionization become the main opacity sources of the envelope.
Moreover, it is worth mentioning that the opacity peak due to iron, in
contrast with the helium peak, remains almost constant over the whole range 
of effective temperatures. The old opacities (Huebner et al. 1977, dashed 
lines), as it is well known, do not show the Z-bump close to log T=5.3.

\noindent
On the basis of these non trivial differences we computed an evolutionary
track by assuming the same input parameters of the "hot" model shown in
Fig. 21 but by adopting the old opacities together with both a coarse and
a fine zoning throughout the envelope regions. Oddly enough, the coarse
zoning evolutionary track evolves toward the white dwarf cooling sequence
and does not show the appearance of GNL, whereas the fine zoning evolutionary
track presents the GNL. In order to provide a comprehensive analysis of the
opacity effects, we performed the same numerical experiments but by
adopting the input parameters of the "cool" model shown in Fig. 8. In
this case we find that both the coarse and the fine zoning evolutionary
tracks do not experience at all the gravonuclear instability. 

\noindent
As a consequence, we think it right to stress that the new 
opacities are not a sufficient condition for the appearance of this 
phenomenon. In fact, several numerical experiments performed to evaluate 
the effect of spatial resolution on the onset and development of gravonuclear
instability disclose that a decrease of the spatial resolution connected
with the iron bump and with the H and He ionization regions causes a
substantial increase of the transparency in the outermost regions.
Therefore the energy released by the He shell is substantially reduced, 
and in turn the radius and luminosity excursions of GNL are reduced 
and/or inhibited. 

\noindent
The increase of spatial resolution in the iron bump and in the
ionization regions causes a decrease of the time step since in the fine
zoning models it is ruled by the pressure gradient in the stellar envelope
and not by nuclear reactions in H and He burning shells.
SH already suggested that the instability we are dealing with will not be
necessarily noted in stellar evolutionary computations, since it will be
passed without notice as long as the adopted time step is twice the
e-folding time of the instability.  Luckily, we approached the phase of
central He exhaustion with time steps of the order of few thousand
years for following in detail the He exhaustion and the ignition of the H 
burning at first and then of the He burning shells. Similar time steps are
much shorter than the Kelvin-Helmotz timescale which governs thermal
runaways, and thus provide a detailed coverage of the onset and development
of the gravonuclear instability. However, if time steps are evaluated
according to the much longer nuclear timescales, the GNL will be
largely smoothed away.
We suggest that the use of old opacities and of a coarse zoning in the 
envelope regions could be the reasons why DRO did not find GNL in their  
recent computations of metal-rich HB structures.

\noindent
In order to speed up the calculations
required to approach the post-AGB evolutionary phases, we decreased the
spatial resolution in the opacity bumps (increase of the time step) for
artificially switching off the GNL. Note that for a proper treatment of
GNL at least 20,000 models for each track would be needed.

\pagebreak

\pagebreak
 
\figcaption [] {Evolutionary parameters for
structures at the RGB tip as a function of the stellar mass. From the
top to the bottom are plotted: the stellar age, the luminosity and the
mass of the He core. The solid and dashed lines are referred to data
obtained by adopting an initial He abundance equal to Y=0.34 and Y=0.37,
respectively.}
 
\figcaption [] {Behavior of the transition
mass $M_{tr}$ as a function of the metallicity. The results shown in this
figure have been derived  by adopting, at lower metallicities, data from
Cassisi \& Castellani (1993), Cassisi et al. (1996), and Paper I.}
 
\figcaption [] {Theoretical isochrones for the H
burning phases for the labeled assumptions about the stellar age and for
two different helium abundances, namely  Y=0.34 (panel a) and Y=0.37
(panel b). The time interval between consecutive isochrones is
2 Gyr, with the exception of the four isochrones corresponding to 0.8, 0.9, 
1.0 and 2.0 Gyr, respectively. The arrows mark the location of the RGB 
clump in both the youngest and the oldest isochrones.}
 
\figcaption [] {ZAHB locations for the He burning stellar models for
the labeled assumptions about the red giant progenitor masses and for
the two different initial He abundances, namely Y=0.34 (panel a) and Y=0.37
(panel b).}
 
\figcaption [] {ZAHB locations for the He burning models at fixed red
giant progenitor mass (\msun =1.8) and for the two quoted initial He
abundances.}
 
\figcaption [] {Evolutionary tracks in the HR
diagram of He burning structures evaluated by assuming fixed metal
content (Z=0.04) and initial helium abundance (Y=0.34). The sets of HB
models have been computed by adopting three different values of the
progenitor mass ($M_{pr}=2.0, 1.8, 1.0 M_{\odot}$) and different
assumptions concerning the efficiency of mass loss along the RGB.}
 
\figcaption [] {Same as Fig. 6, but referred to HB models computed by
adopting a different initial helium abundance, namely Y=0.37.}
 
\figcaption [] {Time behavior of the surface luminosity (top panel) and 
effective temperature (bottom panel) for a "cool" model which presents 
gravonuclear loops. For this model the RG progenitor mass is 
$M_{pr}=1.8 M_{\odot}$. The total mass value, the He core mass at
the He ignition and the chemical composition are labeled.}
 
\figcaption [] {HR diagram showing the linear blue boundaries of both
fundamental and first overtone BL Herculis variable stars. The stability
analysis has been performed by adopting a fixed chemical composition
(Y=0.34, Z=0.04) and three different values of the stellar mass,
namely \msun =0.485, 0.505, 0.52.}
 
\figcaption [] {HR diagram showing the linear blue boundaries of both
fundamental and first overtone BL Herculis variable stars. The stability
analysis has been performed by adopting a fixed stellar mass value
(\msun =0.505) and two different chemical compositions, namely  Y=0.34,
Z=0.04; and Y=0.28, Z=0.02. The dashed lines indicate the blue boundaries
of the instability region connected with type II Cepheids belonging to
globular clusters.}
 
\figcaption [] {Linear work curves per logarithmic temperature versus
the logarithmic temperature for two models computed by adopting the
same luminosity (\lsun =2.0), effective temperature ($T_e$=6400 K),
and stellar mass (\msun =0.505). The different chemical compositions
are labeled. The top panel is referred to the fundamental mode,
whereas the bottom one to the first overtone, surface at right.
The arrows denote the position of the driving regions connected
with the ionization regions (HIR, HEIR) and with the opacity bump
caused by iron. The symbols plotted in the bottom panel mark the location
of the first overtone nodes in radius (diamonds), temperature (triangles),
and luminosity (squares). See text for further explanation.}
 
\figcaption [] {Predicted ratio between the number of RR Lyrae variables 
and the total number of
Horizontal Branch stars as a function of the metal content (see text).}
 
\figcaption {Fundamental bolometric magnitude variation at limiting
amplitude. This sequence of nonlinear models has been computed by adopting
fixed luminosity (\lsun =2.0), stellar mass (\msun =0.505), and chemical
composition (Y=0.34, Z=0.04). The nonlinear periods (days) and the
effective temperatures (K) are labeled.}
 
\figcaption {Same as Fig. 13, but referred to lower temperature models. 
The top panel shows quite clearly that the model located at $T_e$=5700 K 
is a mixed-mode pulsator, and indeed over subsequent periods both the 
shape of the light curve and the pulsational amplitude present substantial 
variations. The value of the nonlinear period labeled in the top panel 
is referred to the pulsation cycle between $\phi=0.0$ and $\phi=1.0$, 
whereas the period of the following cycle is P=2.1903 d.}  

\figcaption [] {Variation of the surface radial velocity at limiting
amplitude. The velocity curves are referred to the sequence of
fundamental nonlinear models described in the text. Positive values
indicate expansion, whereas the negative ones contraction. The labels
and the notation are the same as in Fig. 13.}
 
\figcaption {Same as Fig. 15, but referred to lower temperature models.}  

\figcaption [] {Variation of both bolometric magnitude (top panels) and
surface radial velocity at limiting amplitude. The curves are referred to
two fundamental pulsators computed by adopting the following input
parameters: \lsun =2.2, \msun =0.505, Y=0.34, Z=0.04. Labels and notation
are the same as in Fig. 13.}
 
\figcaption [] {Time behavior along the integration of period
(top panel), bolometric amplitude ($\Delta M = \Delta M_{bol}(max) -
\Delta M_{bol}(min)$; middle panel), and kinetic energy (bottom panel)
for a first overtone model located at $T_e$=6000 K. This model shows
a mode switching to the fundamental mode. The arrow in the top panel
indicates the time at which the radial pulsation switches toward the
fundamental, whereas the arrow in the bottom panel marks the time
at which the model approaches the asymptotic behavior. The initial
velocity profile was obtained by perturbing the radial first overtone
eigenfunction with a constant velocity amplitude of 10 km$s^{-1}$.}
 
\figcaption [] {Comparison in a log-log plane between linear (solid line)
and nonlinear (dotted line) fundamental periods as function of the
effective temperature for the labeled value of stellar mass and luminosity
level. The dashed line shows the fundamental periods obtained by adopting
the analytical relation provided by BCCM.}

\figcaption [] {The evolutionary rates of period change for selected
evolutionary models: (panel a) super-metal-rich BL Herculis; (panel b)
globular cluster type II Cepheid; (panel c) globular cluster RR Lyrae.
The stellar mass and the chemical composition of the adopted
models are labeled.}
 
 
\figcaption [] {Same as Fig. 8, but referred to a "hot" less massive HB 
model, namely M=$0.480M_{\odot}$.}
 
\figcaption [] {Time behavior of gravitational
(panel a), hydrogen (panel b), and helium (panel c) luminosities during
the GNL phase of the HB model in Fig. 21. The dashed lines are referred 
to a numerical experiment performed by assuming a vanishing gravitational 
energy term throughout the stellar structure. See text for further details.}
 
\figcaption [] {Same as Fig. 22, but referred to a less massive HB model 
with M=$0.457M_{\odot}$.} 

\figcaption [] {Time behavior of selected structural  parameters
for the $0.457M_{\odot}$ model from the exhaustion of central He
up to the onset of steady He shell burning. Top to bottom: density, 
temperature, gravitational luminosity, and neutrino luminosity. In the top 
panels the heavy solid line is referred to the center, whereas the solid 
one is referred to the He shell.}
 
\figcaption [] {Opacity distribution as a function of the logarithmic 
temperature for several envelope models computed by adopting a fixed 
luminosity level (\lsun =1.90) and a wide range of effective temperatures. 
The total stellar mass and the chemical composition have been
assumed equal to the HB model adopted in Fig. 21. The solid lines are 
referred to the radiative opacities provided by Rogers \& Iglesias (1992), 
while the dashed lines to the old opacities provided by Huebner et al.  
(1977). It is worth noting that moving from hot envelope models (top) to 
the cooler ones (bottom) the opacity scale increases of almost one  
order of magnitude. The surface effective temperature is labeled.} 
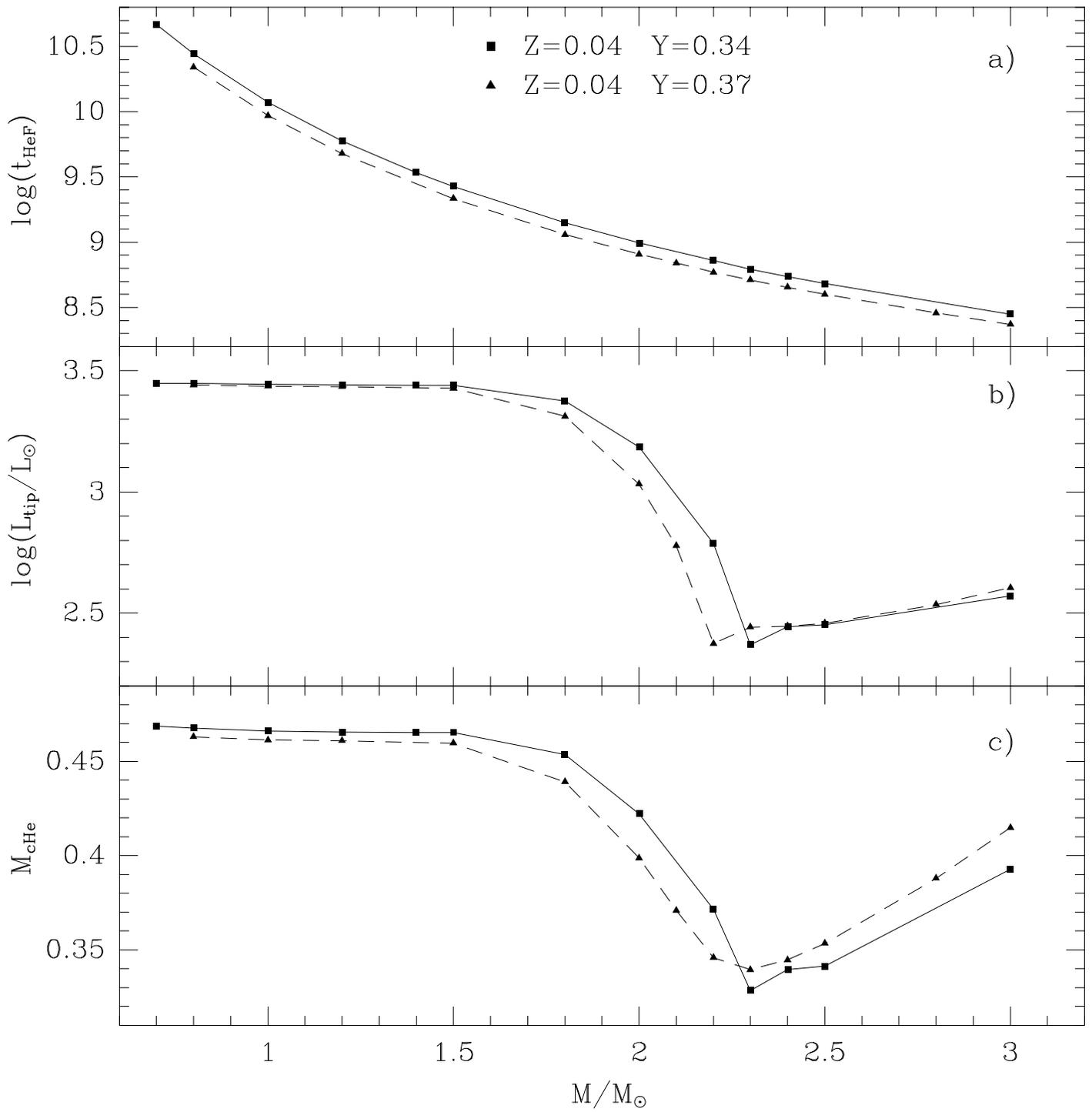
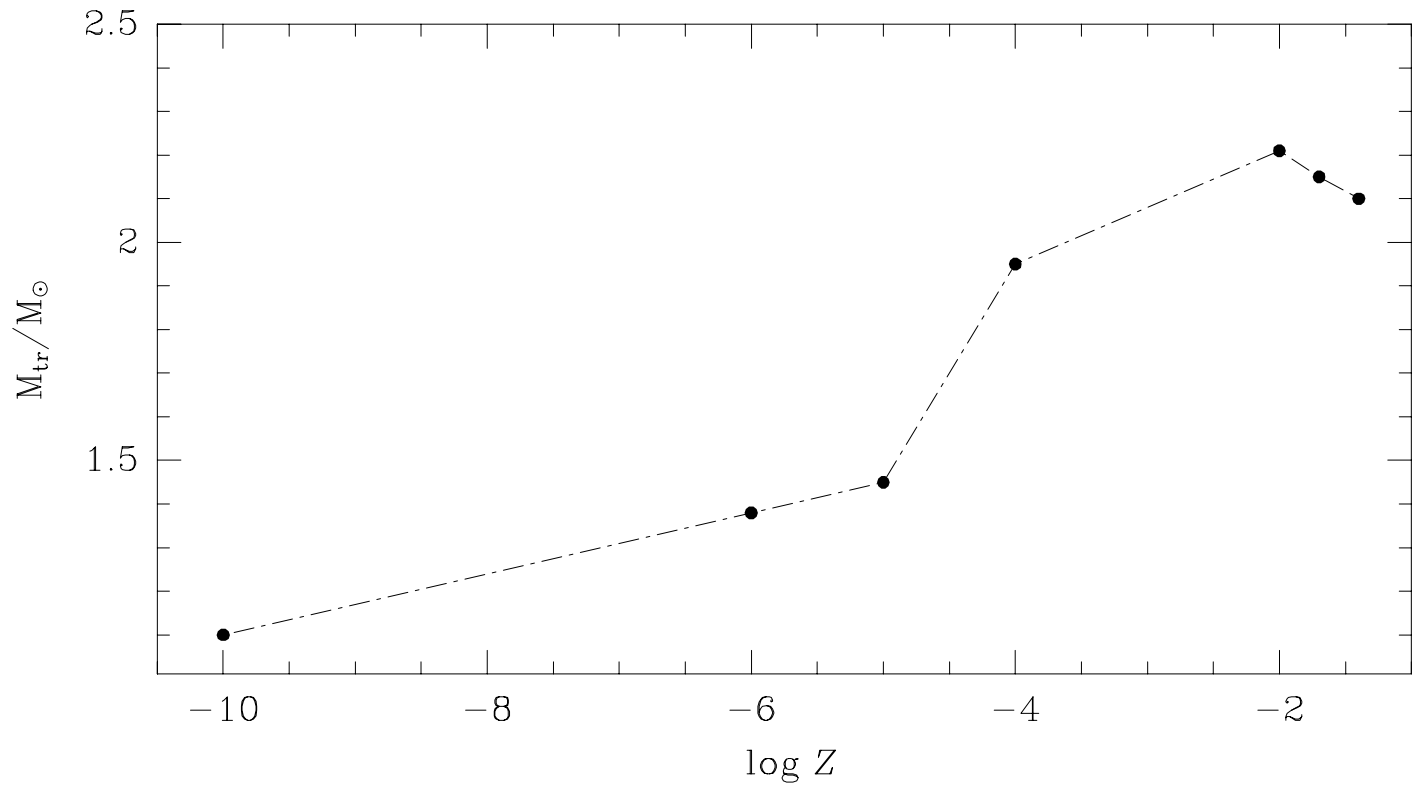
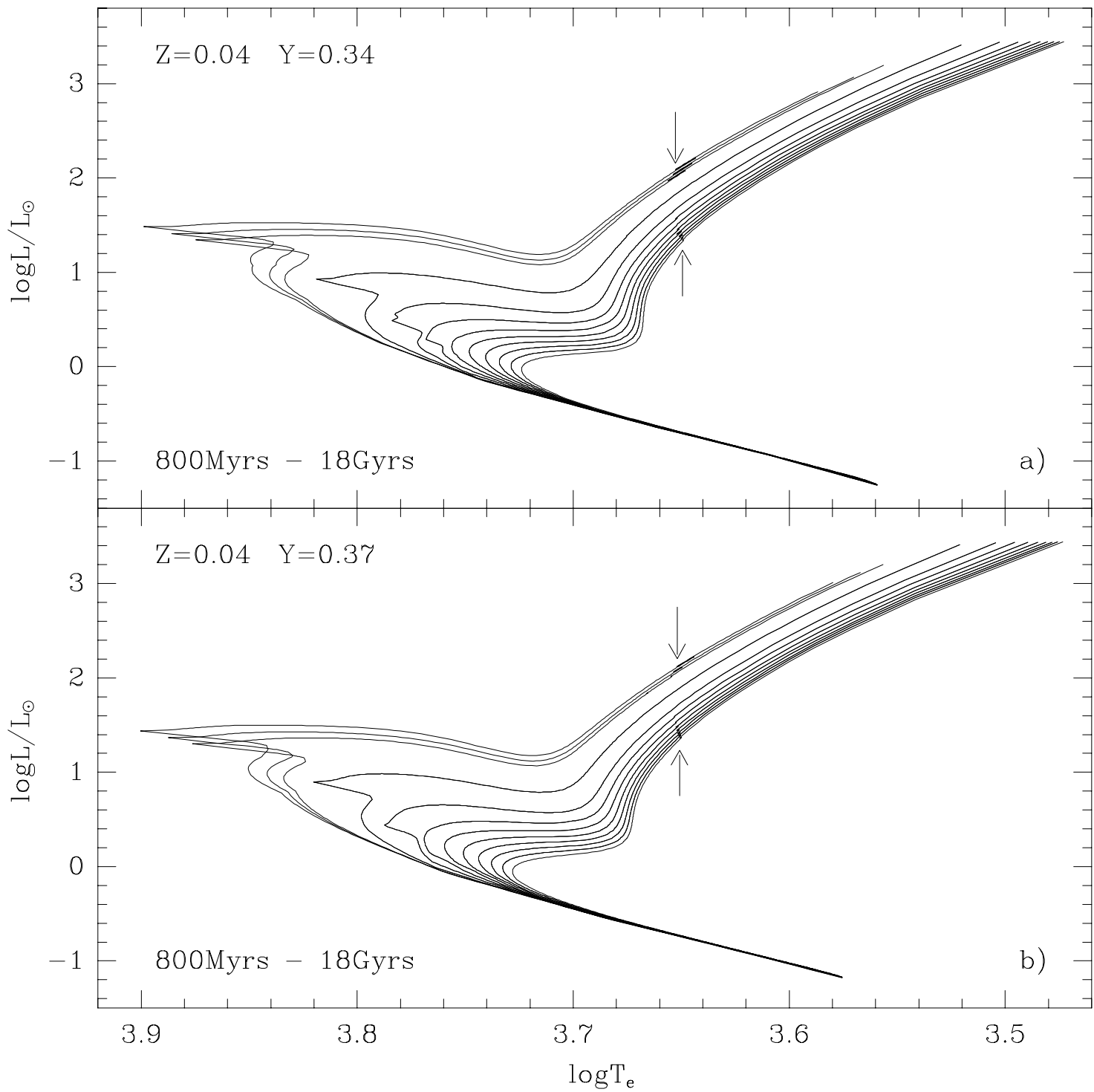
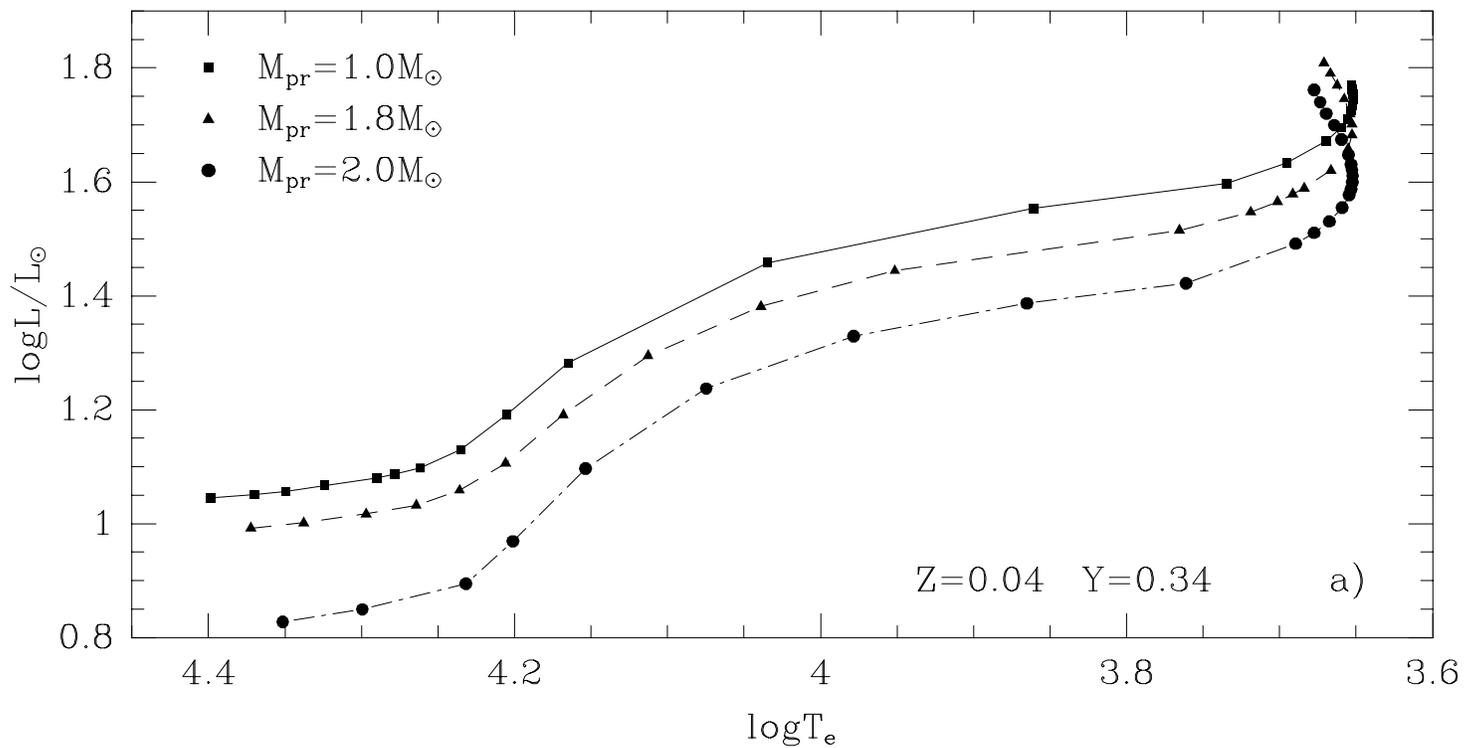
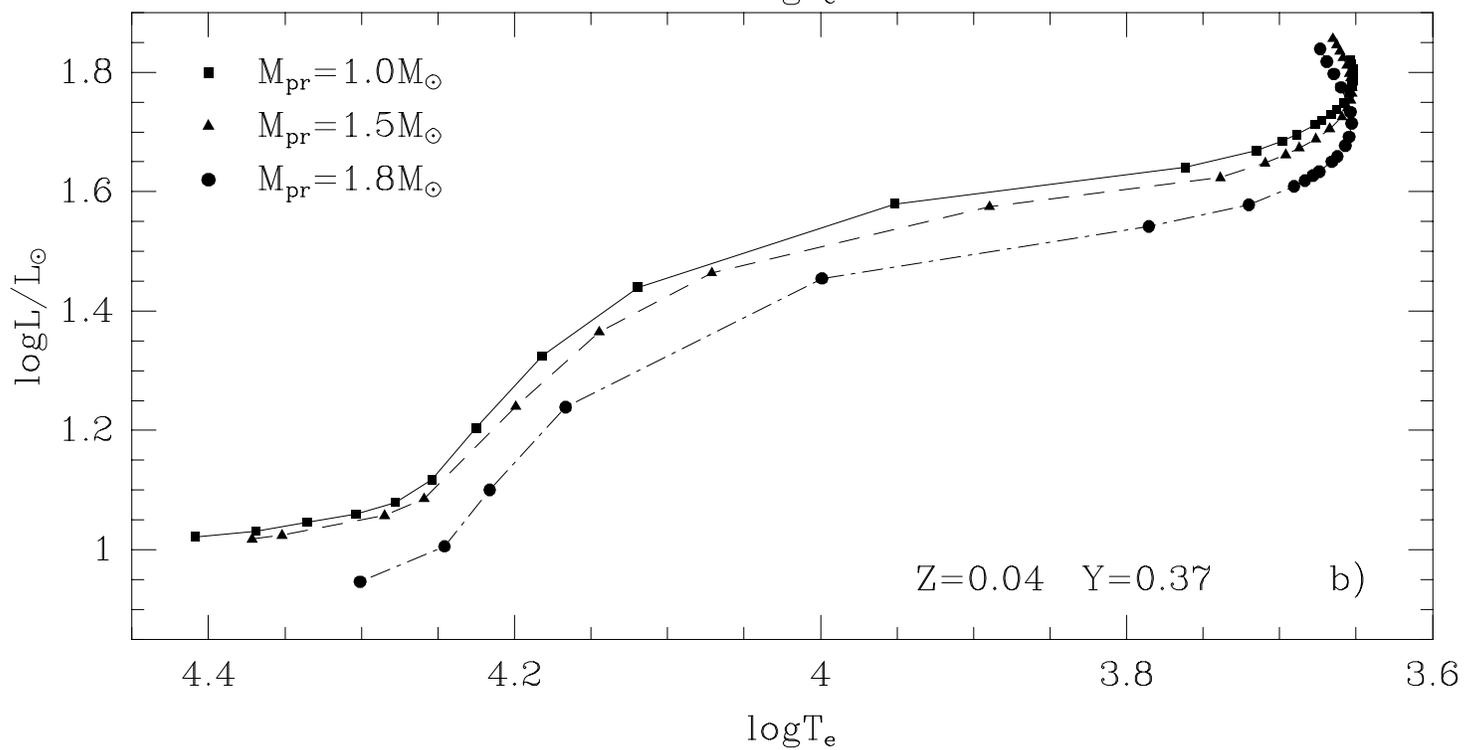
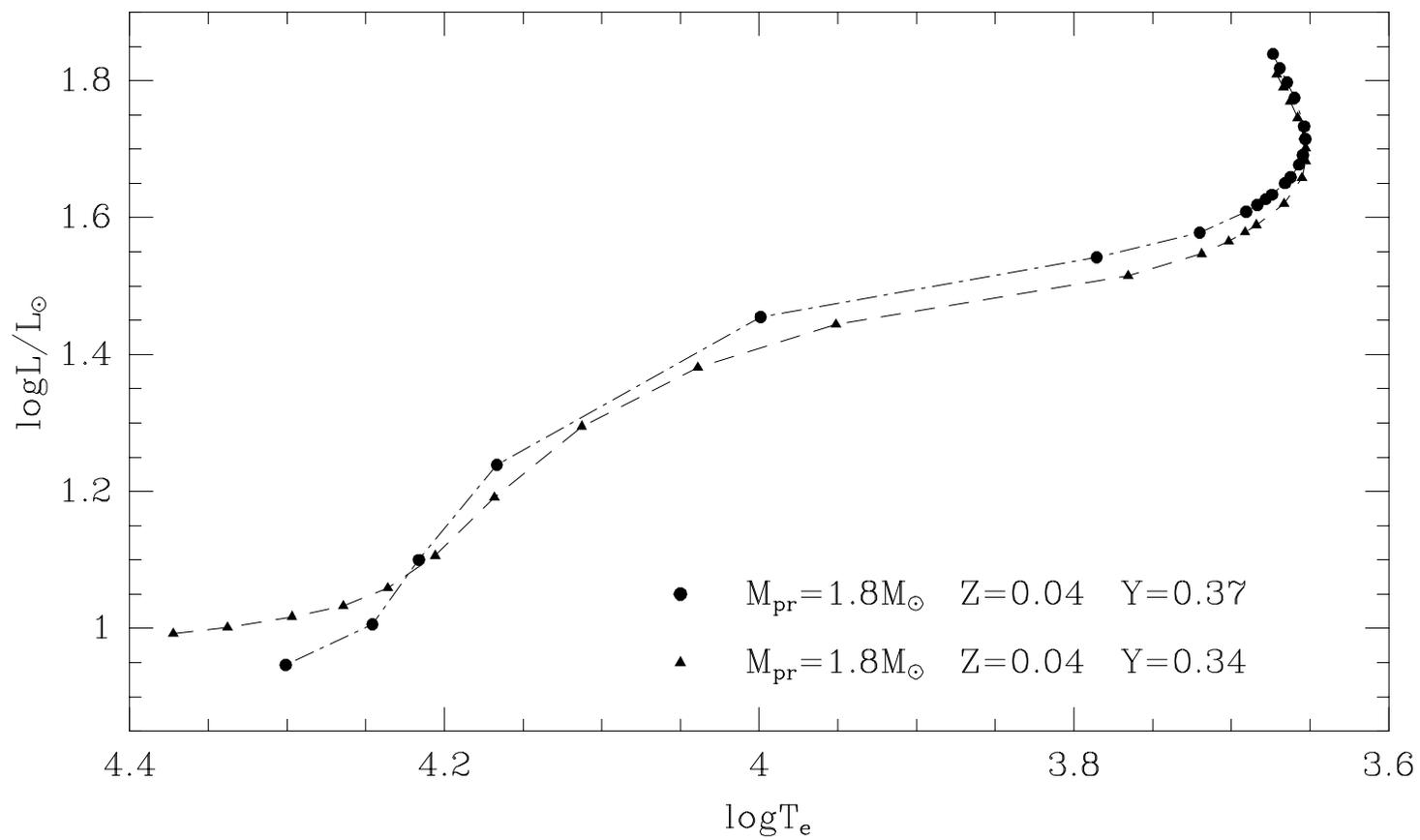
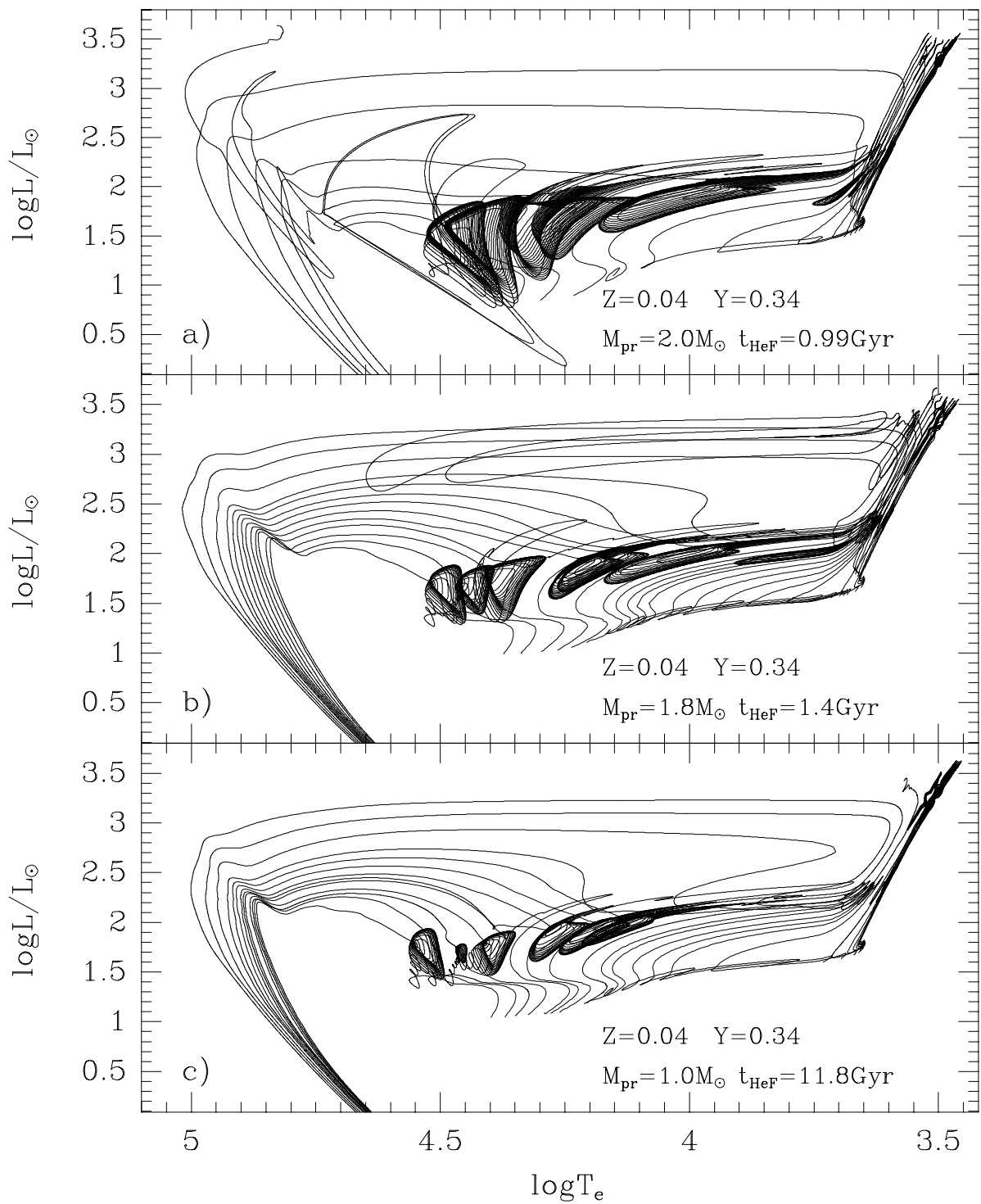
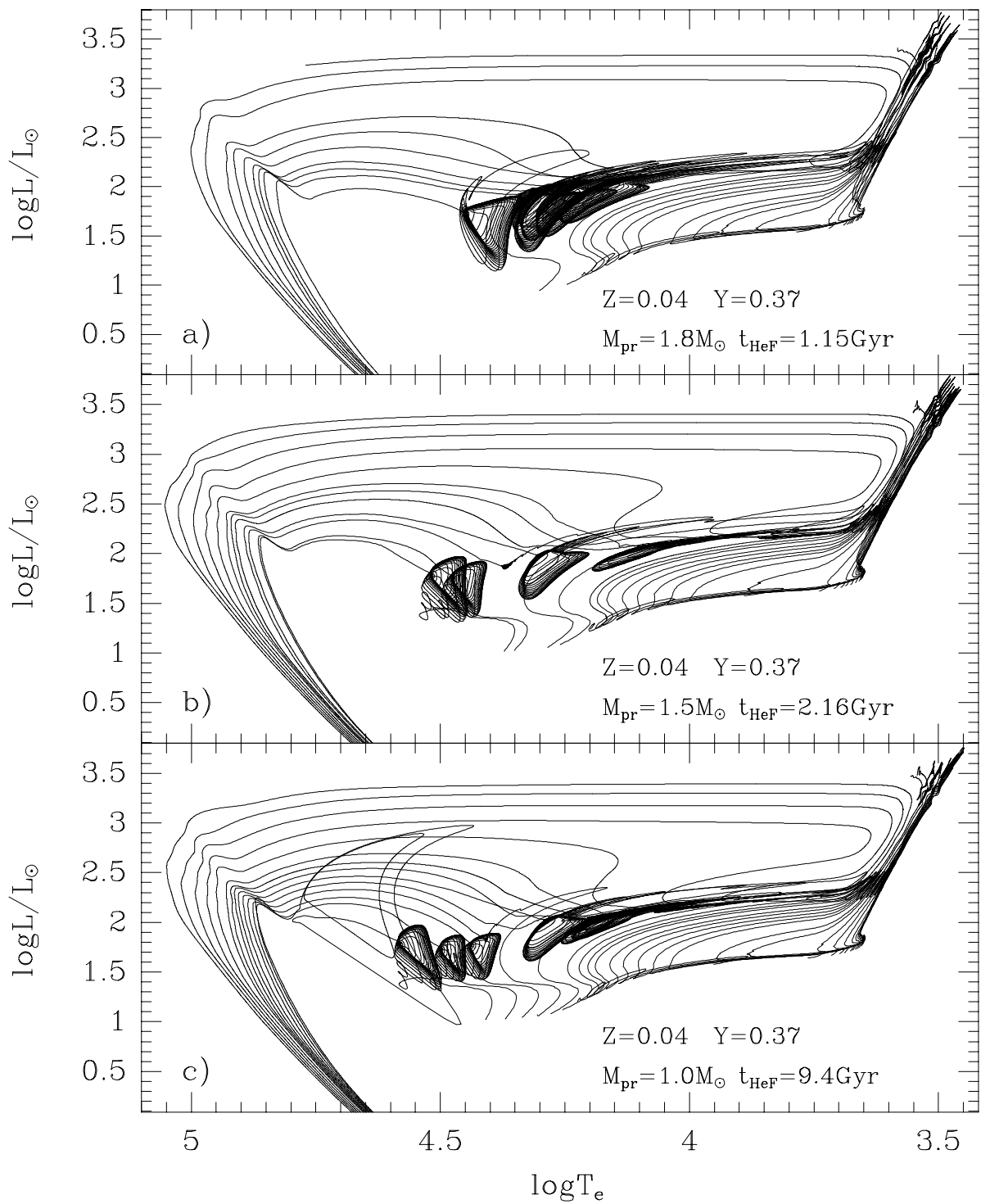
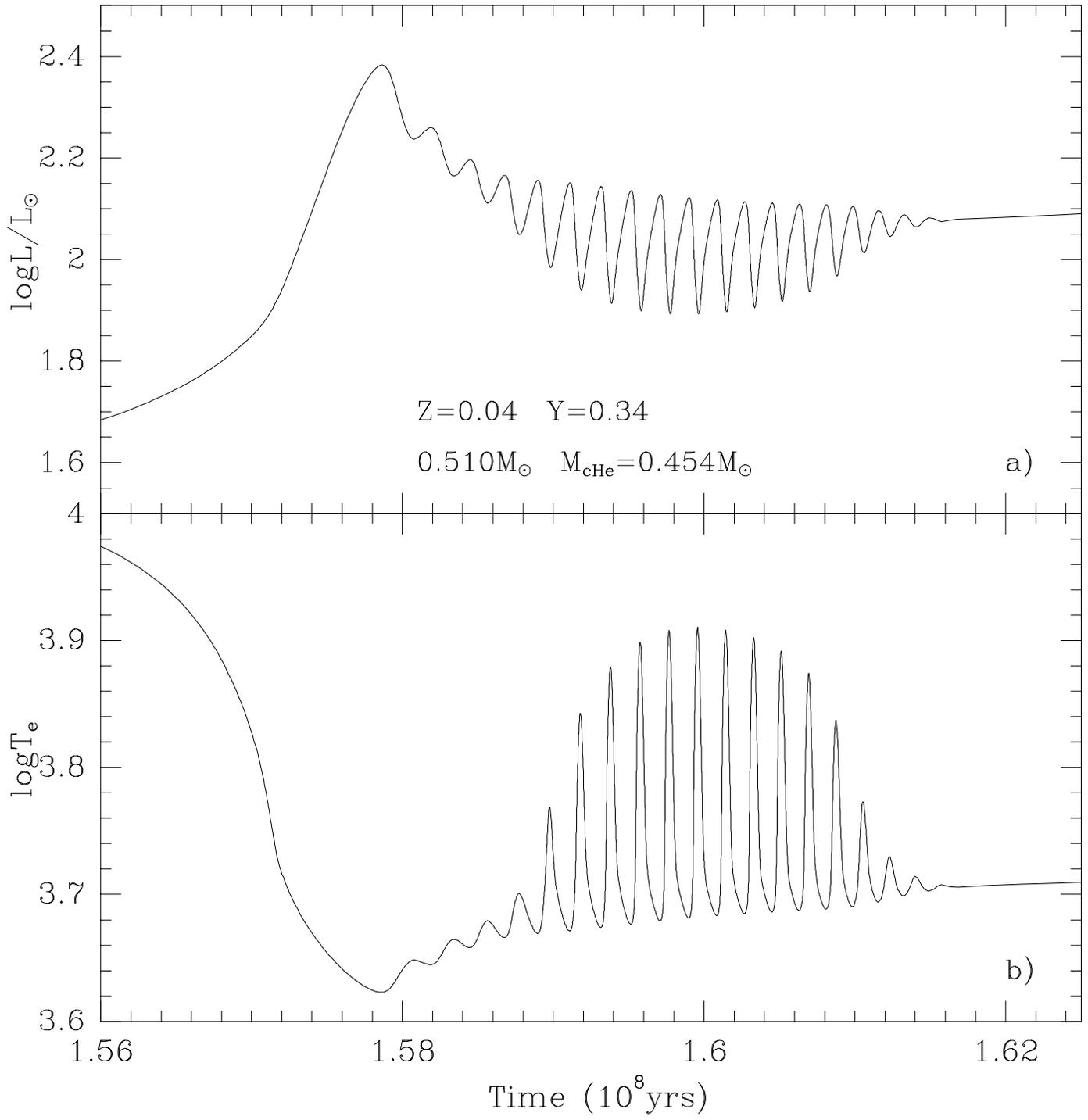
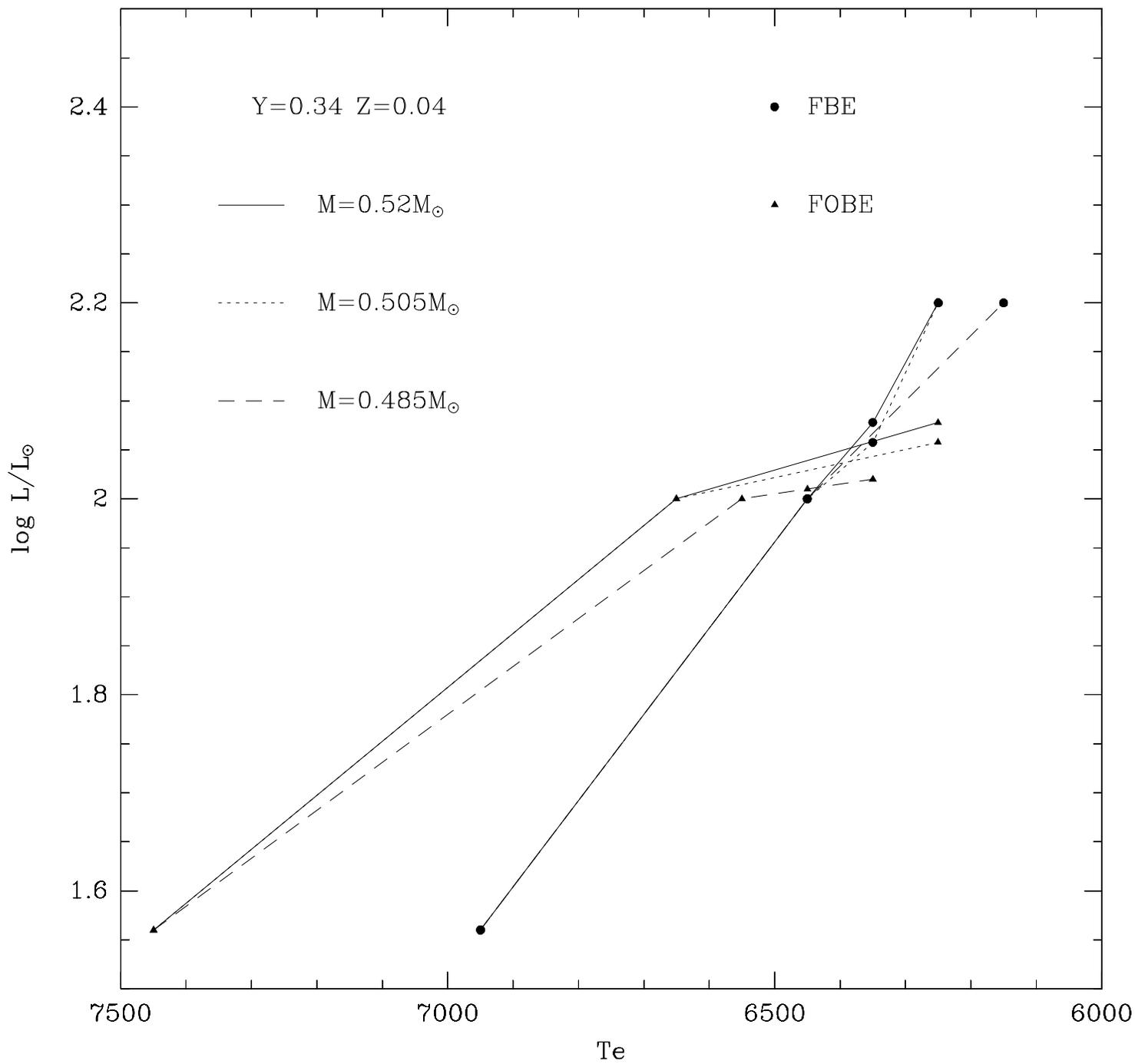
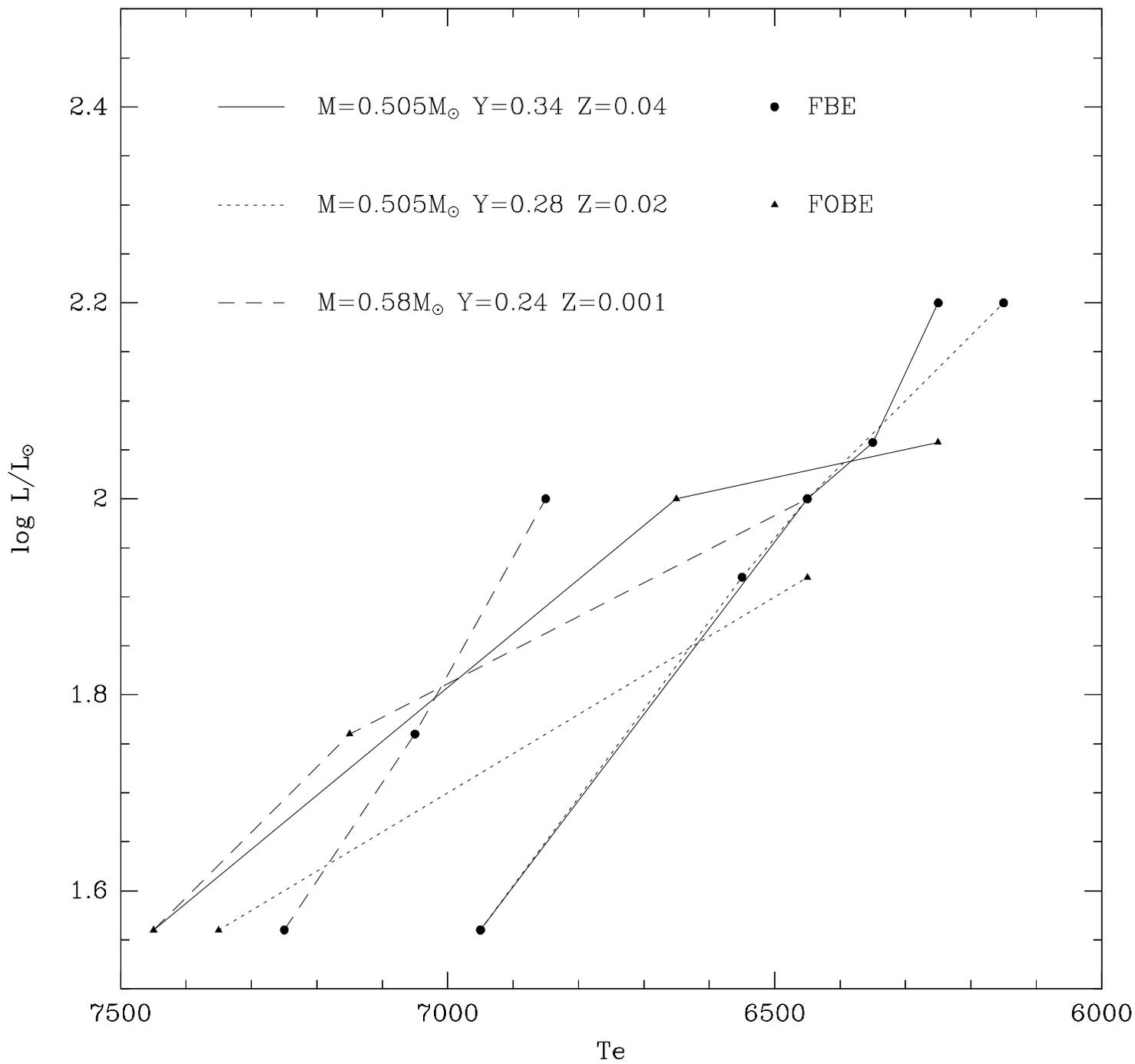
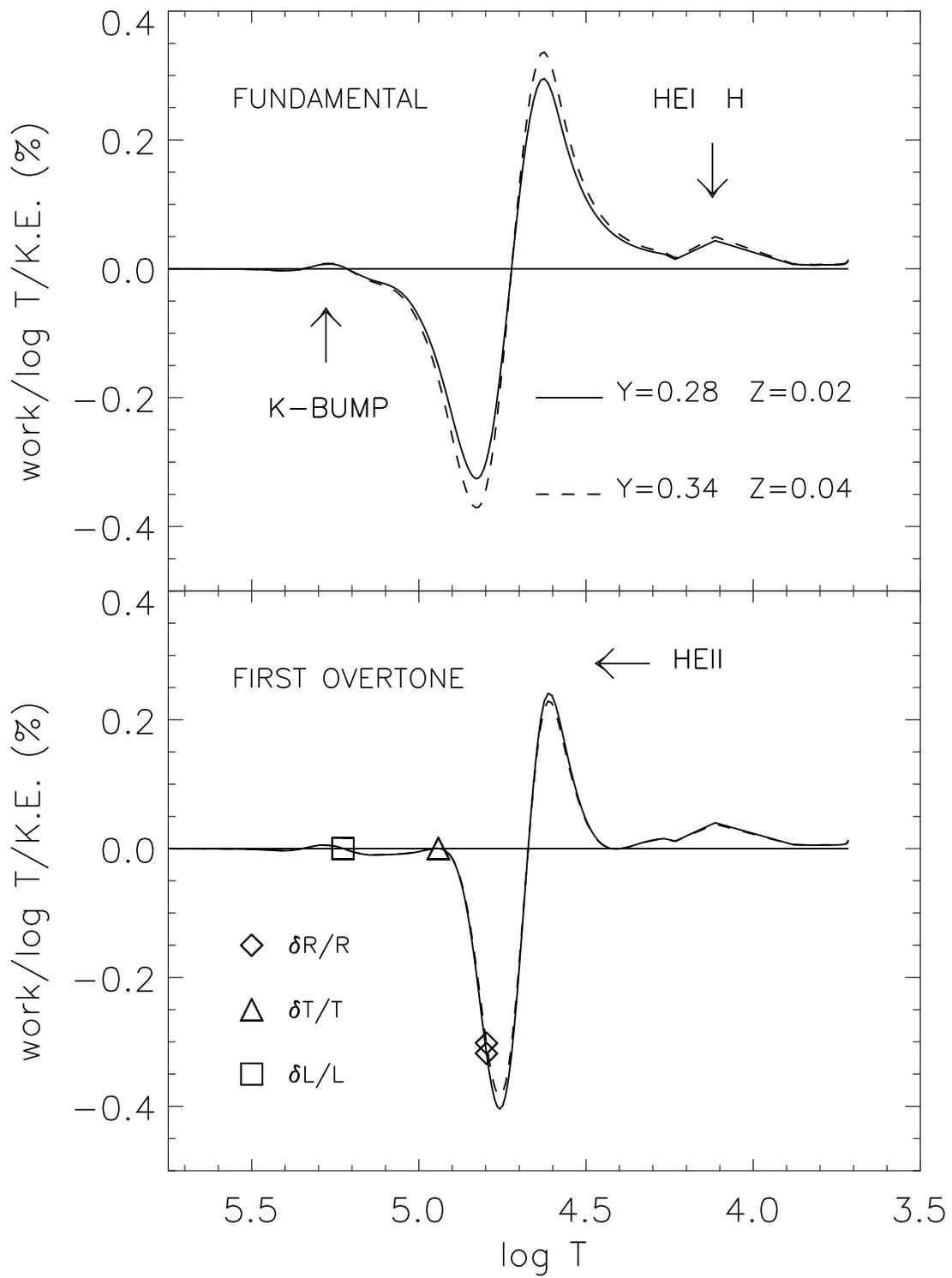
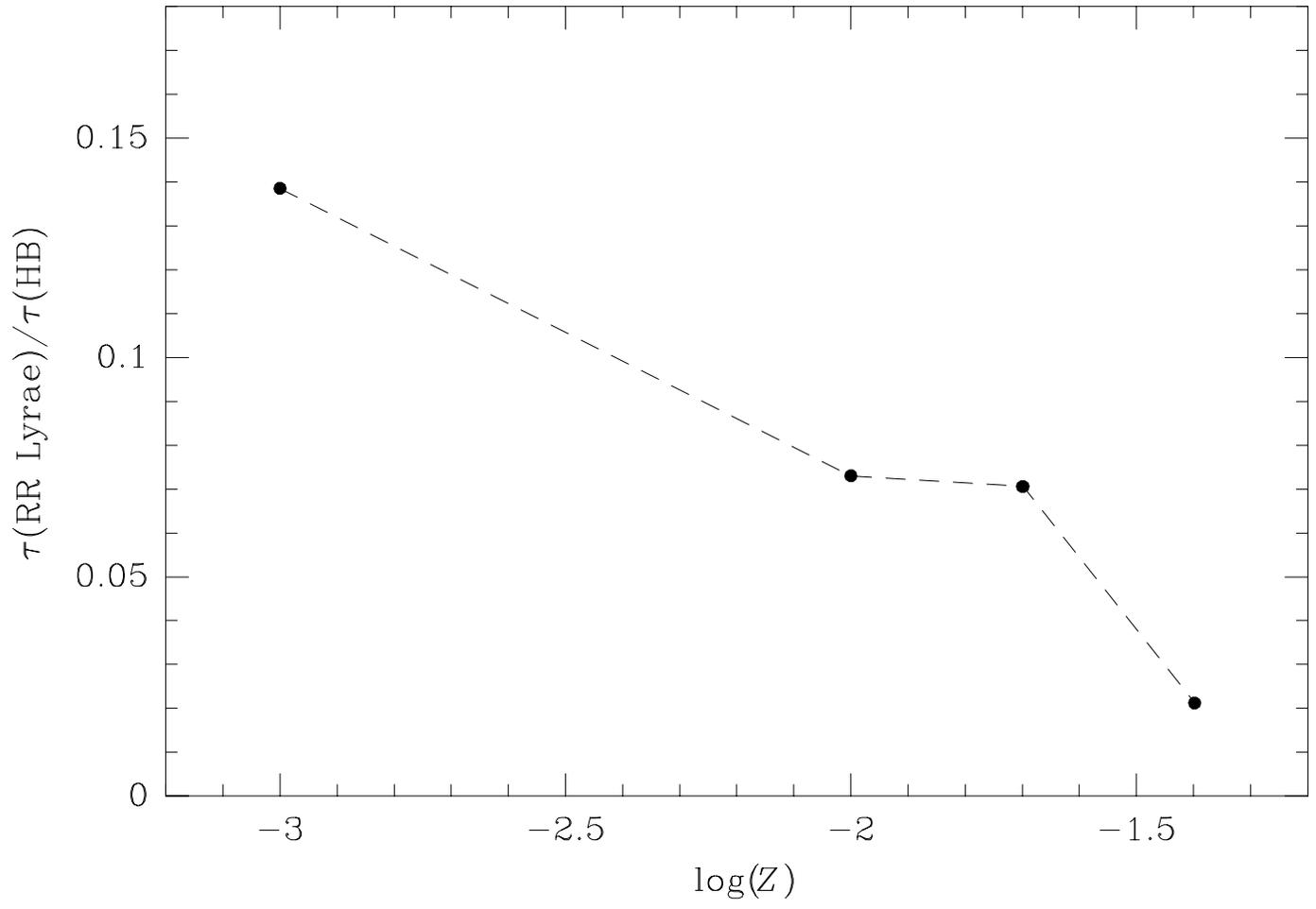
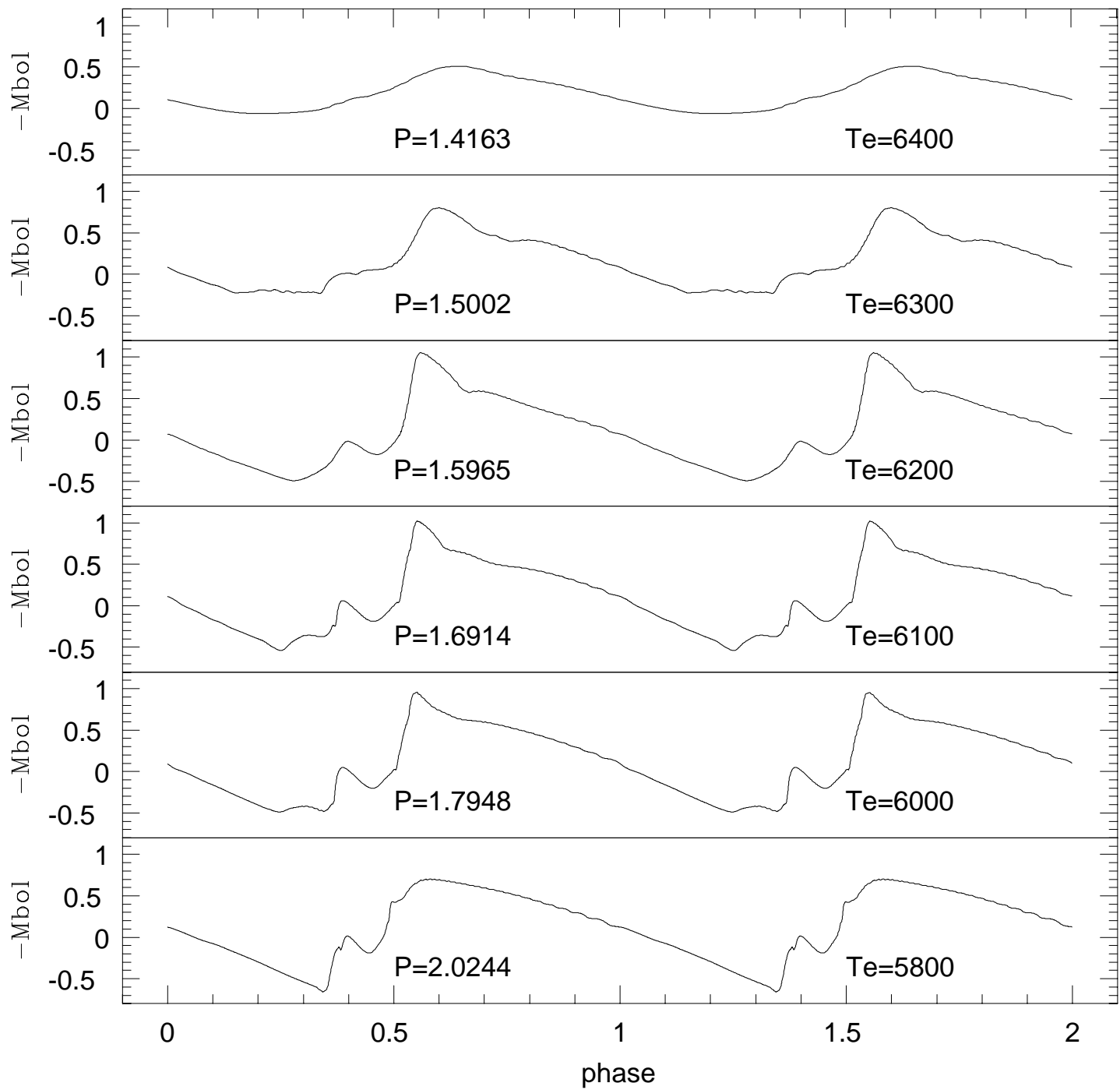
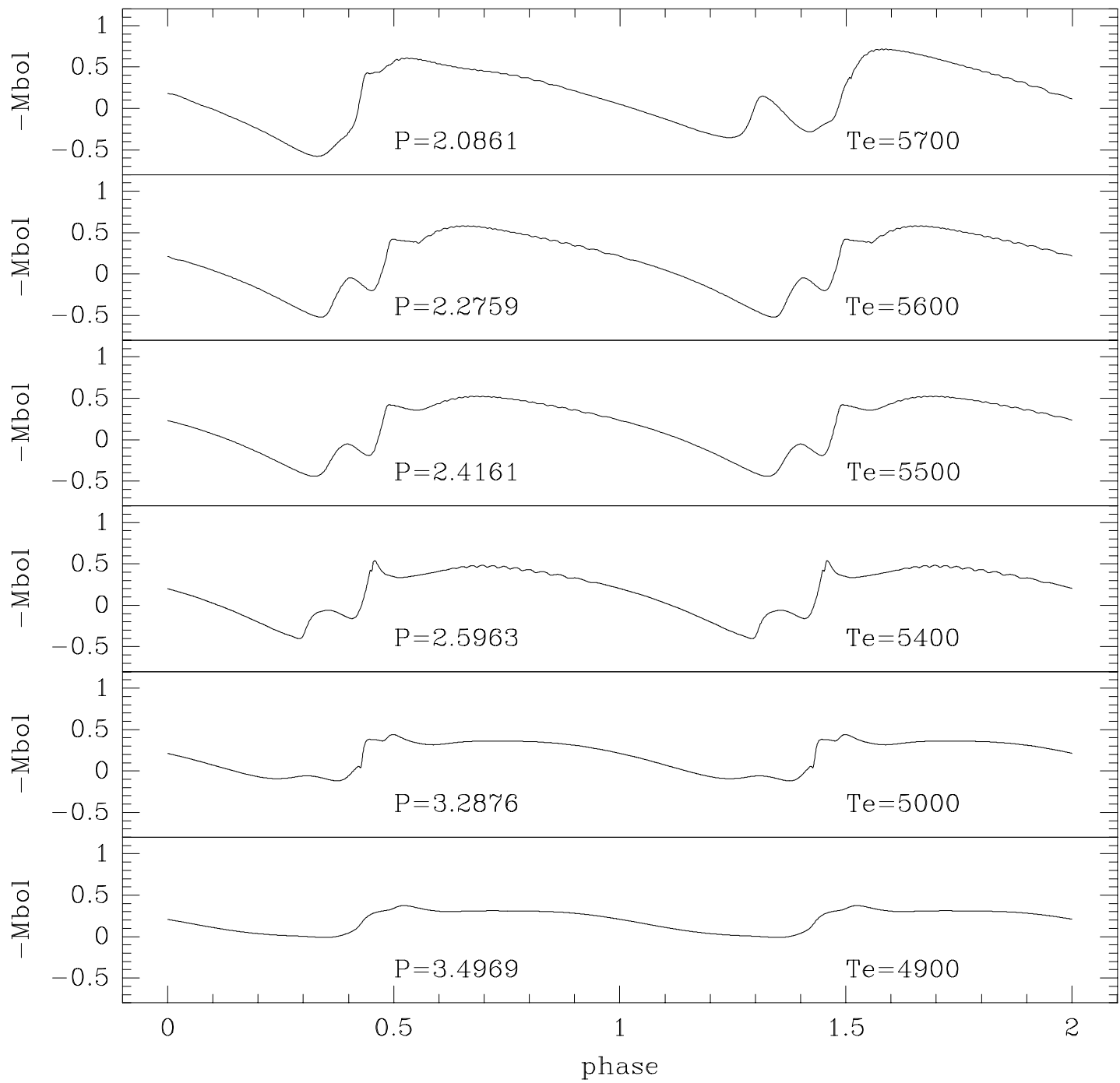
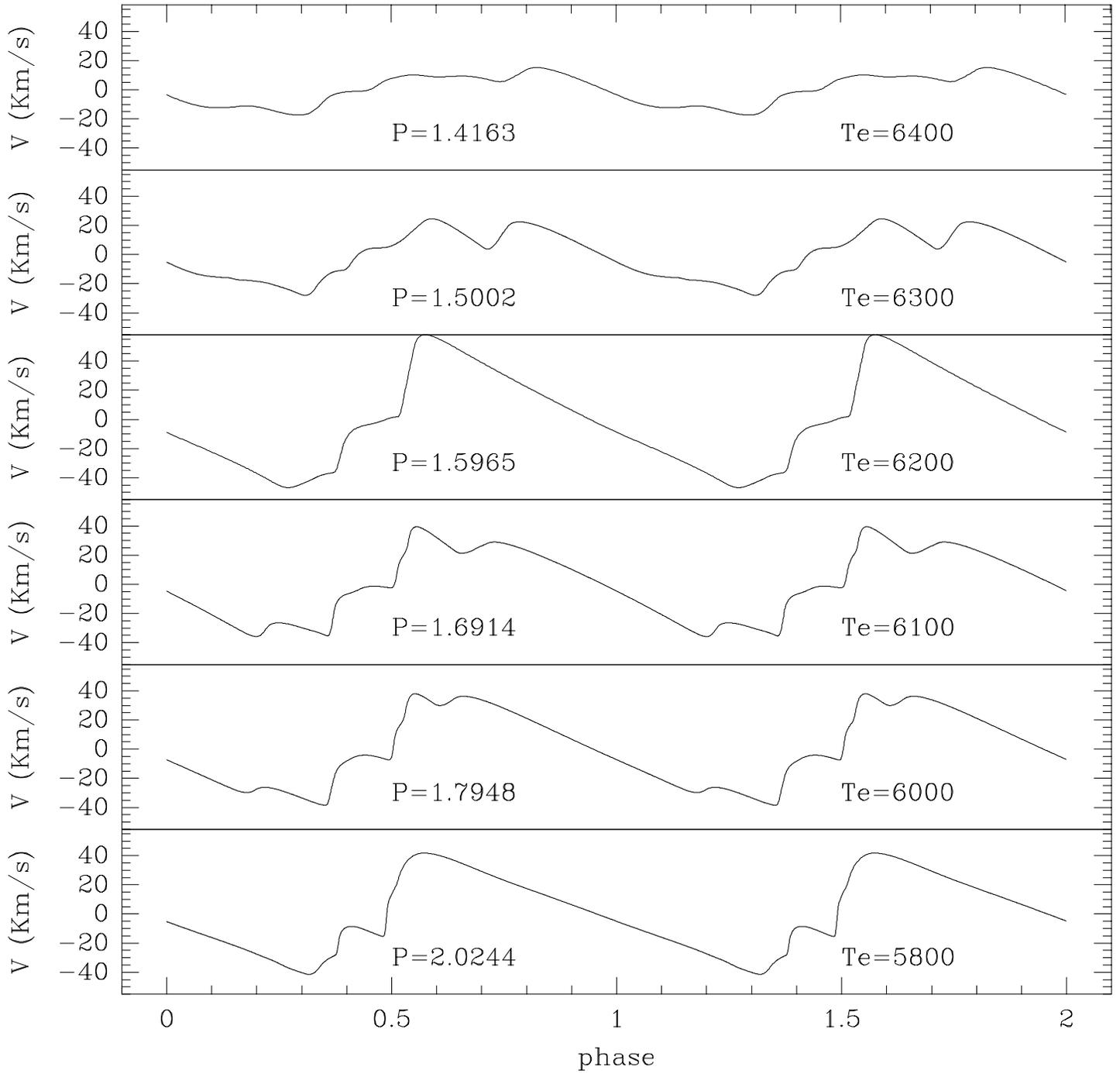
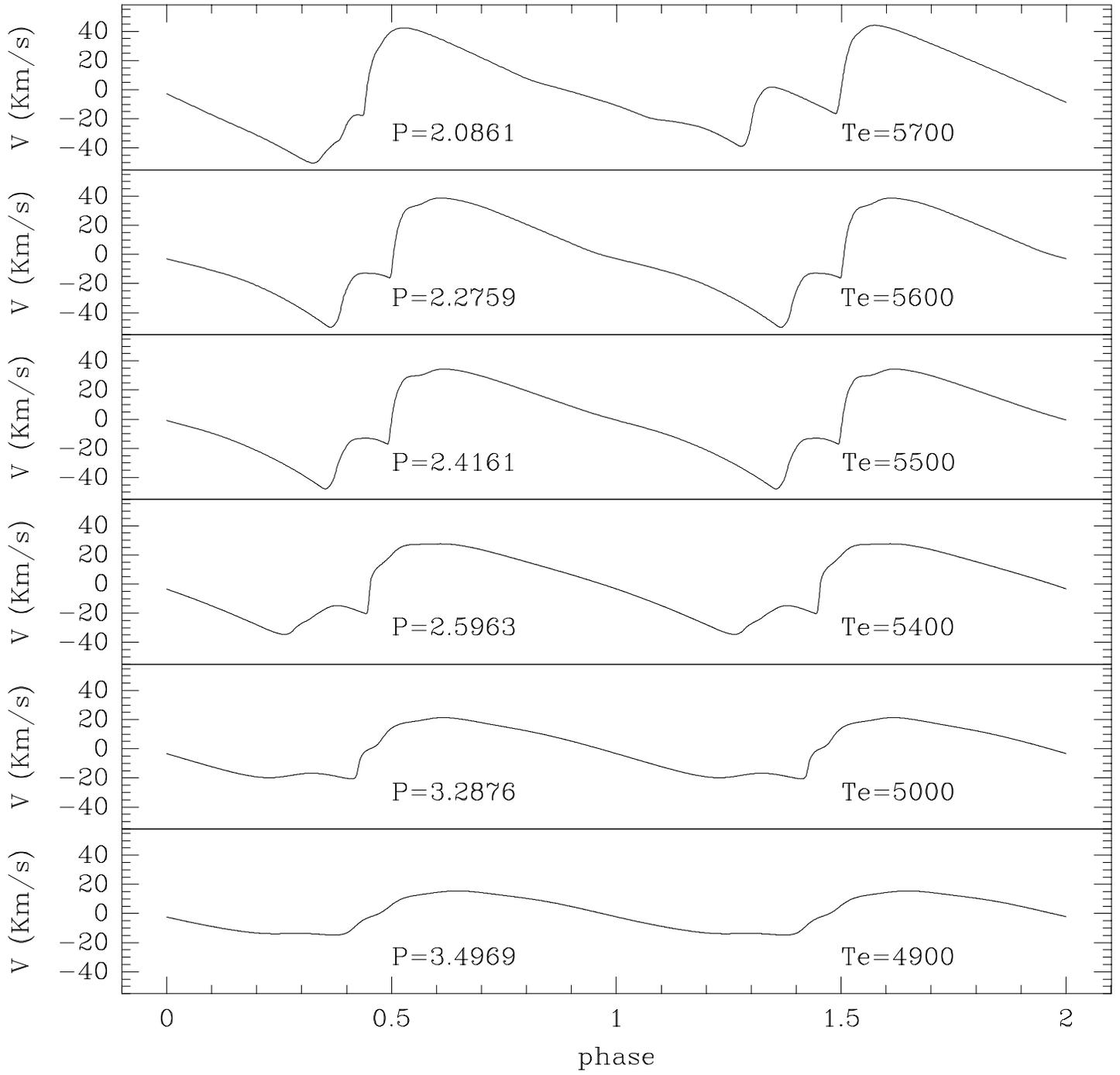
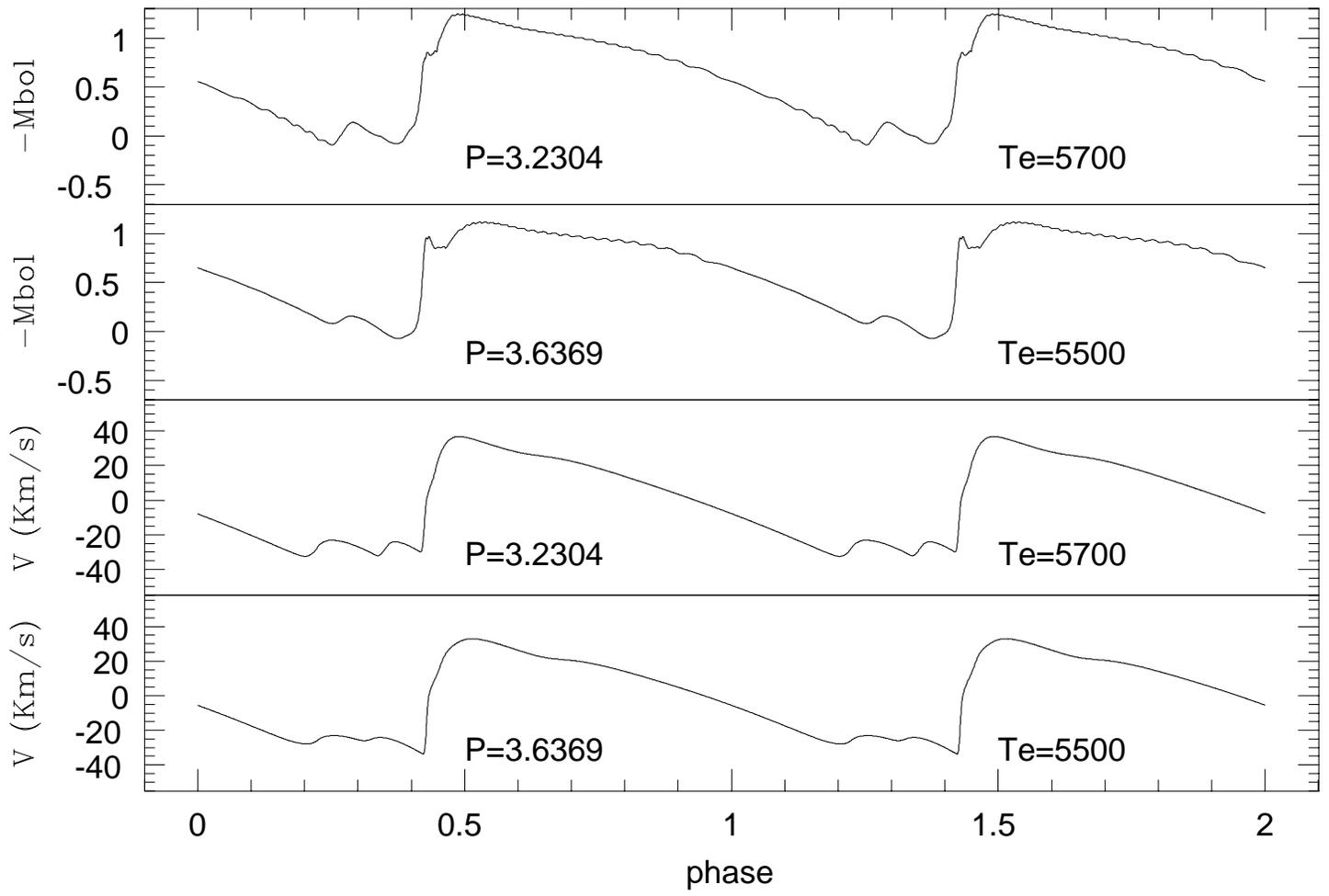
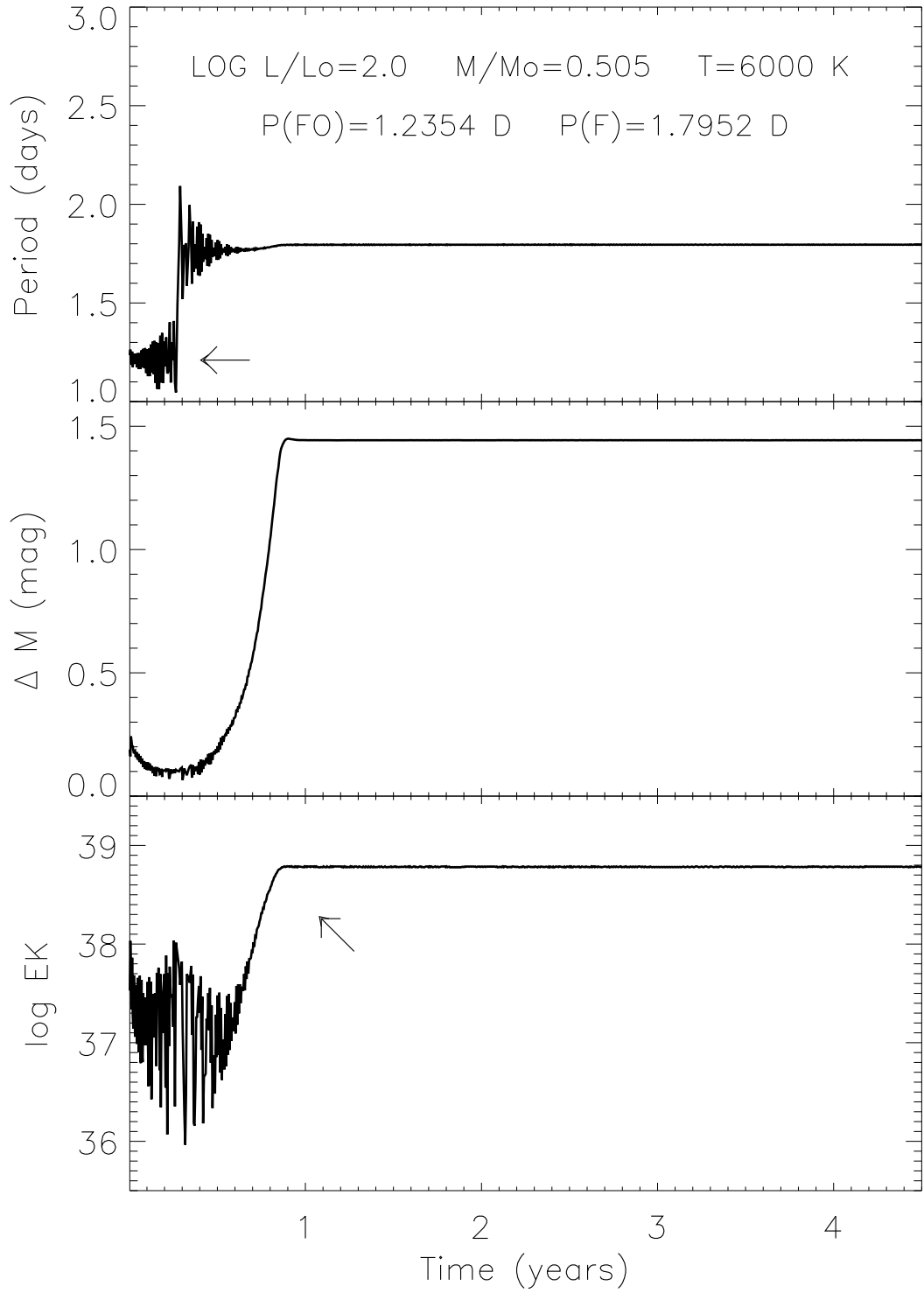
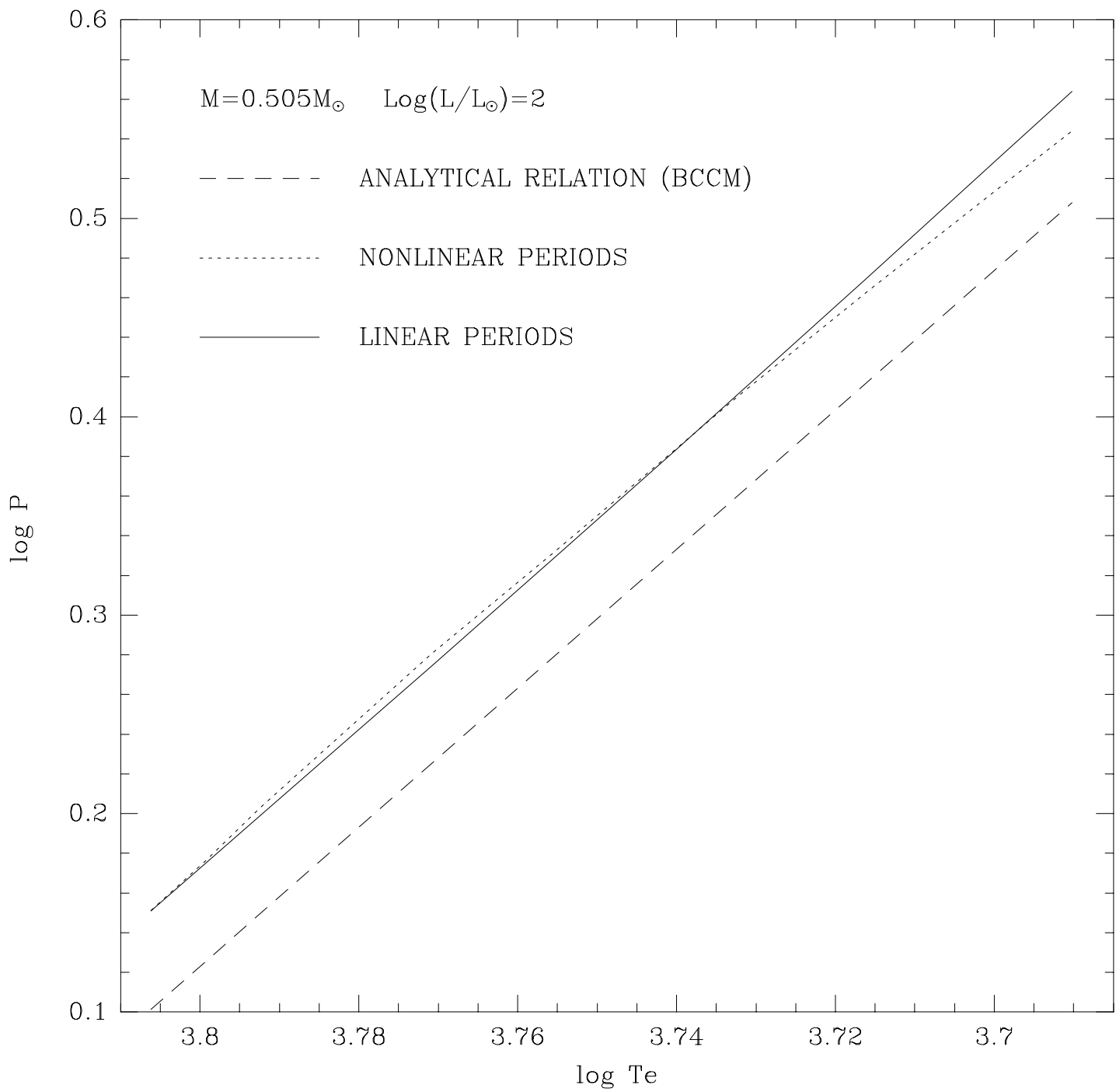
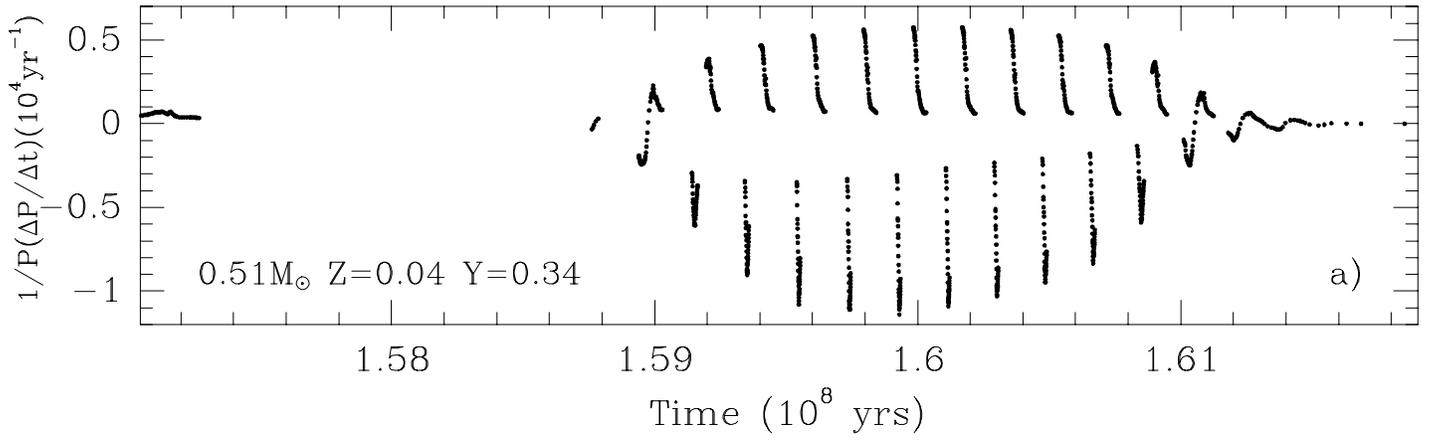
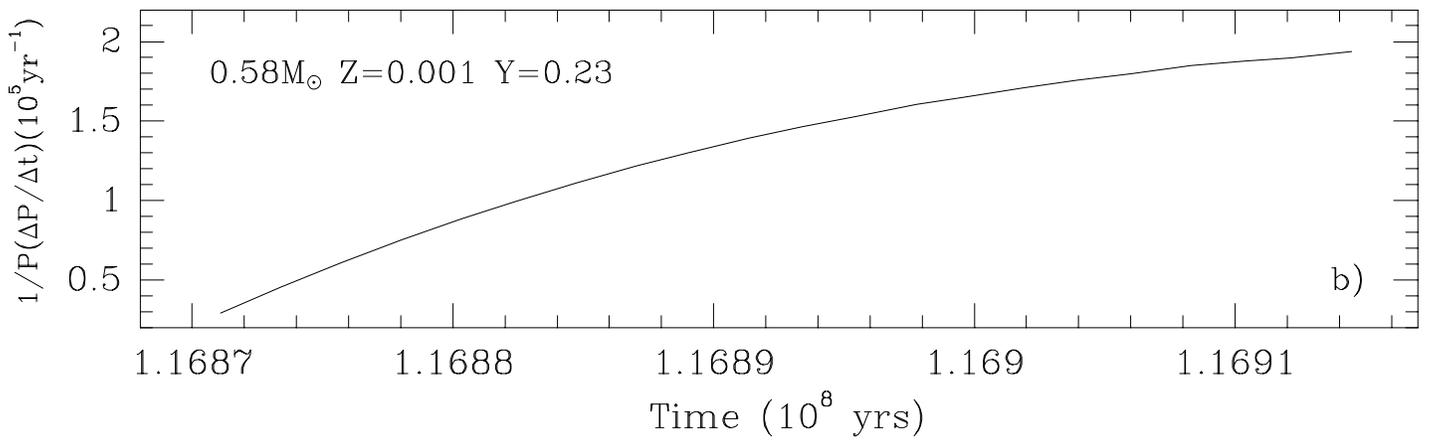
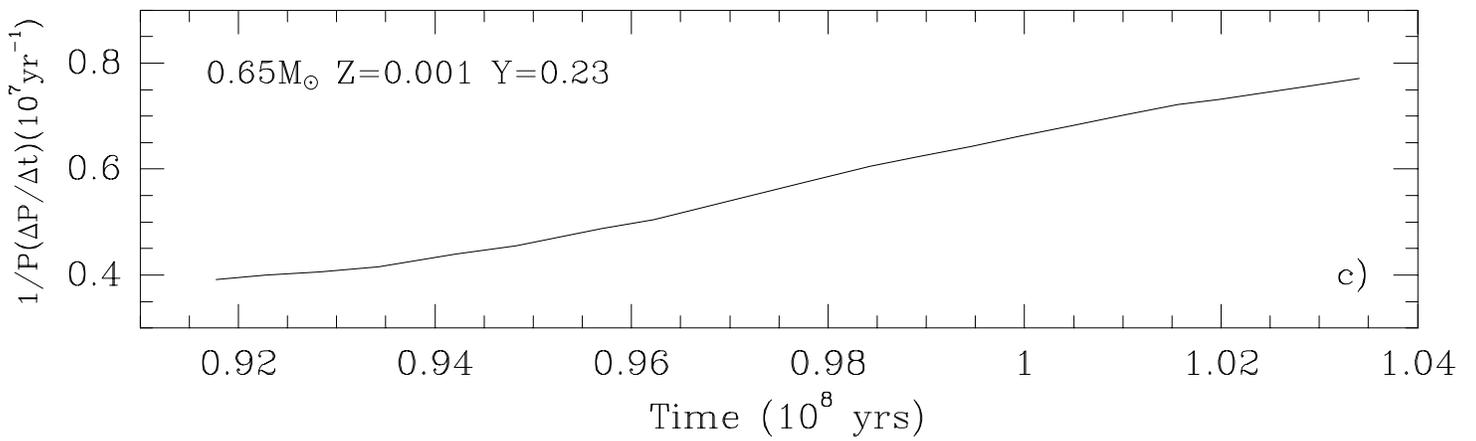
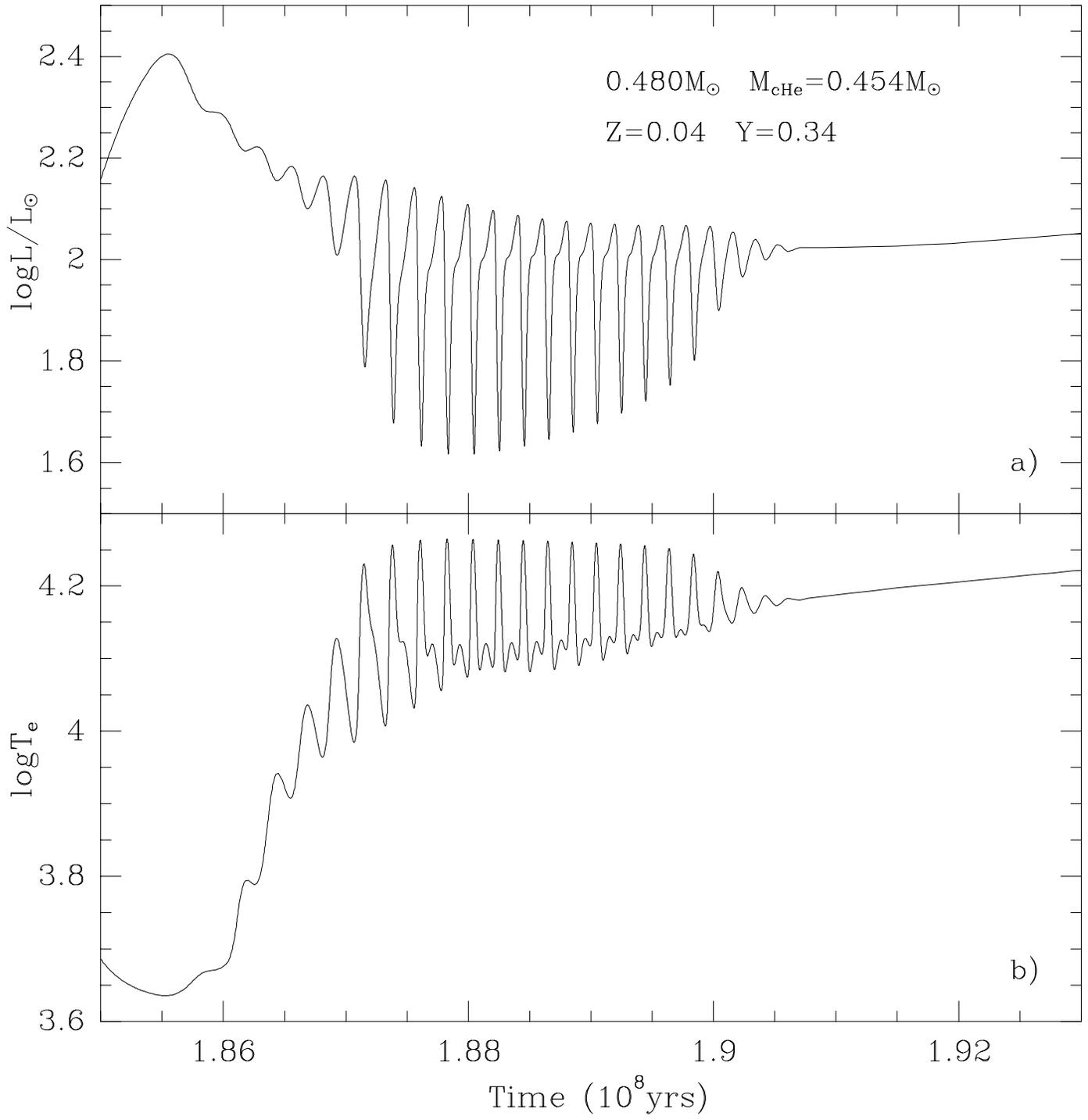
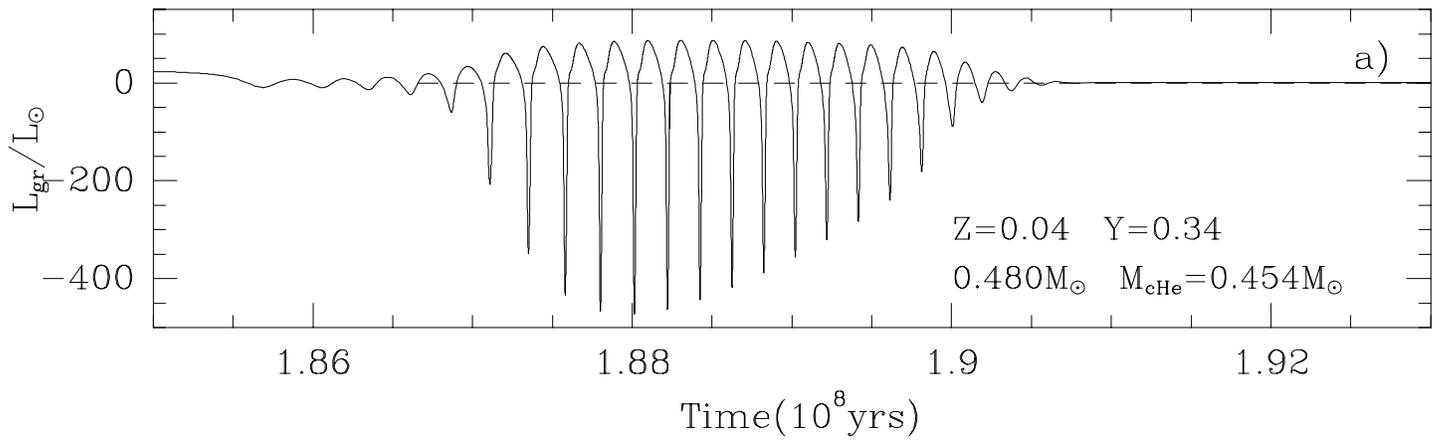
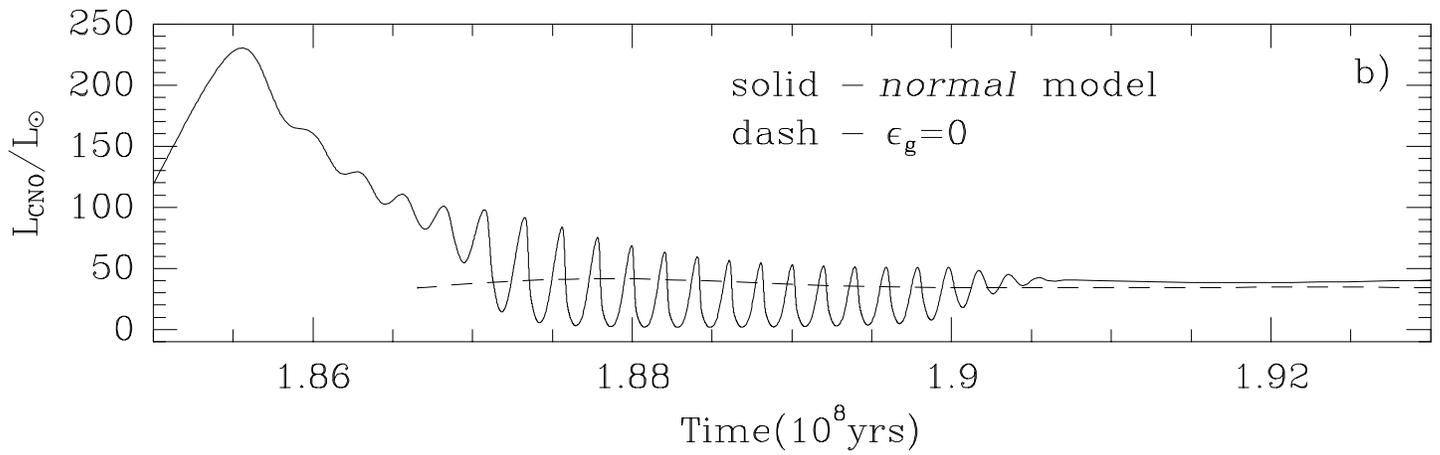
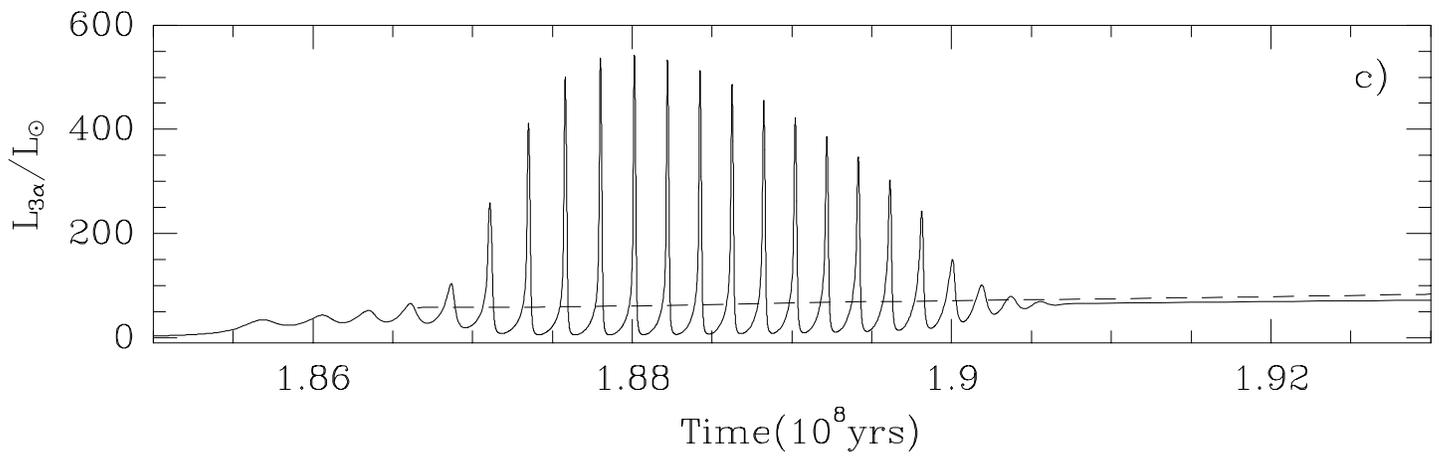
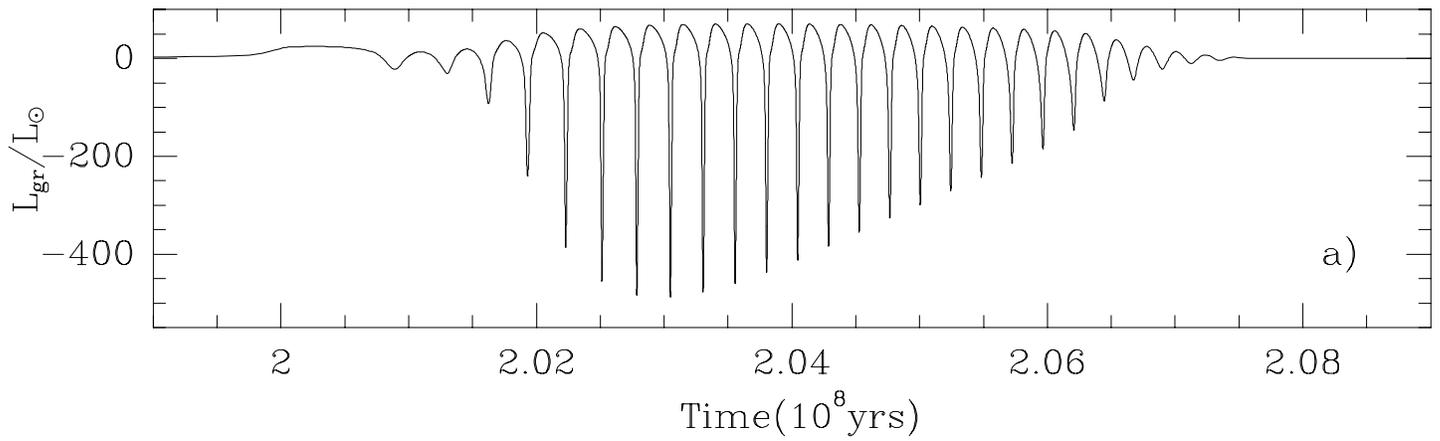
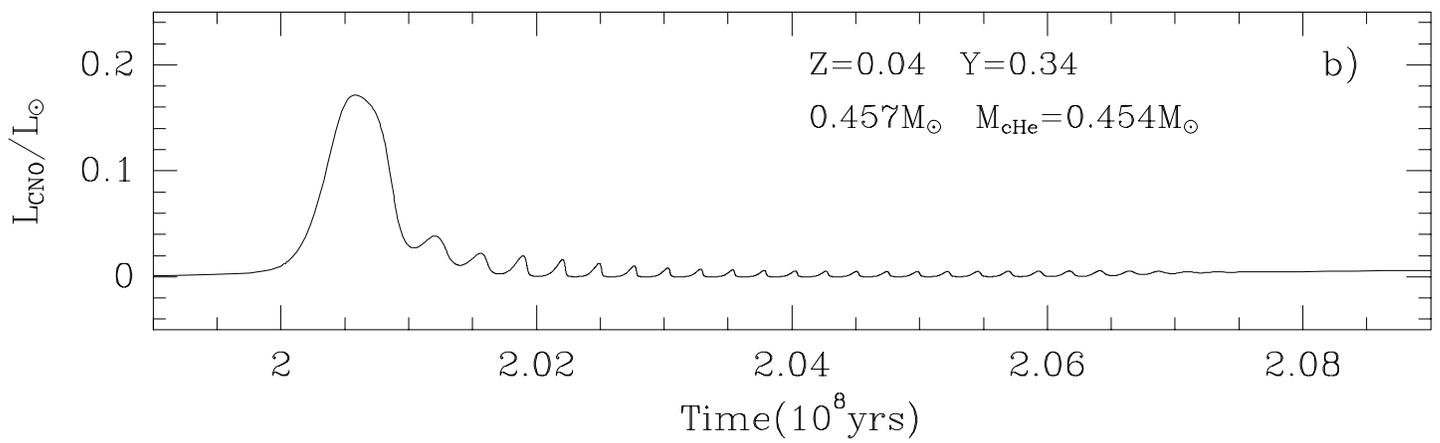
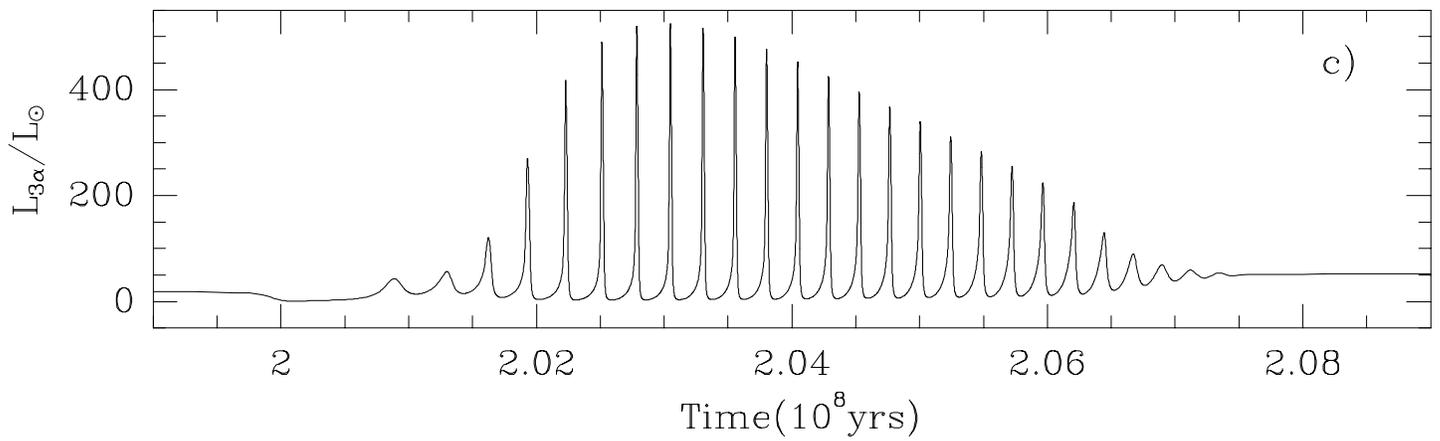
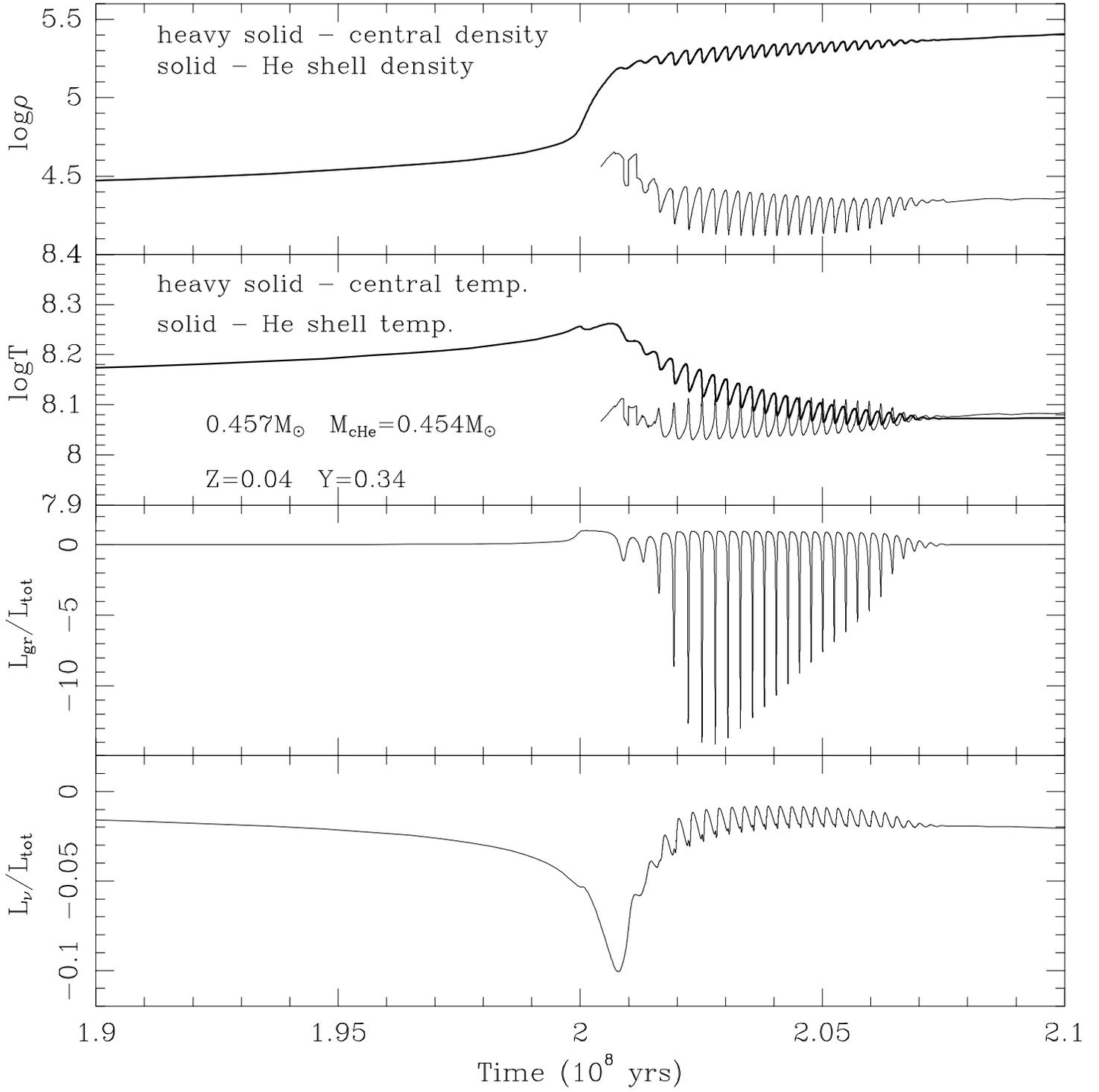
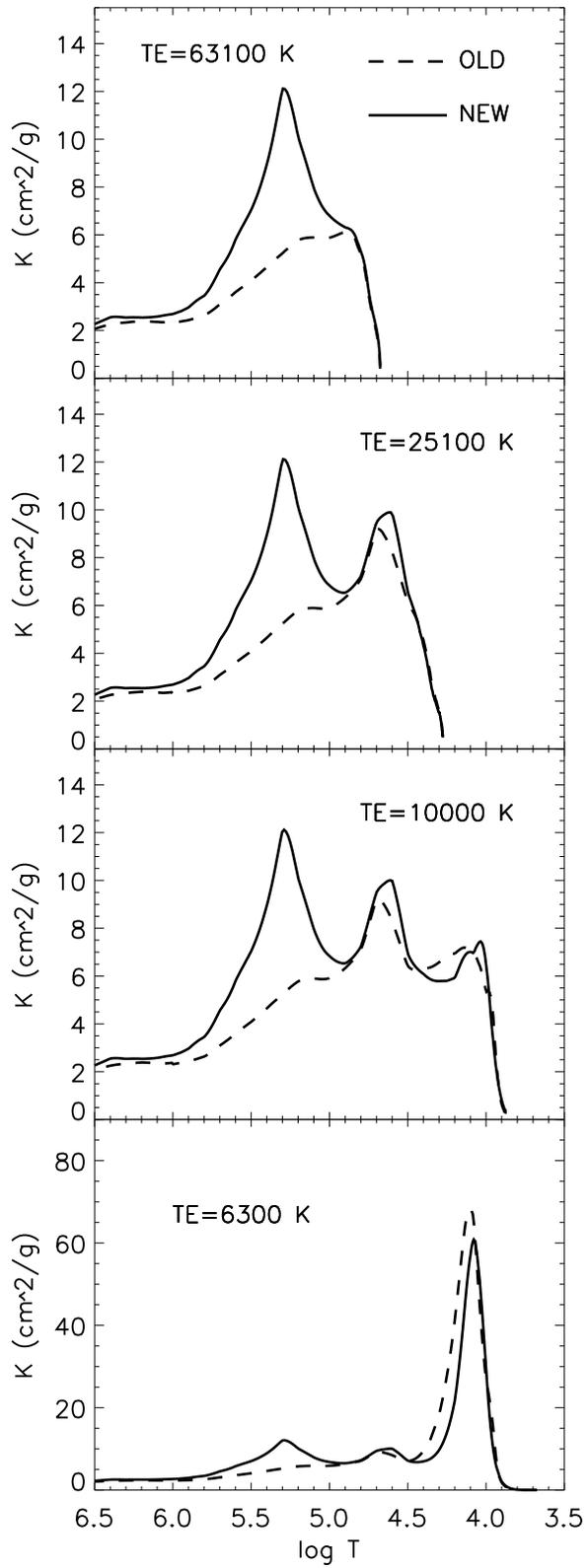
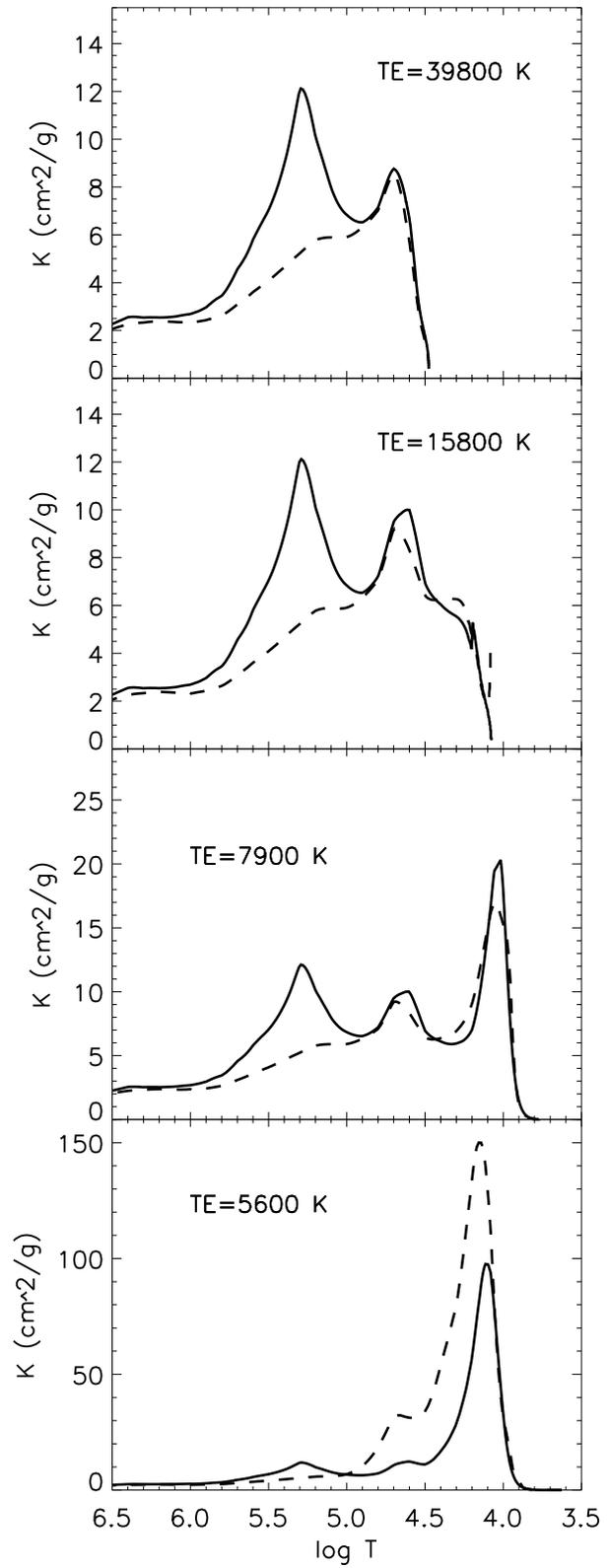
\end{document}



\begin{deluxetable}{ccccc}
\tablecaption{Selected theoretical evolutionary quantities at fixed 
metallicity (Z=0.04) and two different Helium contents.}\label{tbl-1}
\tablehead{
\colhead{$M/M_\odot$\tablenotemark{a}}&
\colhead{$\log(L/L_\odot)_{tip}$\tablenotemark{b}}& 
\colhead{$\log{t_{He}}$\tablenotemark{c}}&
\colhead{$M_{cHe}$\tablenotemark{d}}&
\colhead{${\Delta}Y_{du}$\tablenotemark{e}} }
\startdata
     \multicolumn{5}{c}{Y=0.34}\nl 
        &         &            &          &	 \nl
  0.7	&   3.449 & 	10.668 &   0.469 & 0.014 \nl
  0.8	&   3.448 & 	10.444 &   0.468 & 0.017 \nl
  1.0   &   3.445 &	10.072 &   0.466 & 0.019 \nl	
  1.2	&   3.442 &	 9.775 &   0.466 & 0.015 \nl
  1.4   &   3.441 &      9.533 &   0.465 & 0.011 \nl
  1.5	&   3.440 &	 9.427 &   0.465 & 0.010 \nl
  1.8   &   3.376 &      9.147 &   0.454 & 0.007 \nl
  2.0	&   3.186 &	 8.994 &   0.422 & 0.005 \nl
  2.2   &   2.788 &      8.861 &   0.372 & 0.007 \nl
  2.3   &   2.370 &      8.792 &   0.328 & 0.007 \nl
  2.4   &   2.445 &      8.736 &   0.340 & 0.007 \nl
  2.5   &   2.453 &      8.682 &   0.341 & 0.008 \nl
  3.0   &   2.571 &      8.449 &   0.393 & 0.010 \nl
        &         &            &          &	 \nl
     \multicolumn{5}{c}{Y=0.37}                  \nl 
        &         &            &          &	 \nl
  0.8	&   3.442 & 	10.341 &   0.463 & 0.016 \nl
  1.0   &   3.437 &	 9.970 &   0.461 & 0.016 \nl	
  1.2	&   3.435 &	 9.679 &   0.461 & 0.012 \nl
  1.5	&   3.428 &	 9.334 &   0.460 & 0.008 \nl
  1.8   &   3.313 &      9.059 &   0.439 & 0.005 \nl
  2.0	&   3.033 &	 8.909 &   0.399 & 0.005 \nl
  2.1	&   2.778 &	 8.840 &   0.391 & 0.005 \nl
  2.2   &   2.375 &      8.770 &   0.346 & 0.005 \nl
  2.3   &   2.443 &      8.711 &   0.340 & 0.006 \nl
  2.4   &   2.447 &      8.655 &   0.345 & 0.006 \nl
  2.5   &   2.459 &      8.601 &   0.353 & 0.007 \nl
  2.8   &   2.536 &      8.457 &   0.388 & 0.008 \nl
  3.0   &   2.606 &      8.370 &   0.415 & 0.008 \nl
\enddata
\tablenotetext{a}{ Stellar mass value (solar units).  
\hspace*{0.5mm} $^b$ Logarithmic luminosity (solar units) at the RGB tip.  
\hspace*{0.5mm} $^c$ Logarithmic age (yrs) at the RGB tip.  
\hspace*{0.5mm} $^d$ He core mass values (solar units) at the He ignition.
\hspace*{0.5mm} $^e$ Amount of extra helium dredged up during RGB evolution.}  
\end{deluxetable}

\clearpage
\begin{deluxetable}{ccccccc}
\tablecaption{Selected evolutionary data concerning the isochrones for Z=0.04.}\label{tbl-2}
\tablehead{
\colhead{Age (Gyr)}&
\colhead{$M/M_{\odot}$\tablenotemark{a}}& 
\colhead{$\log(L/L_\odot)$\tablenotemark{b}}&
\colhead{$\log{T_e}$\tablenotemark{c}}& 
\colhead{$M_V$\tablenotemark{d}}& 
\colhead{$(B-V)$\tablenotemark{e}}& 
\colhead{${\Delta}V^{TO}_{HB}$\tablenotemark{f}} }
\startdata
     \multicolumn{7}{c}{Y=0.34}\nl 
     &       &        &       &	      &       &     \nl
 1.0 & 1.908 &  1.344 & 3.875 & 1.429 & 0.266 & 0.15\nl
 2.0 & 1.520 &  0.929 & 3.819 & 2.517 & 0.475 & 1.46\nl
 3.0 & 1.343 &  0.695 & 3.794 & 3.131 & 0.573 & 2.09\nl
 4.0 & 1.220 &  0.488 & 3.784 & 3.664 & 0.615 & 2.64\nl
 5.0 & 1.147 &  0.376 & 3.776 & 3.959 & 0.647 & 2.96\nl
 6.0 & 1.093 &  0.287 & 3.768 & 4.196 & 0.679 & 3.21\nl
 7.0 & 1.058 &  0.244 & 3.761 & 4.317 & 0.707 & 3.35\nl
 8.0 & 1.030 &  0.206 & 3.756 & 4.425 & 0.728 & 3.48\nl
 9.0 & 1.005 &  0.175 & 3.752 & 4.511 & 0.745 & 3.59\nl
10.0 & 0.985 &  0.160 & 3.748 & 4.560 & 0.763 & 3.65\nl
11.0 & 0.963 &  0.131 & 3.744 & 4.644 & 0.779 & 3.76\nl
12.0 & 0.944 &  0.104 & 3.741 & 4.721 & 0.795 & 3.85\nl
13.0 & 0.927 &  0.080 & 3.737 & 4.792 & 0.808 & 3.92\nl
14.0 & 0.910 &  0.050 & 3.734 & 4.877 & 0.821 & 4.01\nl
15.0 & 0.896 &  0.029 & 3.731 & 4.938 & 0.833 & 4.07\nl
16.0 & 0.882 &  0.010 & 3.729 & 4.996 & 0.844 & 4.13\nl
17.0 & 0.870 & -0.009 & 3.726 & 5.051 & 0.854 & 4.18\nl
18.0 & 0.859 & -0.026 & 3.724 & 5.103 & 0.863 & 4.23\nl
     &       &        &       &	      &       &     \nl
     \multicolumn{7}{c}{Y=0.37}                     \nl 
     &       &        &       &	      &       &     \nl
 1.0 & 1.789 &  1.300 & 3.876 & 1.539 & 0.262 & 0.59\nl
 2.0 & 1.435 &  0.898 & 3.820 & 2.593 & 0.470 & 1.80\nl
 3.0 & 1.255 &  0.643 & 3.802 & 3.250 & 0.543 & 2.46\nl
 4.0 & 1.138 &  0.439 & 3.787 & 3.782 & 0.601 & 3.00\nl
 5.0 & 1.074 &  0.329 & 3.775 & 4.078 & 0.650 & 3.30\nl
 6.0 & 1.039 &  0.290 & 3.769 & 4.186 & 0.675 & 3.42\nl
 7.0 & 1.000 &  0.226 & 3.765 & 4.356 & 0.694 & 3.60\nl
 8.0 & 0.979 &  0.223 & 3.750 & 4.375 & 0.714 & 3.62\nl
 9.0 & 0.952 &  0.185 & 3.755 & 4.479 & 0.733 & 3.73\nl
10.0 & 0.927 &  0.142 & 3.751 & 4.597 & 0.751 & 3.85\nl
11.0 & 0.907 &  0.113 & 3.747 & 4.680 & 0.767 & 3.94\nl
12.0 & 0.890 &  0.086 & 3.744 & 4.756 & 0.781 & 4.01\nl
13.0 & 0.873 &  0.062 & 3.741 & 4.826 & 0.794 & 4.08\nl
14.0 & 0.858 &  0.032 & 3.738 & 4.910 & 0.806 & 4.17\nl
15.0 & 0.844 &  0.012 & 3.735 & 4.970 & 0.817 & 4.23\nl
16.0 & 0.831 & -0.014 & 3.733 & 5.041 & 0.828 & 4.30\nl
17.0 & 0.819 & -0.037 & 3.730 & 5.107 & 0.837 & 4.37\nl
18.0 & 0.809 & -0.053 & 3.728 & 5.155 & 0.846 & 4.41\nl
\enddata
\tablenotetext{a}{ Mass value at the TO point (solar units).
\hspace*{0.5mm} $^b$ Logarithmic luminosity of the TO point (solar units).
\hspace*{0.5mm} $^c$ Logarithmic effective temperature (K) of the TO point.
\hspace*{0.5mm} $^d$ V magnitude of the TO point (mag).
\hspace*{0.5mm} $^e$ B-V color of the TO point (mag).
\hspace*{0.5mm} $^f$ V Magnitude difference between the TO point and the ZAHB 
at $\log T_e=3.85$.}
\end{deluxetable}

\begin{deluxetable}{cccccc}
\tablecaption{Selected evolutionary parameters for the He burning phase for 
different assumptions on initial Helium abundance and RG progenitor masses.}\label{tbl-3}
\tablehead{
\colhead{$M_{pr}$\tablenotemark{a}}&
\colhead{$Y_{HB}$\tablenotemark{b}}&
\colhead{$M_{cHe}$\tablenotemark{c}}&
\colhead{$M_{tot}$\tablenotemark{d}}&
\colhead{$M^{AGB}$\tablenotemark{e}}& 
\colhead{$M^{TP}$\tablenotemark{f}} }
\startdata
     \multicolumn{6}{c}{$Y_{ZAMS}$=0.34}\nl 
        &        &       &	      &       &                          \nl
 1.0& 0.359& 0.466& 0.468, 0.470, 0.472, 0.475, 0.480, 0.482& 0.483& 0.54\nl
    &      &      & 0.485, 0.490, 0.495, 0.500, 0.510, 0.520&      &     \nl 
    &      &      & 0.530, 0.550, 0.600, 0.650, 0.700, 0.752&      &     \nl 
    &      &      & 0.800, 0.850, 0.900, 0.950, 1.000\hspace*{9.9mm}&    &  \nl 
 1.8& 0.347& 0.454& 0.457, 0.460, 0.465, 0.470, 0.475, 0.480& 0.473& 0.54\nl
    &      &      & 0.485, 0.490, 0.495, 0.500, 0.510, 0.520&      &     \nl 
    &      &      & 0.530, 0.540, 0.550, 0.600, 0.700, 0.800&      &     \nl 
    &      &      & 0.900, 1.000, 1.200, 1.400, 1.600, 1.800&      &     \nl 
 2.0& 0.345& 0.422& 0.425, 0.430, 0.440, 0.450, 0.470, 0.500& 0.436& 0.54\nl
    &      &      & 0.520, 0.550, 0.600, 0.664, 0.700, 0.750&      &     \nl 
    &      &      & 0.800, 0.850, 0.900, 1.000, 1.200, 1.400&      &     \nl 
    &      &      & 1.600, 1.800, 2.000\hspace*{30.0mm}      &      &    \nl 
        &        &       &	      &       &                          \nl
     \multicolumn{6}{c}{$Y_{ZAMS}$=0.37}                                 \nl 
        &        &       &	      &       &                          \nl
 1.0& 0.386& 0.461& 0.463, 0.466, 0.470, 0.475, 0.480, 0.485& 0.477& 0.55\nl
    &      &      & 0.490, 0.495, 0.500, 0.510, 0.520, 0.530&      &     \nl 
    &      &      & 0.540, 0.550, 0.570, 0.580, 0.600, 0.620&      &     \nl 
    &      &      & 0.650, 0.700, 0.750, 0.800, 0.850, 0.900&      &     \nl 
    &      &      & 0.950, 1.000\hspace*{40.0mm}            &      &     \nl 
 1.5& 0.378& 0.460& 0.463, 0.465, 0.475, 0.480, 0.490, 0.495& 0.476& 0.55\nl
    &      &      & 0.500, 0.510, 0.520, 0.530, 0.540, 0.550&      &     \nl 
    &      &      & 0.570, 0.600, 0.650, 0.700, 0.750, 0.800&      &     \nl 
    &      &      & 0.850, 0.900, 0.950, 1.000, 1.100, 1.200&      &     \nl 
    &      &      & 1.300, 1.400, 1.500\hspace*{30.0mm}     &      &     \nl 
 1.8& 0.375& 0.439& 0.450, 0.460, 0.465, 0.470, 0.480, 0.490& 0.457& 0.55\nl
    &      &      & 0.500, 0.520, 0.530, 0.540, 0.550, 0.580&      &     \nl 
    &      &      & 0.600, 0.650, 0.700, 0.800, 0.900, 1.000&      &     \nl 
    &      &      & 1.200, 1.400, 1.600, 1.800\hspace*{20.0mm}&      &     \nl 
\enddata
\tablenotetext{a}{ Stellar mass value for the RG progenitor (solar units).  
\hspace*{0.5mm} $^b$ Helium abundances adopted for HB models.  
\hspace*{0.5mm} $^c$ Evolutionary He core mass values (solar units) at the He ignition. 
\hspace*{0.5mm} $^d$ Total stellar mass values adopted for HB models (solar units). 
\hspace*{0.5mm} $^e$ Transition mass between AGB-manqu\'e models and models 
which approach the AGB (solar units). 
\hspace*{0.5mm} $^f$ Mass value for the most massive structures which do not 
experience thermal pulses along the AGB (solar units).} 
\end{deluxetable}


\clearpage 
\begin{deluxetable}{cccccccc}
\tablecaption{Linear Nonadiabatic Pulsation Models for $M=0.485M_{\odot}$}
\label{tbl-4} 
\tablehead{
\colhead{$T_e$\tablenotemark{a}}&
\colhead{$\log{(g)}$\tablenotemark{b}}& 
\colhead{$P(F)$\tablenotemark{c}}& 
\colhead{$Q(F)$\tablenotemark{d}}&
\colhead{$\eta(F)$\tablenotemark{e}}& 
\colhead{$P(FO)$\tablenotemark{f}}&  
\colhead{$Q(FO)$\tablenotemark{g}}&
\colhead{$\eta(FO)$\tablenotemark{h}} }  
\startdata
     \multicolumn{8}{c}{\lsun =1.56}\nl 
     &     &       &       &      &	&	&         \nl
7400 & 3.00& .3820& .0382& $-.0097$& .2680& .0268& $.0035$ \nl
7300 & 2.98& .3995& .0383& $-.0080$& .2805& .0269& $.0120$ \nl
7100 & 2.93& .4380& .0386& $-.0041$& .3081& .0272& $.0280$ \nl
7000 & 2.90& .4590& .0388& $-.0020$& .3233& .0273& $.0351$ \nl
6900 & 2.88& .4815& .0390& $.0003$& .3396& .0275& $.0414$ \nl
6700 & 2.83& .5309& .0394& $.0051$& .3754& .0278& $.0506$ \nl
6500 & 2.77& .5871& .0398& $.0101$& .4164& .0282& $.0547$ \nl
6300 & 2.72& .6515& .0402& $.0151$& .4632& .0286& $.0536$ \nl
6100 & 2.66& .7254& .0406& $.0199$& .5169& .0289& $.0471$ \nl
6000 & 2.64& .7665& .0408& $.0222$& .5466& .0291& $.0418$ \nl
5900 & 2.61& .8107& .0411& $.0243$& .5785& .0293& $.0354$ \nl
     &     &       &       &      &	&	&         \nl
     \multicolumn{8}{c}{\lsun =2.0}                       \nl 
     &     &       &       &      &	&	&         \nl
6500 & 2.33& 1.3789& .0437& $-.0073$& .9696& .0307&  $.0023$  \nl
6400 & 2.31& 1.4553& .0440& $.0064$& 1.0218& .0309&  $.0081$   \nl
6300 & 2.28& 1.5375& .0443& $.0204$& 1.0774& .0311&  $.0115$   \nl
6100 & 2.22& 1.7214& .0451& $.0494$& 1.1993& .0314&  $.0116$   \nl
5900 & 2.17& 1.9359& .0458& $.0789$& 1.3376& .0317&  $.0030$   \nl
5700 & 2.11& 2.1878& .0467& $.1079$& 1.4947& .0319&  $-.0135$   \nl
     &     &       &       &      &	&	&              \nl
     \multicolumn{8}{c}{\lsun =2.2}                            \nl 
     &     &       &       &      &	&	&              \nl
6100 & 2.02& 2.5758& .0477& $.0242$ & 1.7445 & .0323 & $-.1014$ \nl
6000 & 2.00& 2.7345& .0482& $.0506$ & 1.8389 & .0324 & $-.1023$ \nl
5800 & 1.94& 3.0929& .0493& $.1042$ & 2.0462 & .0326 & $-.1106$ \nl
5600 & 1.88& 3.5164& .0504& $.1576$ & 2.2812 & .0327 & $-.1272$ \nl
5400 & 1.81& 4.0212& .0517& $.2084$ & 2.5486 & .0328 & $-.1514$ \nl
\enddata
\tablenotetext{a}{ Effective temperature (K).
\hspace*{0.5mm} $^b$ Logarithmic static gravity. 
\hspace*{0.5mm} $^c$ Fundamental period (days).  
\hspace*{0.5mm} $^d$ Fundamental pulsation constant (days).
\hspace*{0.5mm} $^e$ Fundamental growth rate.
\hspace*{0.5mm} $^f$ First overtone period (days).  
\hspace*{0.5mm} $^g$ First overtone pulsation constant (days).   
\hspace*{0.5mm} $^h$ First overtone growth rate.}
\end{deluxetable}

\clearpage 
\begin{deluxetable}{cccccccc}
\tablecaption{Linear Nonadiabatic Pulsation Models for $M=0.505M_{\odot}$}
\label{tbl-5} 
\tablehead{
\colhead{$T_e$\tablenotemark{a}}&
\colhead{$\log{(g)}$\tablenotemark{b}}& 
\colhead{$P(F)$\tablenotemark{c}}& 
\colhead{$Q(F)$\tablenotemark{d}}&
\colhead{$\eta(F)$\tablenotemark{e}}& 
\colhead{$P(FO)$\tablenotemark{f}}&  
\colhead{$Q(FO)$\tablenotemark{g}}&
\colhead{$\eta(FO)$\tablenotemark{h}} }  
\startdata
     \multicolumn{8}{c}{\lsun =1.48}\nl 
     &     &       &       &      &	&	&       \nl
7500 & 3.12& .3074& .0373& $-.0060$ & .2164& .0263 & $.0018$ \nl
7300 & 3.07& .3359& .0376& $-.0039$ & .2366& .0265 & $.0136$ \nl
7100 & 3.02& .3681& .0379& $-.0015$ & .2594& .0267 & $.0250$ \nl
7000 & 3.00& .3857& .0381& $-.0001$ & .2720& .0269 & $.0300$ \nl
6900 & 2.97& .4045& .0383& $.0013$ & .2854& .0270 & $.0345$ \nl
6800 & 2.95& .4245& .0384& $.0028$ & .2997& .0271&  $.0382$ \nl	
6700 & 2.92& .4458& .0386& $.0043$ & .3151& .0273 & $.0411$ \nl
6600 & 2.90& .4685& .0388& $.0058$ & .3315& .0274 & $.0429$ \nl
6100 & 2.76& .6077& .0397& $.0129$ & .4326& .0283 & $.0364$ \nl
5900 & 2.70& .6786& .0401& $.0151$ & .4841& .0286 & $.0263$ \nl
     &     &       &       &      &	&	&           \nl
     \multicolumn{8}{c}{\lsun =1.56}\nl 
     &     &       &       &      &	&	&       \nl
7400 & 3.02& .3724& .0380& $-.0084$ & .2614& .0266 & $.0038$ \nl
7000 & 2.92& .4475& .0386& $-.0017$ & .3150& .0272 & $.0330$ \nl
6900 & 2.90& .4693& .0388& $.0003$ & .3308& .0273 & $.0389$ \nl
6800 & 2.87& .4925& .0390& $.0024$ & .3476& .0275 & $.0438$ \nl
6600 & 2.82& .5438& .0393& $.0067$ & .3848& .0278 & $.0503$ \nl
6400 & 2.76& .6023& .0397& $.0110$ & .4274& .0282 & $.0519$ \nl
6200 & 2.71& .6692& .0401& $.0154$ & .4761& .0285 & $.0485$ \nl
6000 & 2.65& .7463& .0406& $.0193$ & .5320& .0289 & $.0401$ \nl
     &     &       &       &      &	&	&           \nl
     \multicolumn{8}{c}{\lsun =2.0}                         \nl 
     &     &       &       &      &	&	&       \nl
6600 & 2.38& 1.2730& .0431& $-.0171$ & .8968& .0303& $.0037$  \nl
6500 & 2.35& 1.3418& .0434& $-.0053$ & .9448& .0305& $.0119$  \nl
6400 & 2.32& 1.4158& .0437& $.0071$ & .9958&  .0307& $.0178$  \nl 
6300 & 2.30& 1.4954& .0440& $.0198$ & 1.0502& .0309& $.0213$   \nl
6100 & 2.24& 1.6733& .0447& $.0461$ & 1.1697& .0312& $.0215$   \nl
5900 & 2.18& 1.8807& .0454& $.0730$ & 1.3053& .0315& $.0133$   \nl
5700 & 2.12& 2.1240& .0463& $.0993$ & 1.4595& .0318& $-.0029$   \nl
5600 & 2.09& 2.2617& .0467& $.1117$ & 1.5445& .0319& $-.0140$  \nl
5400 & 2.03& 2.5755& .0477& $.1334$ & 1.7321& .0321& $-.0409$  \nl
5200 & 1.96& 2.9509& .0488& $.1489$ & 1.9469& .0322& $-.0732$  \nl
5000 & 1.90& 3.4036& .0501& $.1554$ & 2.1936& .0323& $-.1081$  \nl
4900 & 1.86& 3.6647& .0507& $.1548$ & 2.3305& .0322& $-.1248$  \nl
     &     &       &       &      &	&	&              \nl
     \multicolumn{8}{c}{\lsun =2.2}                            \nl 
     &     &       &       &      &	&	&              \nl
6200 & 2.07& 2.3596& .0468& $.0026$& 1.6160& .0321& $-.0811$ \nl
6000 & 2.01& 2.6547& .0478& $.0508$& 1.7963& .0323& $-.0804$ \nl
5800 & 1.95& 3.0008& .0488& $.1003$& 2.0001& .0325& $-.0885$ \nl
5700 & 1.92& 3.1950& .0493& $.1257$& 2.1101& .0325& $-.0966$ \nl 
5500 & 1.86& 3.6406& .0504& $.1741$& 2.3562& .0327& $-.1172$ \nl 
5450 & 1.85& 3.7667& .0508& $.2389$& 2.4254& .0327& $-.0292$ \nl 
\enddata
\tablenotetext{a}{ Effective temperature (K).
\hspace*{0.5mm} $^b$ Logarithmic static gravity. 
\hspace*{0.5mm} $^c$ Fundamental period (days).  
\hspace*{0.5mm} $^d$ Fundamental pulsation constant (days).
\hspace*{0.5mm} $^e$ Fundamental growth rate.
\hspace*{0.5mm} $^f$ First overtone period (days).  
\hspace*{0.5mm} $^g$ First overtone pulsation constant (days).   
\hspace*{0.5mm} $^h$ First overtone growth rate.}
\end{deluxetable}

\clearpage 
\begin{deluxetable}{cccccccc}
\tablecaption{Linear Nonadiabatic Pulsation Models for $M=0.520M_{\odot}$}
\label{tbl-6} 
\tablehead{
\colhead{$T_e$\tablenotemark{a}}&
\colhead{$\log{(g)}$\tablenotemark{b}}& 
\colhead{$P(F)$\tablenotemark{c}}& 
\colhead{$Q(F)$\tablenotemark{d}}&
\colhead{$\eta(F)$\tablenotemark{e}}& 
\colhead{$P(FO)$\tablenotemark{f}}&  
\colhead{$Q(FO)$\tablenotemark{g}}&
\colhead{$\eta(FO)$\tablenotemark{h}} }  
\startdata
     \multicolumn{8}{c}{\lsun =1.56}\nl 
     &     &       &       &      &	&	&       \nl
7400 & 3.03& .3657& .0378& $-.0076$ & .2568& .0266 & $.0039$ \nl
7300 & 3.01& .3824& .0380& $-.0062$ & .2686& .0267 & $.0112$ \nl
7100 & 2.96& .4191& .0383& $-.0031$ & .2948& .0269 & $.0251$ \nl
6900 & 2.91& .4607& .0386& $.0003$ & .3246& .0272 & $.0369$ \nl
6700 & 2.86& .5078& .0390& $.0041$ & .3587& .0275 & $.0454$ \nl
6500 & 2.80& .5614& .0394& $.0081$ & .3976& .0279 & $.0495$ \nl
6300 & 2.75& .6227& .0397& $.0120$ & .4422& .0282 & $.0488$ \nl
6100 & 2.69& .6930& .0402& $.0157$ & .4933& .0286 & $.0432$ \nl
6000 & 2.67& .7321& .0404& $.0175$ & .5216& .0288 & $.0387$ \nl
     &     &       &       &      &	&	&           \nl
     \multicolumn{8}{c}{\lsun =2.0}                         \nl 
     &     &       &       &      &	&	&       \nl
6600 & 2.39& 1.2484& .0428& $-.0149$& .8799& .0302& $.0097$ \nl
6500 & 2.36& 1.3157& .0431& $-.0040$& .9271& .0304& $.0178$ \nl
6400 & 2.34& 1.3881& .0434& $.0075$& .9773& .0306& $.0237$ \nl
6200 & 2.28& 1.5494& .0441& $.0314$& 1.0878& .0310& $.0284$ \nl
6000 & 2.23& 1.7367& .0448& $.0562$& 1.2132& .0313& $.0246$ \nl
5900 & 2.20& 1.8419& .0452& $.0688$& 1.2822& .0314& $.0196$ \nl
5700 & 2.14& 2.0791& .0460& $.0933$& 1.4342& .0317& $.0037$ \nl
5650 & 2.12& 2.1449& .0462& $.0991$& 1.4755& .0318& $-.0016$ \nl
     &     &       &       &      &	&	&            \nl
     \multicolumn{8}{c}{\lsun =2.2}                          \nl 
     &     &       &       &      &	&	&       \nl
6200 & 2.08& 2.3108& .0465& $.0050$& 1.5878&  .0465& $-.0666$ \nl
6100 & 2.05& 2.4491& .0470& $.0275$& 1.6741&  .0470& $-.0649$ \nl
6000 & 2.03& 2.5986& .0474& $.0504$& 1.7658&  .0474& $-.0657$ \nl
5800 & 1.97& 2.9360& .0484& $.0972$& 1.9670&  .0484& $-.0736$ \nl
5600 & 1.91& 3.3346& .0495& $.1438$& 2.1954&  .0495& $-.0902$ \nl
5450 & 1.86& 3.6825& .0504& $.1771$& 2.3873&  .0504& $-.1077$ \nl
\enddata
\tablenotetext{a}{ Effective temperature (K).
\hspace*{0.5mm} $^b$ Logarithmic static gravity. 
\hspace*{0.5mm} $^c$ Fundamental period (days).  
\hspace*{0.5mm} $^d$ Fundamental pulsation constant (days).
\hspace*{0.5mm} $^e$ Fundamental growth rate.
\hspace*{0.5mm} $^f$ First overtone period (days).  
\hspace*{0.5mm} $^g$ First overtone pulsation constant (days).   
\hspace*{0.5mm} $^h$ First overtone growth rate.}
\end{deluxetable}

\clearpage 
\begin{deluxetable}{ccccccccc}
\tablecaption{Nonlinear Survey for $M=0.505M_{\odot}$}
\label{tbl-7} 
\tablehead{
\colhead{$T_e$\tablenotemark{a}}&
\colhead{$P$\tablenotemark{b}}& 
\colhead{$\Delta R/R_{ph}$\tablenotemark{c}}& 
\colhead{$\Delta u$\tablenotemark{d}}&
\colhead{$\Delta M_{bol}$\tablenotemark{e}}& 
\colhead{$\Delta {\log{g_s}}$\tablenotemark{f}}&  
\colhead{$\Delta {\log{g_{eff}}}$\tablenotemark{g}}&
\colhead{$\Delta T$\tablenotemark{h}}&
\colhead{$\Delta T_{e}$\tablenotemark{i}} }  
\startdata
     \multicolumn{9}{c}{\lsun =1.48\hspace*{1.25cm} First Overtone} \nl 
     &     &       &       &      &	&	&      &     \nl
7000 & .2712& .036& 25.94& .310&  .03& .32& 450& 550 \nl
6900 &  .2847&  .043&	 31.44&	 .345&  .04&  .40&  450&  600	\nl
6700 & .3138& .040& 28.25& .265& .03& .36& 350& 400 \nl
     &     &       &       &      &	&	&      &     \nl
     \multicolumn{9}{c}{\lsun =1.48\hspace*{1.25cm} Fundamental} \nl 
     &     &       &       &      &	&	&      &     \nl
6800 & .4246& .120& 69.12& 1.023&  .11& .71& 1350& 1650 \nl
6700 & .4460& .126& 69.63& .971& .11& .79& 1250& 1500 \nl
6600 & .4681& .036& 24.25& .214& .03& .32& 300& 350 \nl
6100 & .6068& .090& 39.18& .334& .08& .46& 450& 550 \nl
6000 & .6405& .069& 28.51& .218& .06& .33& 300& 400\nl
     &     &       &       &      &	&	&      &     \nl
     \multicolumn{9}{c}{\lsun =2.0\hspace*{1.275cm} Fundamental}\nl 
     &     &       &       &      &	&	&      &     \nl
6400 & 1.4164& .098& 32.59& .571&  .08& .40& 750& 900 \nl
6300 & 1.4997& .156& 52.49& 1.033& .13& .69& 1550& 1750 \nl
6200 & 1.5959& .300& 104.75& 1.545& .25& 1.57& 2100& 2550 \nl
6100 & 1.6910& .236& 75.26& 1.564& .20& 1.34& 2050& 2500 \nl
5900 & 1.9077& .293& 95.52& 1.388& .24& 1.30& 1750& 2150 \nl 
5800 & 2.0258& .276& 83.15& 1.361& .24& 1.34& 1500& 1800 \nl
5700~\tablenotemark{j} & 2.1392& .312& 87.85& 1.130& .28& 1.41& 1400& 1750\nl
5600 & 2.2761& .292& 88.67& 1.103& .25& 1.29& 1250& 1550 \nl
5500 & 2.4144& .280& 82.21& .968& .24& 1.26& 1150& 1450 \nl
5400 & 2.5701& .254& 62.05& .943&  .22& 1.28& 1200& 1450 \nl  
5200 & 2.9040& .240& 52.96& .742&  .21& 1.21& 1000& 1250 \nl
5000 & 3.2861& .207& 41.99& .559& .18& .96& 750& 950 \nl
4900 & 3.5000& .155& 30.31& .385& .14& .54& 550& 650 \nl
     &     &       &       &      &	&	&      &     \nl
     \multicolumn{9}{c}{\lsun =2.2\hspace*{1.275cm} Fundamental}\nl 
     &     &       &       &      &	&	&      &     \nl
5700 & 3.2305    & .324      & 69.28      & 1.342     &	.28&1.51	& 1700     & 2100    \nl
5500 & 3.6369    & .313      & 66.60      & 1.194     &	.27&	1.54&    1300  & 1600    \nl
\enddata
\tablenotetext{a}{ Effective temperature (K).
\hspace*{0.5mm} $^b$ Period (days). 
\hspace*{0.5mm} $^c$ Fractional radius variation.  
\hspace*{0.5mm} $^d$ Radial velocity amplitude (Km/s).
\hspace*{0.5mm} $^e$ Bolometric amplitude (mag.).
\hspace*{0.5mm} $^f$ Amplitude of logarithmic static gravity.  
\hspace*{0.5mm} $^g$ Amplitude of logarithmic effective gravity.
\hspace*{0.5mm} $^h$ Surface temperature variation.   
\hspace*{0.5mm} $^i$ Effective temperature variation.
\hspace*{0.5mm} $^j$ Mixed-mode pulsator. The amplitudes referred to this 
model are the average values over the last twenty cycles.}
\end{deluxetable}